\pdfoutput=1
\documentclass{JINST}

\usepackage{lineno}

\title{Hadron shower decomposition in the highly granular CALICE analogue hadron calorimeter}

\author{\centering 
\LARGE\bf The CALICE Collaboration
}

\author{\centering
G.~Eigen
\\ \it
University of Bergen, 
Inst.~ of Physics, 
Allegaten 55, N-5007 Bergen, Norway
}

\author{\centering
T.\,Price, 
N.\,K.\,Watson 
\\ \it
University of Birmingham,
School of Physics and Astronomy,
Edgbaston, Birmingham B15 2TT, UK
}

\author{\centering 
J.\,S.\,Marshall,
M.\,A.\,Thomson, 
D.\,R.\,Ward
\\ \it
University of Cambridge, Cavendish Laboratory, J J Thomson Avenue, CB3 0HE, UK
}

\author{\centering 
D.\,Benchekroun, 
A.\,Hoummada, 
Y.\,Khoulaki
\\ \it
Universit\'{e} Hassan II A\"{\i}n Chock, Facult\'{e} des sciences.\, B.P. 5366 Maarif, Casablanca, Morocco
}

\author{\centering 
J.\,Apostolakis,
A.\,Dotti$^1$, %\footnote{Now at SLAC},
G.\,Folger, 
V.\,Ivantchenko,
A.\,Ribon, 
V.\,Uzhinskiy
\\ \it 
CERN, 1211 Gen\`{e}ve 23, Switzerland
}

\author{\centering 
J.\,-Y.\,Hostachy, 
L.\,Morin 
\\ \it
Laboratoire de Physique Subatomique et de Cosmologie - Universit\'{e}
Grenoble-Alpes, CNRS/IN2P3, Grenoble, France
}

\author{\centering 
E.\,Brianne,
A.\,Ebrahimi,
K.\,Gadow,
P.\,G\"{o}ttlicher,
C.\,G\"{u}nter,
O.\,Hartbrich,
B.\,Hermberg,
A.\,Irles,
F.\,Krivan,
K.\,Kr\"{u}ger,
J.\,Kvasnicka,
S.\,Lu,
B.\,Lutz,
V.\,Morgunov$^2$, %\thanks{Deceased}
C.\,Neub\"user,
A.\,Provenza,
M.\,Reinecke,
F.\,Sefkow,
S.\,Schuwalow,
H.L.\,Tran
\\ \it
DESY, Notkestrasse 85,
D-22603 Hamburg, Germany
}

\author{\centering
E.\,Garutti, 
S.\,Laurien, 
M.\,Matysek,    
M.\,Ramilli,
S.\,Schroeder
\\ \it
Univ. Hamburg,
Physics Department,
Institut f\"ur Experimentalphysik,
Luruper Chaussee 149,
22761 Hamburg, Germany
}

\author{\centering 
 K.\,Briggl, 
 P.\,Eckert,  
 Y.\,Munwes, 
 H.\,-Ch.\,Schultz-Coulon, 
 W.\,Shen, 
 R.\,Stamen
\\ \it
 University of Heidelberg, Fakultat f\"ur Physik und Astronomie,
Albert Uberle Str. 3-5 2.OG Ost,
D-69120 Heidelberg, Germany
}

\author{\centering
B.\,Bilki,
E.\,Norbeck$^2$,  %\thanks{Deceased},
D.\,Northacker,
Y.\,Onel
\\ \it
University of Iowa, Dept. of Physics and Astronomy,
203 Van Allen Hall, Iowa City, IA 52242-1479, USA
}

\author{\centering 
B.\,van\,Doren,
G.\,W.\,Wilson
\\ \it
University of Kansas, Department of Physics and Astronomy,
Malott Hall, 1251 Wescoe Hall Drive, Lawrence, KS 66045-7582, USA
}

\author{\centering
K.~Kawagoe,
H.~Hirai,
Y.~Sudo,
T.~Suehara,
H.~Sumida,
S.~Takada,
T.~Tomita,
T.~Yoshioka
\\ \it
Department of Physics, Kyushu University, Fukuoka 812-8581, Japan
}

\author{\centering 
M.\,Wing$^3$ %\thanks{Also at DESY and Univ. Hamburg}
\\ \it
Department of Physics and Astronomy, University College London,
Gower Street,
London WC1E 6BT, UK
}

\author{\centering 
A.\,Bonnevaux,
C.\,Combaret, 
L.\,Caponetto,
G.\,Grenier, 
R.\,Han, 
J.C.\,Ianigro,
R.\,Kieffer, 
I.\,Laktineh,  
N.\,Lumb, 
H.\,Mathez, 
L.\,Mirabito,
A.\,Steen
\\ \it
Universit\'{e} de Lyon, Universit\'{e} Lyon 1, 
CNRS/IN2P3, IPNL 4 rue E Fermi 69622,
Villeurbanne CEDEX, France
}

\author{\centering 
J.\,Berenguer~Antequera,
E.\,Calvo~Alamillo, 
M.-C.\,Fouz, 
J.\,Marin,
J.\,Puerta-Pelayo, 
A.\,Verdugo
\\ \it
CIEMAT, Centro de Investigaciones Energeticas, Medioambientales y Tecnologicas, Madrid, Spain 
}

\author{\centering 
B.\,Bobchenko$^4$, %\thanks{Also at National Research Nuclear University MEPhI},
O.\,Markin$^4$,  
E.\,Novikov, 
V.\,Rusinov$^4$,  
E.\,Tarkovsky$^4$  
\\ \it
Institute of Theoretical and Experimental Physics, B. Cheremushkinskaya ul. 25,
RU-117218 Moscow, Russia
}

\author{\centering 
N.\,Kirikova,  
V.\,Kozlov, 
P.\,Smirnov, 
Y.\,Soloviev 
\\ \it
P.\,N.\, Lebedev Physical Institute,
Russian Academy of Sciences,
117924 GSP-1 Moscow, B-333, Russia
}

\author{\centering 
D.\,Besson, 
P.\,Buzhan, 
M.\,Chadeeva$^{5\star}$,  
M.\,Danilov$^{5,6}$, %\thanks{Also at Moscow Institute of Physics and Technology MIPT},
A.\,Drutskoy$^5$, 
A.\,Ilyin, 
D.\,Mironov$^{5,6}$,  
R.\,Mizuk$^5$, 
E.\,Popova 
\\ \it
National Research Nuclear University 
MEPhI (Moscow Engineering Physics Institute)
31, Kashirskoye shosse,
115409 Moscow, Russia
}

\author{\centering 
M.\,Gabriel, 
P.\,Goecke,
C.\,Kiesling,
N.\,van\,der\,Kolk, 
F.\,Simon, 
M.\,Szalay 
\\ \it
Max Planck Inst. f\"ur Physik,
F\"ohringer Ring 6,
D-80805 Munich, Germany
}

\author{\centering
S.\,Bilokin, 
J.\,Bonis, 
P.\,Cornebise, 
R.\,P\"oschl, 
F.\,Richard, 
A.\,Thiebault,
D.\,Zerwas
\\ \it
Laboratoire de L'acc\'elerateur Lin\'eaire,
Centre d'Orsay, Universit\'e de Paris-Sud XI,
BP 34, B\^atiment 200,
F-91898 Orsay CEDEX, France
}

\author{\centering 
M.\,Anduze,
V.\,Balagura,
E.\,Becheva,
V.\,Boudry, 
J-C.\,Brient,
J-B.\,Cizel,
C.\,Clerc, 
R.\,Cornat,
M.\,Frotin,
F.\,Gastaldi,
F.\,Magniette,
P.\,Mora de Freitas, 
G.\,Musat,
S.\,Pavy,
M.\,Rubio-Roy,
M.\,Ruan$^7$, %\thanks{Now at IHEP, Beijing and CERN},
H.\,Videau
\\ \it
 Laboratoire Leprince-Ringuet (LLR)  -- \'{E}cole Polytechnique,
 CNRS/IN2P3,
 Palaiseau, F-91128 France
}

\author{\centering 
S.\,Callier,
F.\,Dulucq,  
G.\,Martin-Chassard, 
L.\,Raux,
N.\,Seguin-Moreau,
Ch.\,de la Taille 
\\ \it
Laboratoire OMEGA -- \'{E}cole Polytechnique-CNRS/IN2P3, 
Palaiseau, F-91128 France
}

\author{\centering 
J.\,Cvach, 
P.\,Gallus, 
M.\,Havranek, 
M.\,Janata, 
D.\,Lednicky, 
M.\,Marcisovsky, 
I.\,Polak, 
J.\,Popule, 
L.\,Tomasek, 
M.\,Tomasek,
P.\,Sicho, 
J.\,Smolik, 
V.\,Vrba, 
J.\,Zalesak 
\\ \it
Institute of Physics, Academy of Sciences of the Czech Republic, Na Slovance 2,
CZ-18221 Prague 8, Czech Republic
}

\author{\centering              
K.\,Kotera, 
H.\,Ono$^8$, %\footnote{Now at Nippon Dental University, 1-8 Hamaura-cho
T.\,Takeshita
\\ \it
Shinshu Univ.\,,
Dept. of Physics,
3-1-1 Asaki,
Matsumoto-shi, Nagano 390-861,
Japan
}

\author{\centering              
S.\,Ieki,
Y.\,Kamiya,
W.\,Ootani, 
N.\,Shibata
\\ \it
ICEPP, The University of Tokyo, 7-3-1 Hongo, Bunkyo-ku, Tokyo
113-0033, Japan}

\author{{\centering              
D.\,Jeans,
S.\,Komamiya, 
H.\,Nakanishi 
\\ \it
Department of Physics, Graduate School of Science, The University of
Tokyo, 7-3-1 Hongo, Bunkyo-ku, Tokyo 113-0033, Japan
}

\it
$^\star$ Corresponding author\newline
E-mail: \email{marina.chadeeva@lebedev.ru}
}

\author{  \\
\llap{$^1$}Now at SLAC. \\
\llap{$^2$}Deceased.\\
\llap{$^3$}Also at DESY and Univ. Hamburg.\\
\llap{$^4$}Also at National Research Nuclear University MEPhI.\\
\llap{$^5$}Also at P.\,N.\, Lebedev Physical Institute, Russian Academy of Sciences. \\
\llap{$^6$}Also at Moscow Institute of Physics and Technology MIPT.\\
\llap{$^7$}Now at IHEP, Beijing and CERN.\\
\llap{$^8$}Now at Nippon Dental University, 1-8 Hamaura-cho Chuo-ku, Niigata, 951-8580, Japan.\\

}
\abstract{The spatial development of hadronic showers in the CALICE scintillator-steel analogue hadron calorimeter is studied using test beam data collected at CERN and FNAL for single positive pions and protons with initial momenta in the range of 10--80~GeV/$c$. Both longitudinal and radial development of hadron showers are parametrised with two-component functions. The parametrisation is fit to test beam data and simulations using the {\sffamily QGSP\_BERT} and {\sffamily FTFP\_BERT} physics lists from {\scshape Geant4} version 9.6. The parameters extracted from data and simulated samples are compared for the two types of hadrons. The response to pions and the ratio of the non-electromagnetic to the electromagnetic calorimeter response, $h/e$, are estimated using the extrapolation and decomposition of the longitudinal profiles.}


\begin{document}
%\linenumbers

%%%%%%%%%%%%%%%%%%%%%%%%%%%%%%%%%%%%%%%%%%%%%%%%%%%%%%%%%%%%%%
\section{Introduction}
\label{sec:intro}

The development of hadronic showers in calorimeters is a complicated process characterised by significant fluctuations of the reconstructed energy as well as of both longitudinal and radial shower sizes~\cite{Wigmans:2000,Gabriel:1994}. 
Hadronic showers induced by mesons and baryons are observed to induce different response in the calorimeter~\cite{CMS:2008barrelResp,ATLAS:2010pionProton}. This behaviour can be explained by the conservation of baryon number, which results in different energy available for the production of secondaries. 

The highly granular CALICE calorimeters, like the scintillator-steel analogue hadron calorimeter (Fe-AHCAL)~\cite{AHCAL:2010cc}, provide a unique opportunity to study the development of hadronic showers in fine detail. The comparison of shapes of hadron showers induced by different types of particles is hampered by the fact that the observed average profiles are convolved with the distributions of the shower start position, which also differ due to the dependence of nuclear interaction lengths on particle type. The high granularity of the Fe-AHCAL allows the position of the first inelastic interaction to be identified, thereby disentangling the spatial shower development from the distribution of the shower start position. The deconvolution of the shower start distribution from the shower development helps to exclude a significant component of fluctuations of the spatial energy density distribution from shower to shower. 
Understanding of hadronic shower development and parametrisation of the energy density distribution are also important for the estimation of leakage from calorimeters, validation of hadronic shower models in simulations, and for the improvement of particle flow algorithms. 

A detailed study of the global parameters of hadronic showers, such as calorimeter response, resolution, shower radius and longitudinal centre of gravity in the highly granular CALICE Fe-AHCAL is presented in refs.~\cite{Valid:2013,AHCAL:2015pionProton}, including a comparison between data and simulations using the {\scshape Geant4}  toolkit~\cite{Geant4:2003}. In the studied energy range from 10 to 80~GeV, proton showers were found to be on average $\sim$5\% longer and $\sim$10\% wider than pion showers. The {\scshape Geant4} physics lists used in the comparison give better predictions for protons than for pions and predict a steeper energy dependence of the calorimeter response compared to that observed in data. The prediction for the mean shower radius is significantly improved for the physics list {\sffamily FTFP\_BERT} in {\scshape Geant4}~9.6, where the agreement with data is within 5\% for the entire energy range studied compared to 10\% disagreement for {\scshape Geant4}~9.4. 
 
Global observables cannot reveal the details of spatial differences.
A typical hadronic shower consists of a relatively narrow core with high energy density surrounded by an extended halo. The core is commonly interpreted as being formed by electromagnetic cascades initiated by photons from decays of $\pi^{0}$s produced in hard interactions in the shower start phase~\cite{Wigmans:2000,Gabriel:1994}. The spatial distribution of the energy density within a hadronic shower can be represented as the sum of two contributions: an electromagnetic and a hadronic component. The electromagnetic component tends to develop near the shower axis and close to the point of the first inelastic interaction of the incoming hadron. The hadronic component is in turn  produced by charged secondary hadrons from the cascade; it dominates in the shower tail, thereby making hadronic showers wider and longer than their electromagnetic counterparts.

The high longitudinal and transverse granularity of the CALICE Fe-AHCAL allows detailed measurements of shower development. Both longitudinal and radial profiles of hadron-induced showers can be parametrised with two-component functions using the phenomenological approach proposed in ref.~\cite{Bock:1981}.
In  this study,  longitudinal and transverse shower profiles are decomposed into core and halo contributions by fitting an empirical parametrisation to the energy density distributions extracted from both data and simulations.  The fitted parameters and the extracted core shower fractions are compared to simulations using two {\scshape Geant4} physics lists. Longitudinal shower profiles are represented as the sum of a ``short'' and a ``long'' component. It appears that the parameters of the ``short'' component are similar to those of electromagnetic showers. Following this observation, the parametrisation of longitudinal profiles can be roughly interpreted as a decomposition of showers into electromagnetic and hadronic components.

The parametrisation of profiles is one of the possible ways to study calorimeter properties. For instance, the depth of the Fe-AHCAL is $\sim$5.3$\lambda_{\mathrm{I}}$, which is not enough to fully contain all hadronic showers, so its response is systematically shifted. However, a correction to the average response can be obtained by extrapolating the longitudinal profiles outside of the range covered by calorimeter. 
In previous studies, the response of the Fe-AHCAL to hadrons has been already estimated directly by using the combined CALICE setup that is long enough to accommodate nearly all showers of the studied energies~\cite{AHCAL:2012res}. The comparison of the two approaches can help to understand the reliability of estimates based on the extrapolation of the longitudinal profiles.

In addition, the decomposition of profiles allows an estimation of the $h/e$ ratio, where $h$ characterises the  calorimeter response to the hadronic component of hadron-induced showers, and $e$ is the  response to electromagnetic showers.  The traditional way to estimate the characteristic $h/e$ ratio of a calorimeter is based on the assumption that the mean hadronic fraction scales with energy according to a power law~\cite{CDF:1997,ATLAS:2009}. To extract the value of $h/e$ from the power-law approximation of the energy dependence of the response, an assumption about the so-called ``scale energy'' is also necessary~\cite{Groom:2007}, at which multiple pion production becomes significant. The decomposition of profiles allows the extraction of the ratio of responses $h/e$ without any assumptions about the energy dependence of the calorimeter response and about the ``scale energy''. The only assumption is that the parametrisation can be extrapolated outside the fit range used.

Section~\ref{sec:testbeam} describes the test beam data, event selection procedure, simulations and systematic uncertainties. The ratios of the simulated shower profiles to the data shower profiles are shown in section~\ref{sec:ratios}. The parametrisation of the shower profiles is described in section~\ref{sec:para} followed by the comparison of extracted parameters in section~\ref{sec:paraComp}. The estimate of the  response of the extended calorimeter from the extrapolation of the longitudinal profile is discussed in section~\ref{sec:resp}. Section~\ref{sec:pheno} contains a description of the new approach to the extraction of the calorimeter characteristic $h/e$ using the proposed decomposition of the longitudinal profiles.

%%%%%%%%%%%%%%%%%%%%%%%%%%%%%%%%%%%%%%%%%%%%%%%%%%%%%%%%%%%%%%
\section{Test beam data and simulation}
\label{sec:testbeam}

%==============================
\subsection{Experimental setup}
\label{sec:setup}

The presented analysis is based on the positive hadron data collected during the CALICE test beam campaigns at CERN in 2007 and at FNAL in 2009. The CALICE setup at CERN is described in detail in ref.~\cite{AHCAL:2012res}. The setup comprises the Si-W electromagnetic calorimeter (ECAL)~\cite{ECAL:2008}, the Fe-AHCAL, and the scintillator-steel tail catcher serving also as a muon tracker (TCMT)~\cite{TCMT:2012}. Positive hadron beams in the momentum range from 30 to 80~GeV were delivered from the CERN SPS H6 beam line. 
Test beam data for positive hadrons with initial momenta of 10 and 15~GeV were collected at FNAL using the Fe-AHCAL and TCMT~\cite{FeegeDis:2011}. 

The setup at CERN included a threshold gaseous \u{C}erenkov counter, while the setup at FNAL included a differential gaseous \u{C}erenkov counter. The \u{C}erenkov counter was placed upstream of the calorimeter setup and was used for offline discrimination between pions and protons on an event-by-event basis. All data used for the present analysis were taken at normal incidence of the beam particles with respect to the calorimeter front plane.

The Fe-AHCAL is a sandwich structure of 38 active layers interleaved with steel plates. The total absorber thickness amounts to 21~mm of stainless steel per layer. Each active layer has a transverse size of 90$\times$90~cm$^2$ and is assembled from 0.5~cm thick scintillator tiles (cells) with transverse sizes of 3$\times$3~cm$^2$ in the central, 6$\times$6~cm$^2$ in the intermediate and 12$\times$12~cm$^2$ in the peripheral regions. The spatial position of each calorimeter cell is defined in a right-handed Cartesian coordinate system with the $z$-axis oriented along the beam direction, perpendicular to the calorimeter front plane, and the $y$-axis pointing up. 
 
The light in each scintillator tile is individually readout by a silicon photomultiplier (SiPM). The visible signal in each calorimeter cell was obtained in units of minimum-ionising particle (MIP). The calibration and energy reconstruction procedures in the Fe-AHCAL are described in refs.~\cite{AHCAL:2010cc,AHCAL:2011em,WAHCAL:2014}. Only cells with a signal above 0.5~MIP were considered for further analysis and are called hits hereafter.  

The total depth of the Fe-AHCAL is $\sim$5.3$\lambda_{\mathrm{I}}$ (38 physical layers with $\sim$0.14$\lambda_{\mathrm{I}}$ per layer). The first section of the TCMT consists of 9 physical layers comprised of 2~cm thick steel absorber plates and 0.5~cm thick scintillator strips and has the same sampling as the Fe-AHCAL ($\sim$0.14$\lambda_{\mathrm{I}}$ per layer).

%==============================
\subsection{Event selection}
\label{sec:evtsel}

The event selection procedure includes the rejection of muons, double particle events and positrons from the data samples collected without the electromagnetic calorimeter. The gas pressure of the \u{C}erenkov counter was set between the pion and proton thresholds. The inefficiency of the \u{C}erenkov counter at the level of several percent results in pion contamination of the proton samples. The procedure and approaches to particle identification and the estimation of the proton sample purities are described in detail in ref.~\cite{AHCAL:2015pionProton}. The estimated sample purities for each data sample are shown in table~\ref{tab:runList}.

Events well contained in the Fe-AHCAL were selected from both data and simulated samples for further analysis of shower profiles. For this purpose, the longitudinal position of the first inelastic interaction of the incoming hadron (shower start) was identified on an event-by-event basis using a dedicated algorithm~\cite{AHCAL:2015pionProton}. The difference between the reconstructed shower start layer and the true one in the simulated samples does not exceed $\pm$1 layer for more than 80\% of events.  
The distribution of this difference is much more peaked and has fatter tails than the Gaussian distribution of the same width as shown in ref.~\cite{AHCAL:2015pionProton}, so the value of $\pm$1 layer is taken as the uncertainty of the shower start identification.

For the samples taken with the ECAL in front of the  Fe-AHCAL, events were required to have the identified shower start in the physical layers 2, 3, 4, 5, 6. The first physical layer was excluded due to higher uncertainties of the shower start identification. The depth of one layer of the Fe-AHCAL is $\sim$1.2 radiation lengths. For data samples taken without the ECAL, we selected events with the shower start identified in the physical layers 3, 4, 5, 6 in order to reduce the fraction of remaining positrons in the sample. The exclusion of events with the shower start beyond the sixth layer helps to minimise the leakage into the TCMT. In this paper, shower profiles are analysed with respect to the identified shower start.

The analysed data samples and number of selected events are shown in table~\ref{tab:runList}. The resulting contamination of the selected hadron samples by muons does not exceed 0.1\% and the admixture of double particle events is less than one percent for all samples. The contamination of the selected samples by positrons is negligible, except for the two pion samples taken without the electromagnetic calorimeter. The fraction of positrons in the selected pion samples at 10~GeV and 15~GeV does not exceed $\sim$2.5\% and $\sim$0.4\%, respectively.

\begin{table}
 \caption{The total number of collected events, the number of selected pion and proton events for data samples used in the analysis, and the estimated purity of the selected samples.}
 \label{tab:runList}
 \begin{center}
  \begin{tabular}{|c|c|c|c|c|c|c|}
   \hline
    Beam   & Total  & Setup & \multicolumn{2}{c|}{$\pi^{+}$} & \multicolumn{2}{c|}{protons} \\
    \cline{4-7}
    momen- & number & with  & Number      &        & Number      &        \\
    tum    & of     & ECAL  & of selected & Purity & of selected & Purity \\
    GeV    & events &       & events      &        & events      &        \\
   \hline
   %\hline
    10 &  45839 & no  &  5275 & 0.975$\pm$0.015 & 1239 & 0.74$\pm$0.13 \\
   %\hline
    15 &  46323 & no  &  6660 & 0.99$\pm$0.01   & 2122 & 0.80$\pm$0.09 \\
   %\hline
    30 & 192066 & yes & 12888 & $>$0.99         & 9076 & 0.95$\pm$0.01 \\
   %\hline
    40 & 201069 & yes & 24756 & $>$0.99         & 5682 & 0.84$\pm$0.07 \\
   %\hline
    50 & 199829 & yes & 25039 & $>$0.99         & 4914 & 0.79$\pm$0.08 \\
   %\hline
    60 & 208997 & yes & 25136 & $>$0.99         & 6731 & 0.86$\pm$0.06 \\
   %\hline
    80 & 197062 & yes & 20169 & $>$0.99         &10001 & 0.83$\pm$0.04 \\
   \hline
  \end{tabular}
 \end{center}
\end{table}

%==============================
\subsection{Monte Carlo simulations}
\label{sec:sim}

Samples of single pion and single proton events were simulated using two physics lists, {\sffamily QGSP\_BERT} and {\sffamily FTFP\_BERT}, from the package {\scshape Geant4} version~9.6 patch~1 in the Mokka framework~\cite{Geant4Note:2010,Mokka}. The size of each sample is 50000 events per energy point and per particle type.

The {\sffamily QGSP\_BERT} physics list is widely used for simulation in the LHC experiments and has demonstrated the best agreement with data in earlier versions, for instance in version 9.2~\cite{Geant4Note:2010}. The {\sffamily QGSP\_BERT} physics list employs the Bertini cascade model ({\sffamily BERT}) below 9.5~GeV, the quark-gluon string precompound model ({\sffamily QGSP}) above 25~GeV, and the low energy parametrised model ({\sffamily LEP}) in the intermediate energy region. The transition regions between models are from 9.5 to 9.9~GeV and from 12 to 25~GeV. 

Since 2013, the {\sffamily FTFP\_BERT} physics list is recommended for HEP simulations by the {\scshape Geant4} collaboration~\cite{Geant4:2013proc} as it was significantly improved in version 9.6. The {\sffamily FTFP\_BERT} physics list uses the Bertini cascade model for low energies and the Fritiof precompound model ({\sffamily FTFP}) for high energies with a transition region from 4 to 5~GeV.

The simulated samples were digitised taking into account the SiPM response, optical crosstalk between neighbouring scintillator tiles in the same layer, and calorimeter noise extracted from data. The digitisation was validated using the electromagnetic response of the Fe-AHCAL~\cite{AHCAL:2011em}. The beam profile and its position on the calorimeter front face in each test beam data sample were reproduced in the simulations. 

%==============================
\subsection{Observables}
\label{sec:observ}

The current analysis is dedicated to the study of shower profiles with respect to the identified shower start position. The longitudinal profiles from the shower start are presented as distributions of visible energy $\Delta E(z)$ per layer, where the energy $E$ is given in units of MIP and $z$ is the longitudinal depth of the layer. The first bin corresponds to the physical layer in which the shower start is identified. 
The longitudinal depth $z$ is measured in units of effective nuclear interaction length $\lambda_{\mathrm{I}}$ that 
was calculated using data on material properties from PDG tables~\cite{PDG:2012}. For the compound structure of the CALICE Fe-AHCAL $\lambda_{\mathrm{I}}$ is estimated to be 231~mm, which was confirmed by studies of proton showers in the same prototype~\cite{AHCAL:2015pionProton}.
Although the data from the TCMT is not used for fits to avoid problems with intercalibration, the longitudinal shower development (except for profile ratios) is shown in both Fe-AHCAL and the first part of the TCMT up to $\sim$6.5$\lambda_{\mathrm{I}}$ (47 physical layers in total). The effective radiation length, $X_{0}$, is also used where appropriate;  $X_{0}=25.5$~mm for the Fe-AHCAL. 

The radial shower profile is the distribution of the energy density, $\frac{\Delta E}{\Delta S}(r)$, at distance $r$ from the shower axis. The visible energy $\Delta E$ in units of MIP is measured in the ring of width $\Delta r$ and of area $\Delta S = 2\pi r \Delta r$, assuming the  integration along the longitudinal direction. The shower axis is extracted either from the event centre of gravity in the Fe-AHCAL or from track coordinates if the hadron track in the Si-W ECAL can be reconstructed.\footnote{The following algorithm is used to find a primary track on an event-by-event basis: a layer by layer search of a single hit candidate per layer is performed in the beam direction, that is, along the normal to the calorimeter front plane, using the nearest neighbour criterion. The search starts from the seed in the first non-empty layer of the calorimeter and ends one layer before the identified shower start. A minimum length of four hits is required for the identified primary track.} The coordinate vector of the event centre of gravity, $\vec x_{\mathrm{cog}}$ is defined as the energy weighted sum of the coordinates of all hits in the Fe-AHCAL as

\begin{equation}
 \vec x_{\mathrm{cog}} = \left(\sum_{i=1}^{N}{e_{i} \: \vec x_{i}}\right)/\left(\sum_{i=1}^{N}{e_{i}}\right),
\label{eq:cog}
\end{equation}
 
\noindent where the sum runs over the $N$ hits of the Fe-AHCAL, $e_i$ is the energy of hit $i$, and $\vec{x_i}$ is the transverse position coordinate of the cell. The transverse centre of gravity can be defined with an accuracy of $\sim$3~mm while the uncertainty of the incoming track transverse coordinates is $\sim$2~mm. 

The hits in physical layers before the shower start layer are not included in the radial profiles.
In contrast to longitudinal profiles, transverse profiles cannot be calculated in the TCMT because the latter is built of strips which do not allow simultaneous determination of both transverse coordinates. 
It should be mentioned that the effective Moli\`{e}re radius for the Fe-AHCAL, $R_{\mathrm{M}}$, is 24.5~mm. Identical procedures for identification of the shower start layer and the shower axis on an event-by-event basis are applied to data and simulated samples.

%==============================
\subsection{Systematic uncertainties}
\label{sec:systematics}

The following sources of systematic uncertainties, which can affect the shape of the shower profiles, were investigated: 

\begin{itemize}
  \item layer-to-layer variations of the response;
  \item identification of the shower start layer;
  \item identification of the shower axis;  
  \item pion contamination of the proton samples;
  \item positron contamination of samples collected without electromagnetic calorimeter;
  \item leakage from the Fe-AHCAL.
\end{itemize}

The estimation of the uncertainties is discussed in detail in appendix~\ref{app:sys}. The contributions from the identification of the shower start layer and the leakage from the Fe-AHCAL were found to have a negligible impact on the profile parameters and do not affect the comparison of data and simulations. The layer-to-layer variations give the most significant contribution to the uncertainty of the longitudinal profiles, they increase with energy and lead to an uncertainty of $\sim$12\% around the shower maximum at 80~GeV. The identification of the shower axis for the data samples taken without the electromagnetic calorimeter gives the biggest contribution to the uncertainty of the radial profiles, which amounts up to 10\% at the energies of 10 and 15~GeV. 

The contamination of the pion samples by positrons is relatively low (see table~\ref{tab:runList}) and gives negligible contribution to the uncertainties, in contrast to the pion contamination of the proton samples. The latter introduces a noticeable bias and affects the shape of the proton profiles, which is corrected using the known sample purities as described in appendix~\ref{app:sysProton}.       

The calculation of the reconstructed energy and extraction of the $h/e$ ratio require a conversion from the units of MIP to the GeV energy scale. The conversion coefficient from MIP to GeV for the Fe-AHCAL (electromagnetic calibration) was extracted from dedicated positron runs with a systematic uncertainty of 0.9\%~\cite{AHCAL:2011em}. Other contributions, such as the uncertainty due to the saturation correction of the SiPM response, are discussed in detail in ref.~\cite{AHCAL:2011em}; they were analysed by varying the calibration constants within allowed limits (11\% for the re-scaling factor of the saturation correction) and were found to be negligible for hadrons in the energy range studied.

%%%%%%%%%%%%%%%%%%%%%%%%%%%%%%%%%%%%%%%%%%%%%%%%%%%%%%%%%%%%%%
\section{Simulations to data ratio of hadron shower profiles}
\label{sec:ratios}

\begin{figure}
 \centering
 \includegraphics[width=7.5cm]{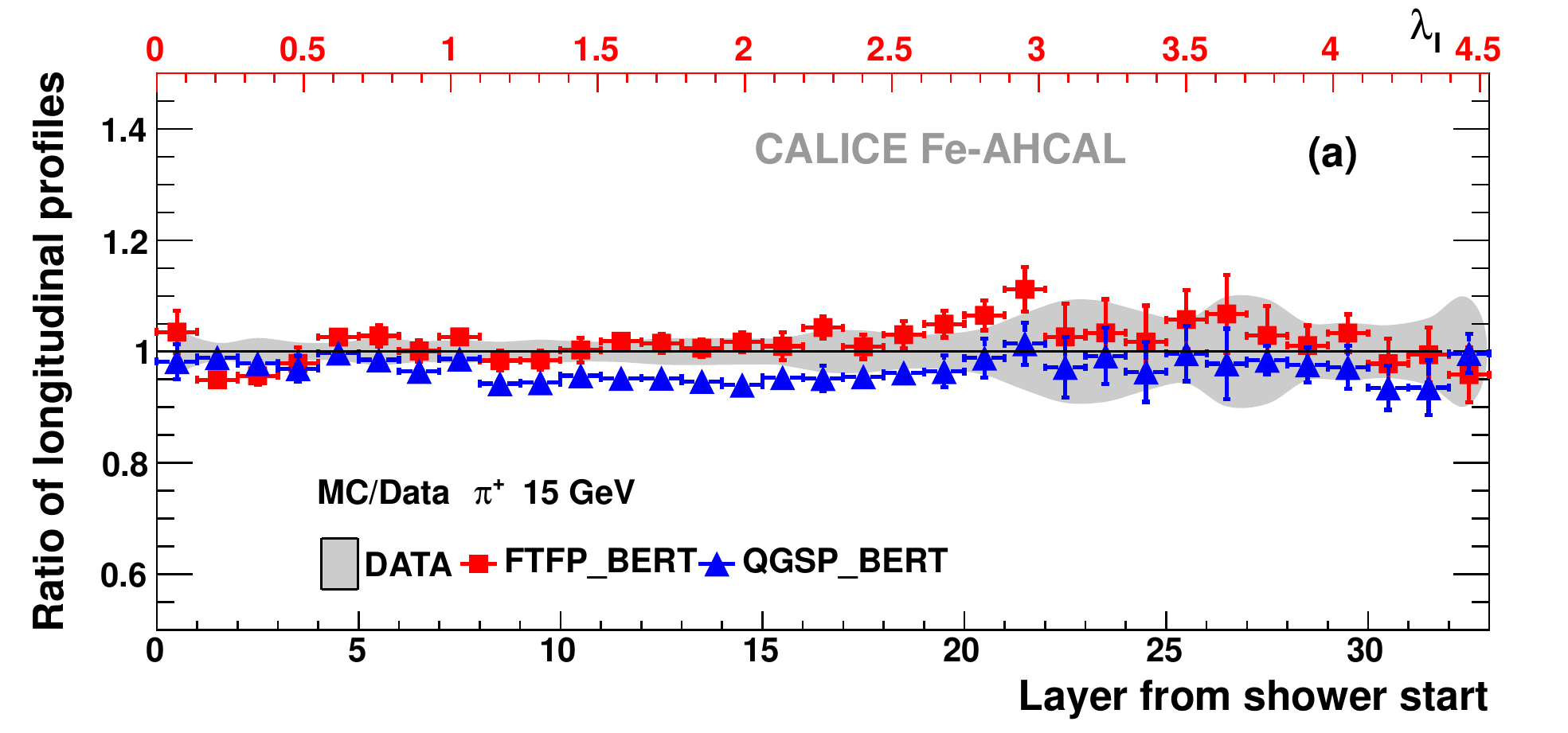}
 \includegraphics[width=7.5cm]{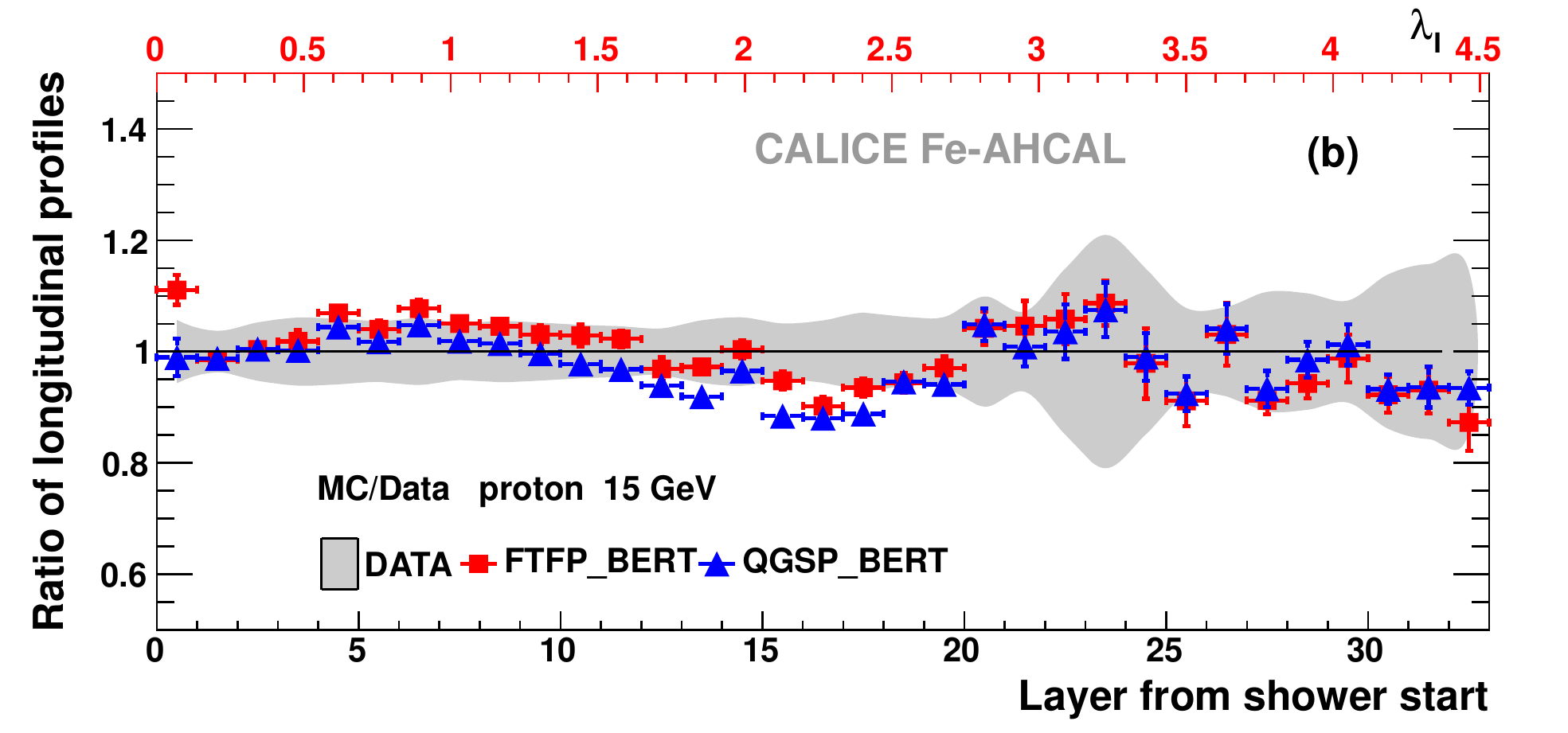}
 \newline
 \includegraphics[width=7.5cm]{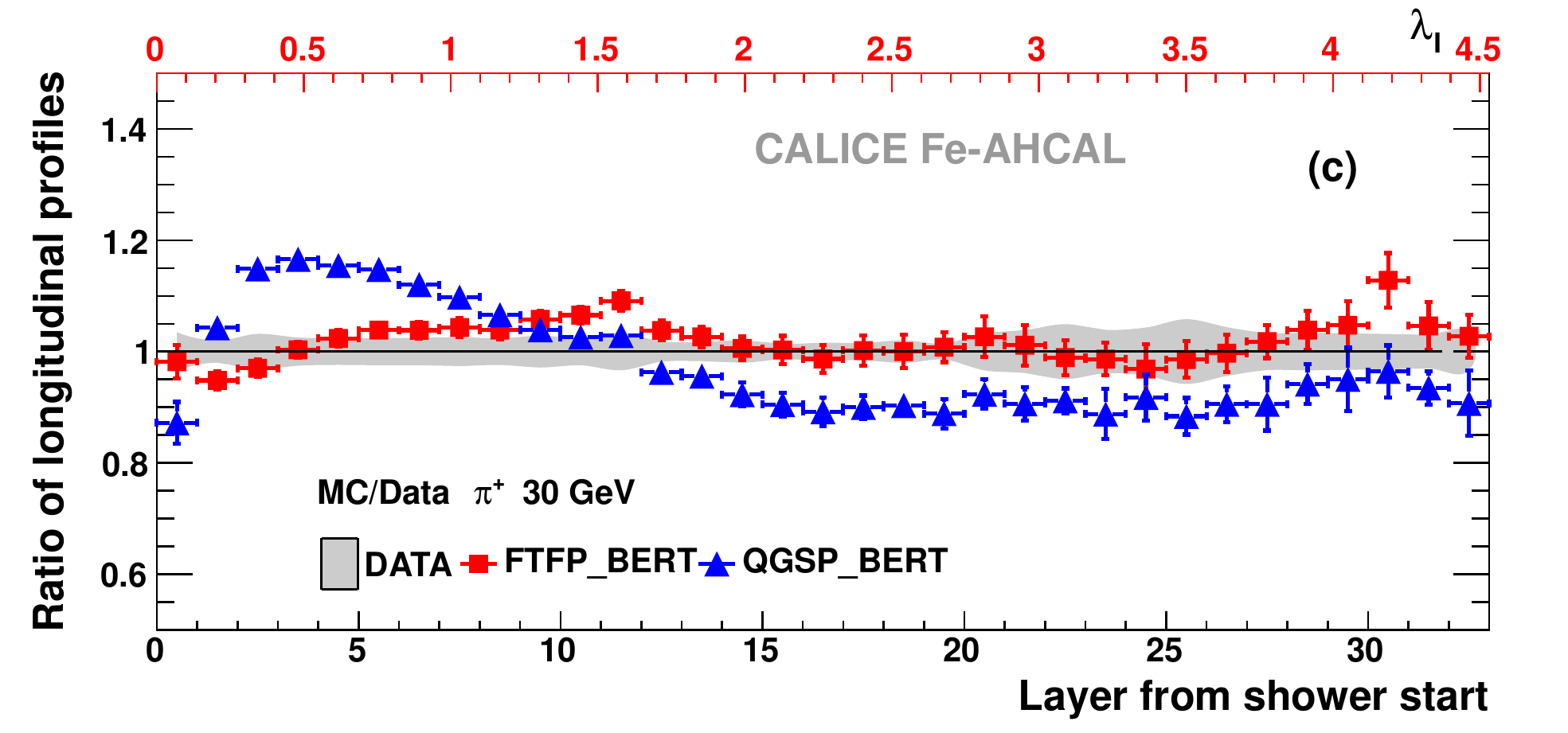}
 \includegraphics[width=7.5cm]{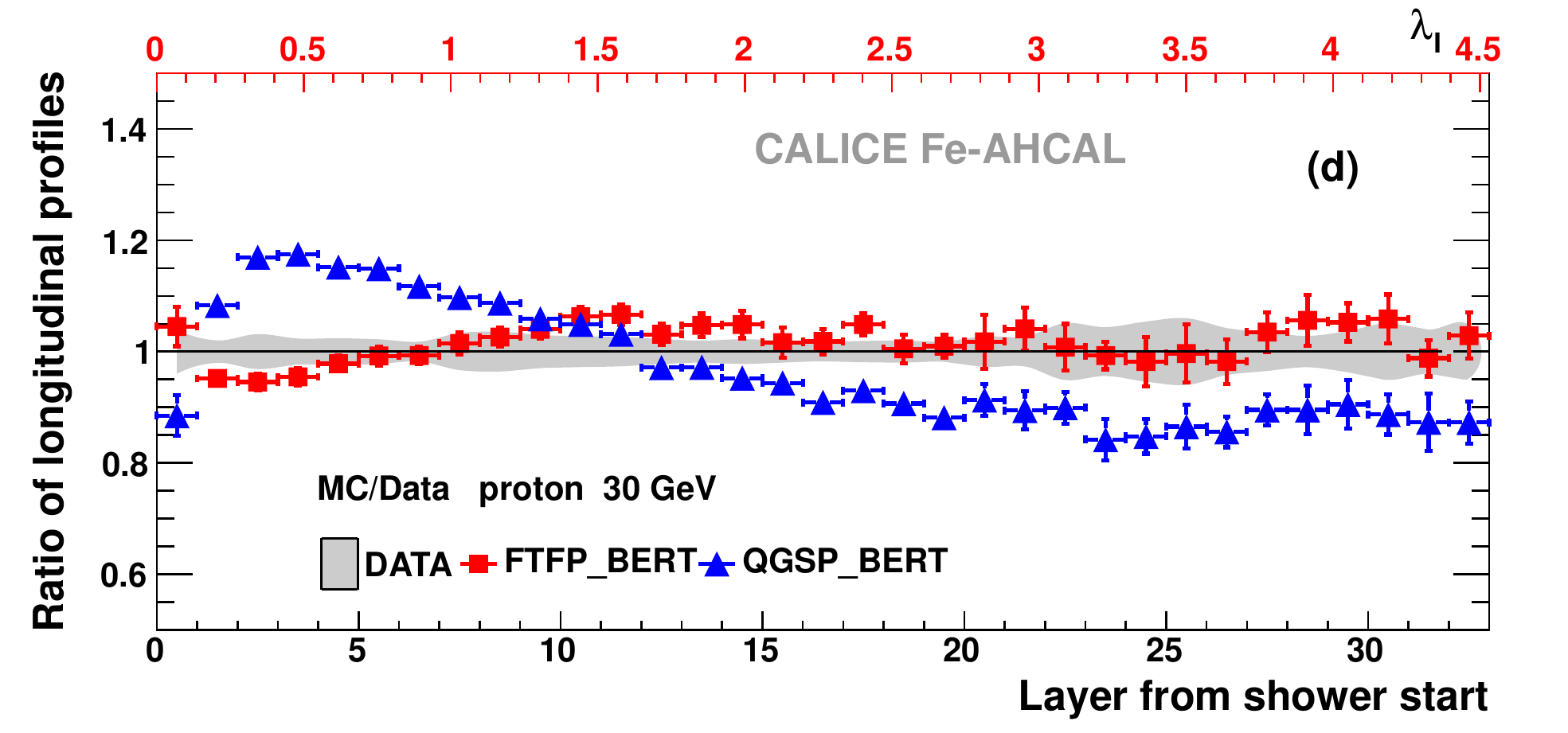}
 \newline
 \includegraphics[width=7.5cm]{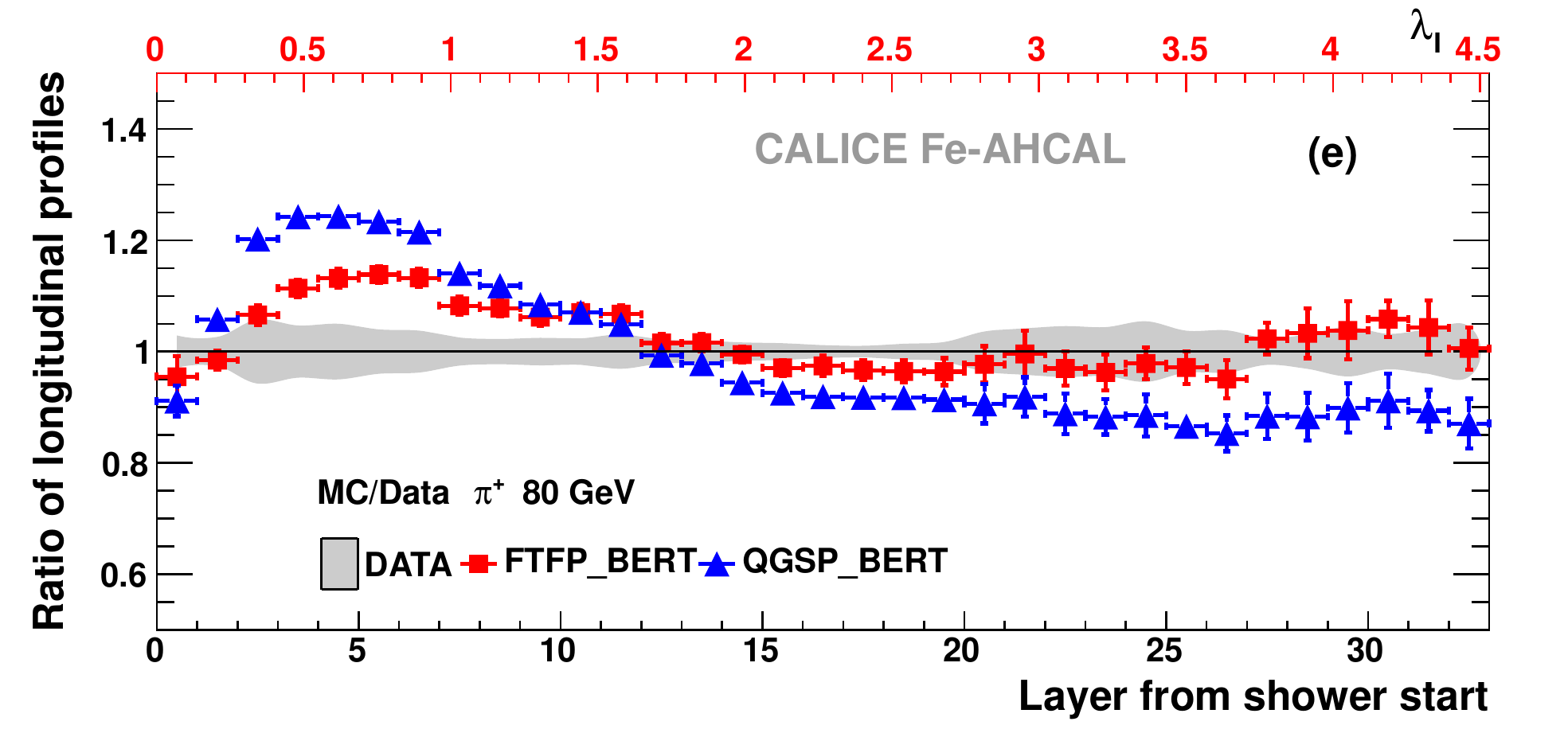}
 \includegraphics[width=7.5cm]{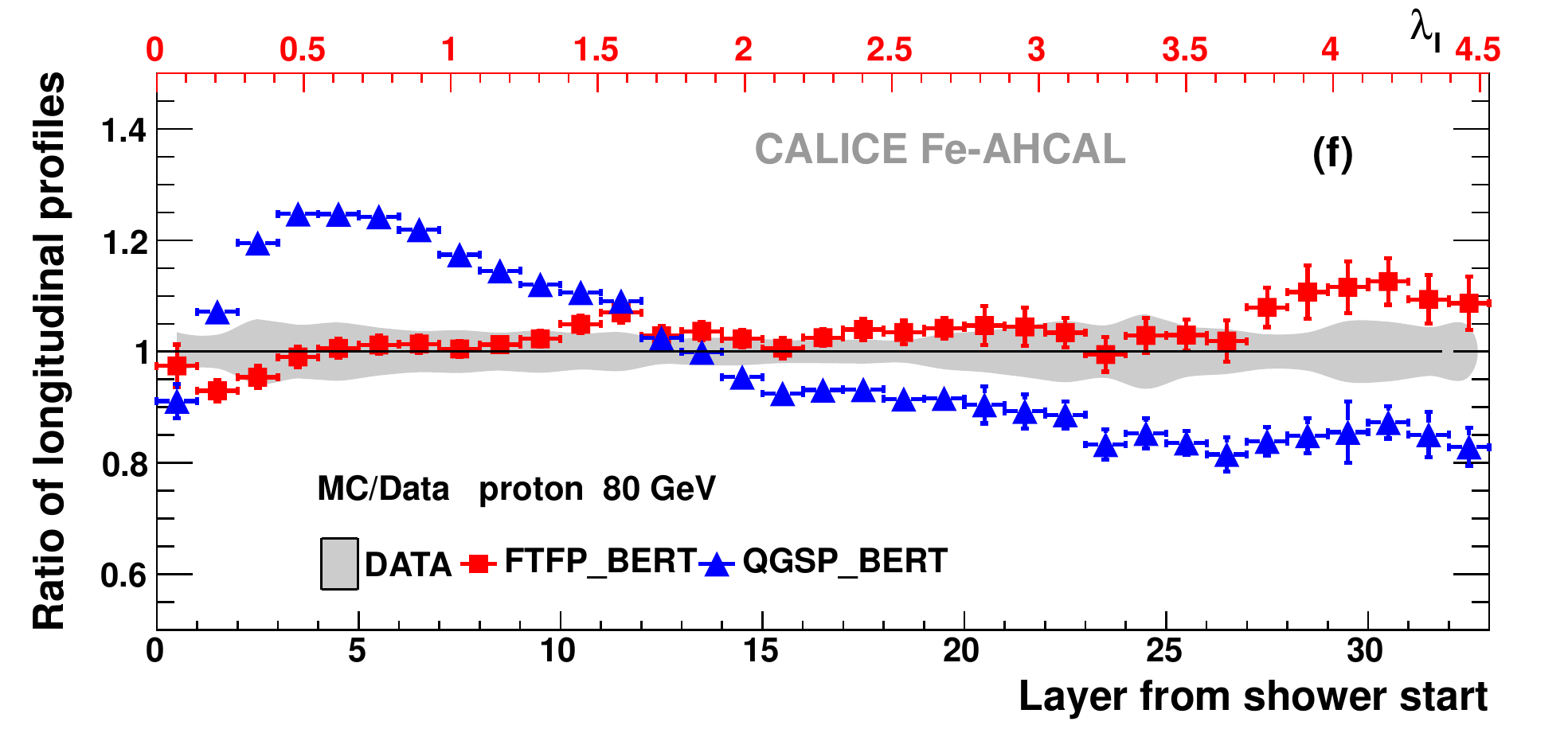}
 \caption{Ratio of longitudinal profiles of showers induced by 15, 30 and 80~GeV hadrons from simulated samples to those from data samples for (left) pions and (right) protons. The grey band and the error bars show the uncertainty for data and simulations, respectively. The upper red axis show the longitudinal depth in units of $\lambda_{\mathrm{I}}$.}
 \label{fig:ratioLng}
\end{figure}

The quality of Monte Carlo predictions for shower development can be illustrated using the ratios of shower profiles extracted from simulated events to those extracted from test beam data. The ratios of longitudinal profiles that represent the measured visible energy per layer (see section~\ref{sec:observ}) are shown in figure~\ref{fig:ratioLng}.\footnote{Such a comparison with the physics lists from {\scshape Geant4} version 9.4 was done in ref.~\cite{Valid:2013}}
The longitudinal profiles can be extracted from the CALICE Fe-AHCAL with a bin size of $\sim$0.14$\lambda_{\mathrm{I}}$. 

The proton profiles are well reproduced by the {\sffamily FTFP\_BERT} physics list at all studied energies. The profiles of pion showers are reproduced by {\sffamily FTFP\_BERT} within 5\% at 15 and 30~GeV. The overestimation of deposition around the shower maximum increases with energy up to $\sim$12\% at 80~GeV. The tail of the shower is well reproduced at all energies.  

The {\sffamily QGSP\_BERT} physics list underestimates the energy deposition for pions at 10~GeV, gives a good prediction at 15~GeV and significantly overestimates the amount of energy deposited around the shower maximum for both pions and protons at higher energies, showing more than 20\% excess at 80~GeV. The deposition in the tail of the shower is underestimated by this physics list for energies of 30~GeV and above.  

\begin{figure}
 \centering
 \includegraphics[width=7.5cm]{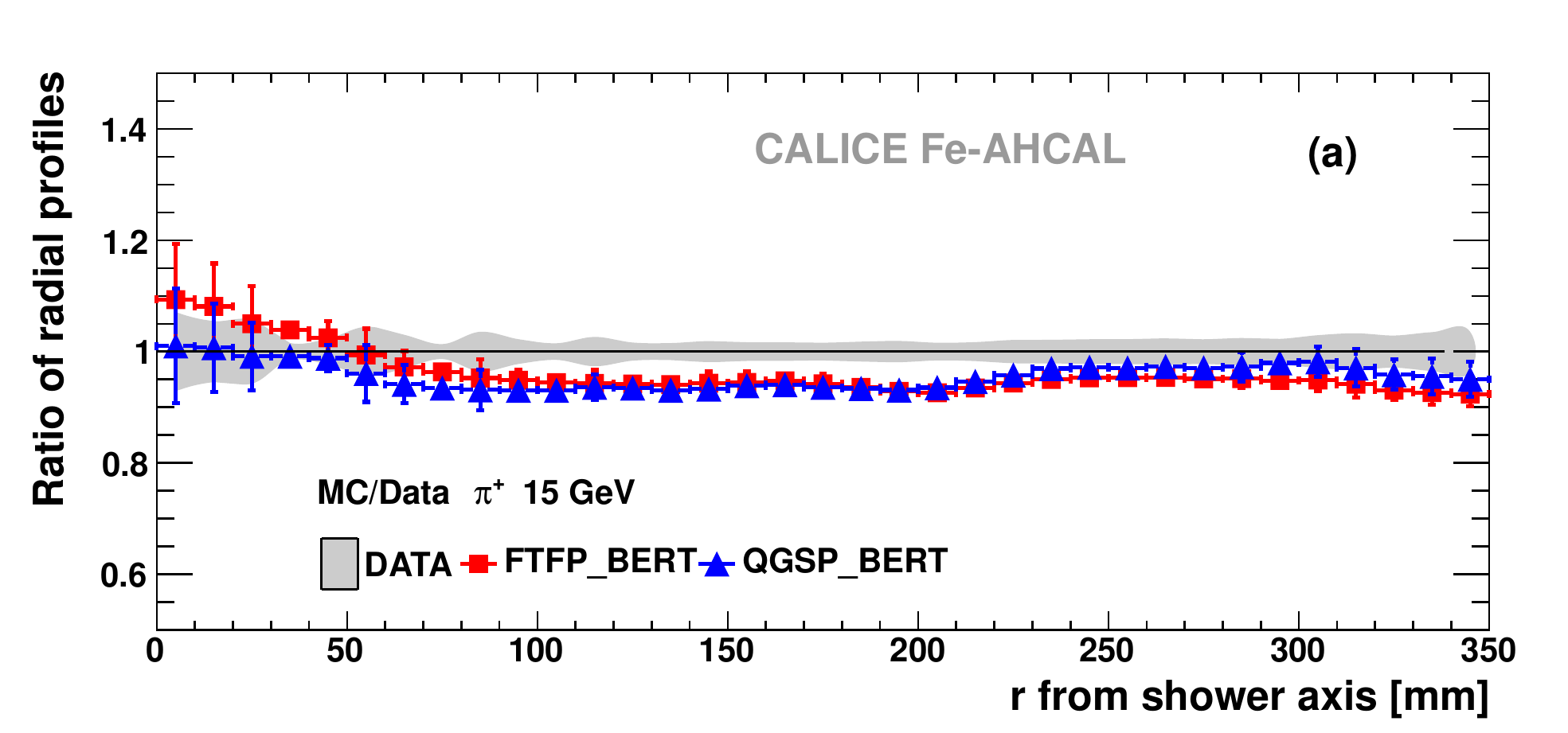}
 \includegraphics[width=7.5cm]{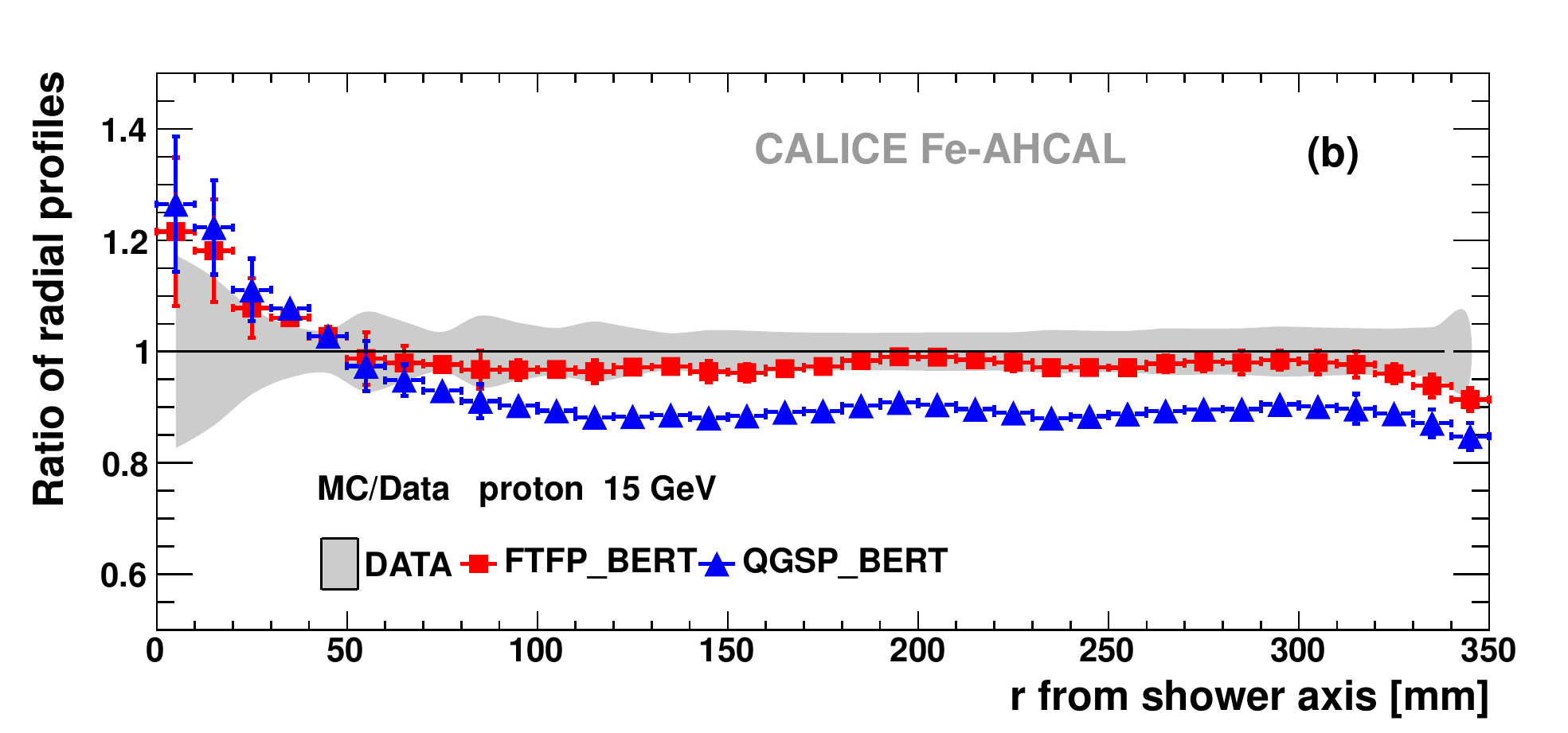}
 \newline
 \includegraphics[width=7.5cm]{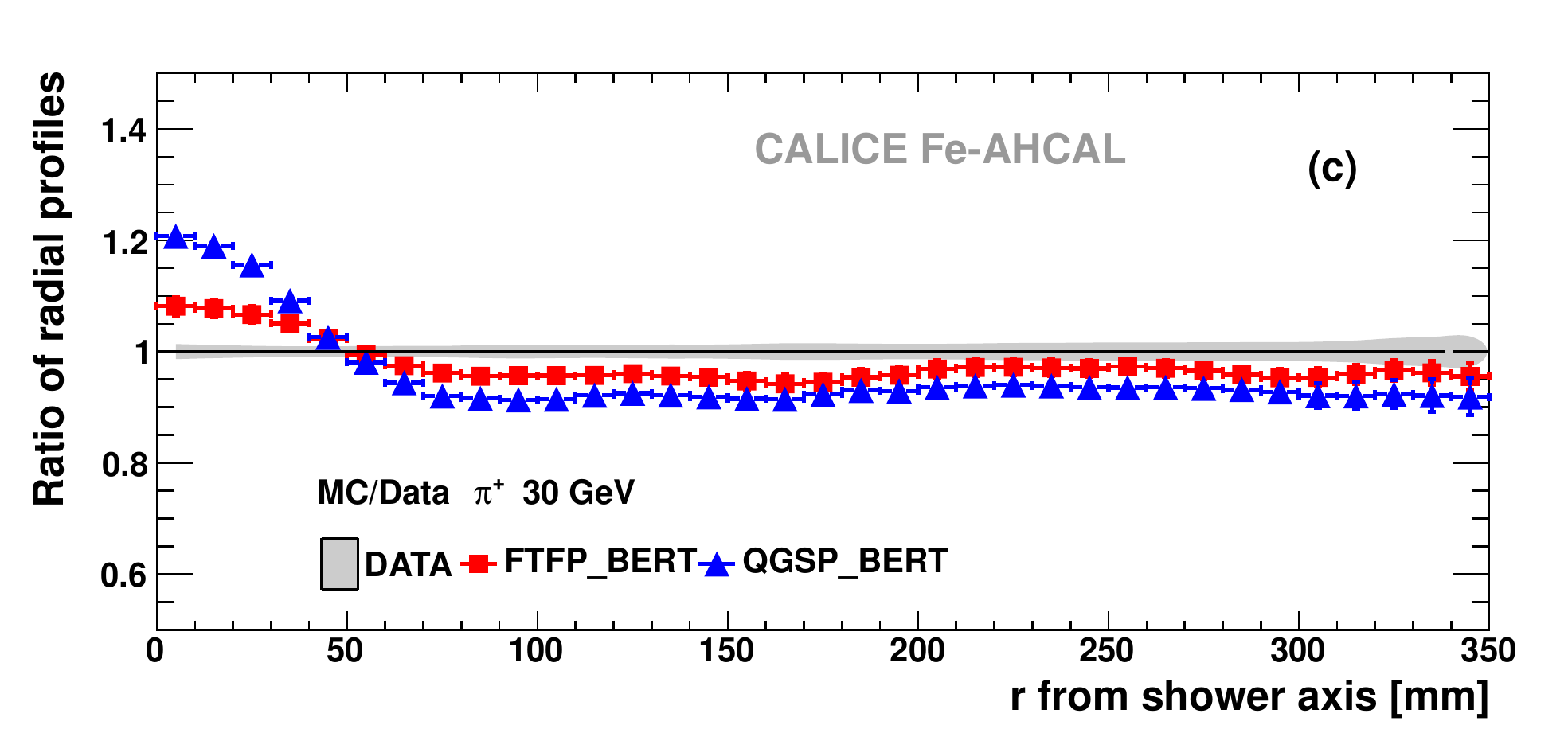}
 \includegraphics[width=7.5cm]{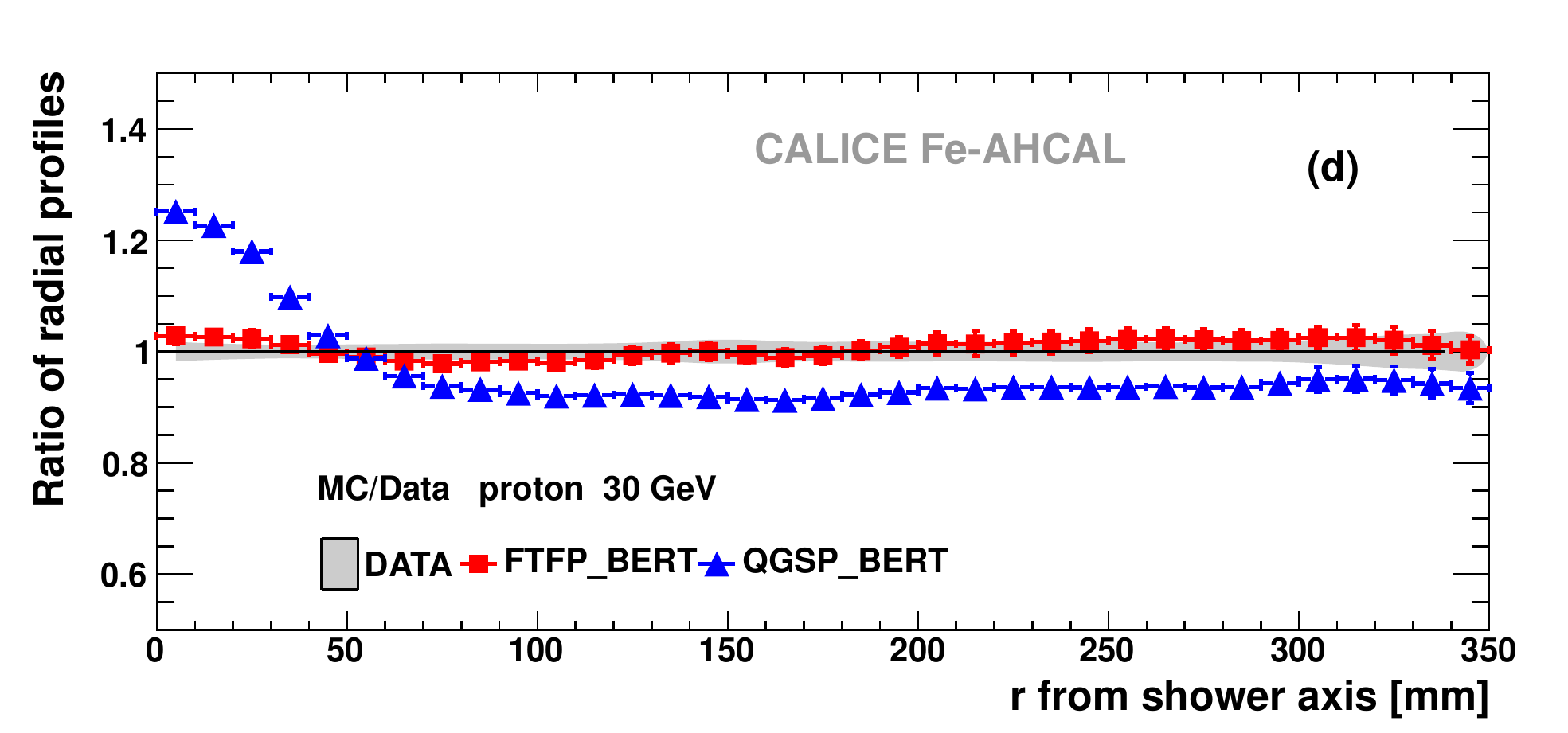}
 \newline
 \includegraphics[width=7.5cm]{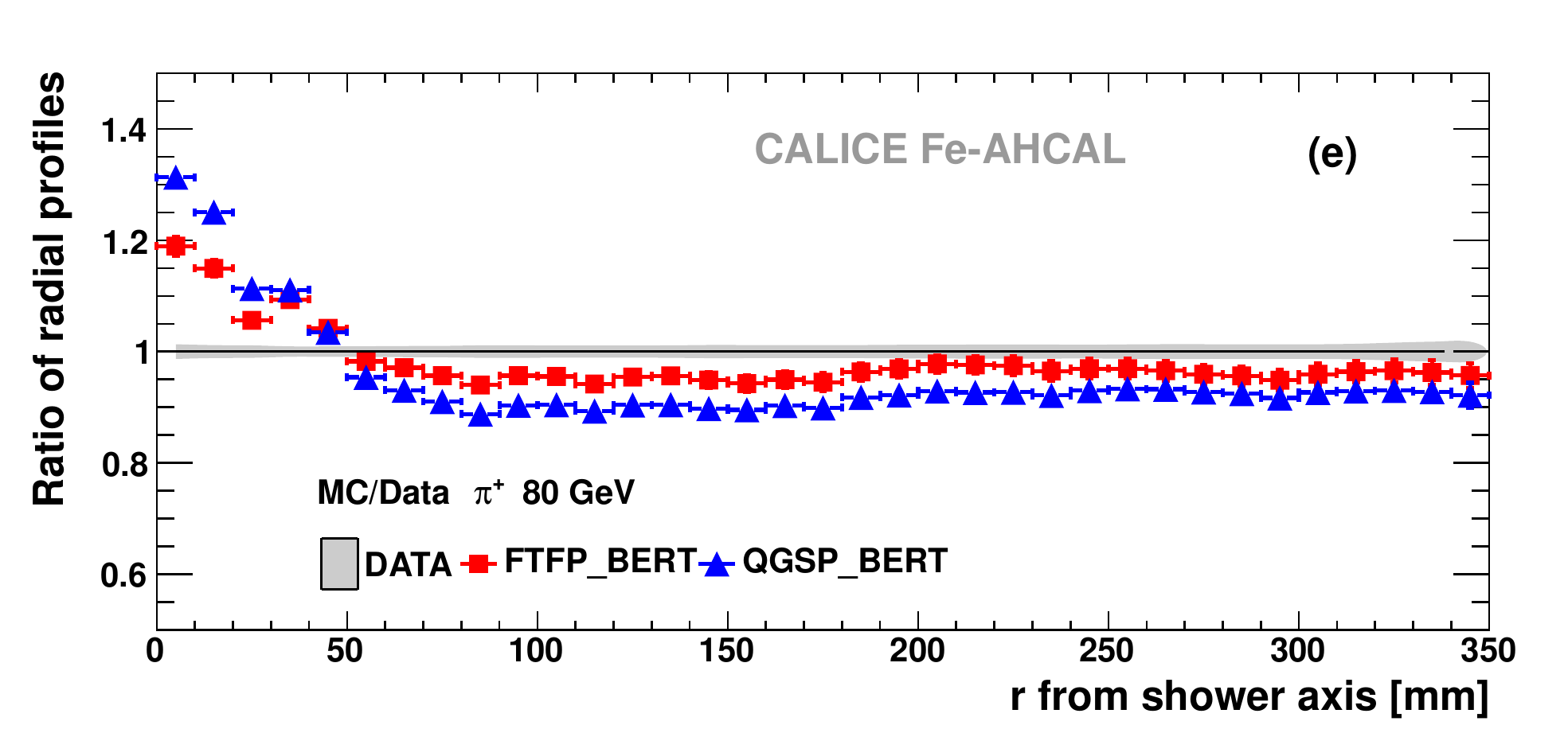}
 \includegraphics[width=7.5cm]{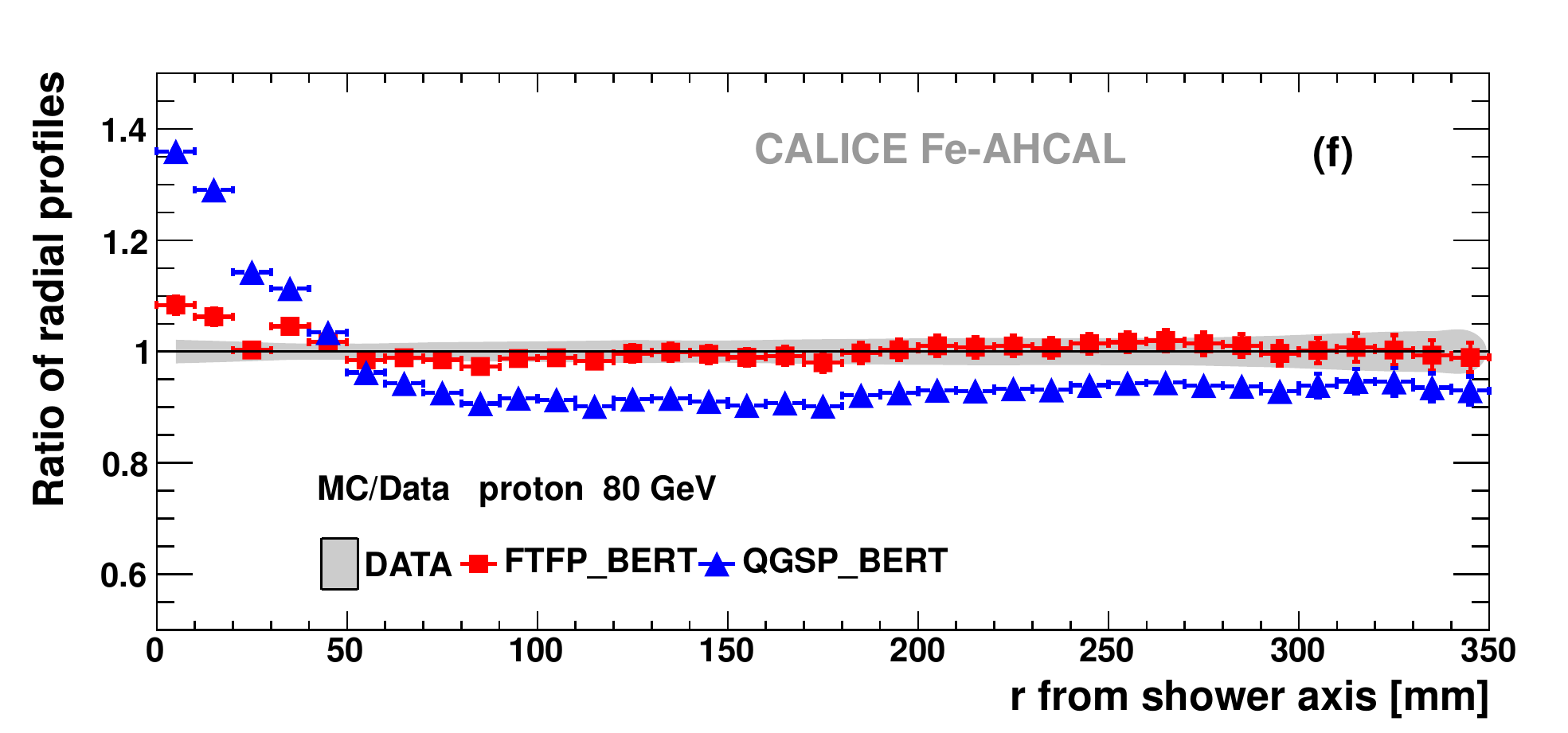}
 \caption {Ratio of radial profiles of showers induced by 15, 30 and 80~GeV hadrons from simulated samples to those from data samples for (left) pions and (right) protons. The grey band and the error bars show the uncertainty for data and simulations, respectively.}
 \label{fig:ratioRad}
\end{figure}

The accuracy of the shower axis estimate on an event-by-event basis, as described in section~\ref{sec:observ}, justifies the use of radial bins with a width of 10 mm, corresponding to one third of the transverse size of the central calorimeter cell. Figure~\ref{fig:ratioRad} shows the ratios of radial profiles obtained from simulated samples to those extracted from CALICE test beam data. The comparison of radial profiles demonstrates a tendency similar to that observed for longitudinal profiles. The radial development of proton showers is predicted by the {\sffamily FTFP\_BERT} physics list within systematic uncertainties.  The energy deposition near the shower axis is overestimated by {\sffamily FTFP\_BERT} for pions by up to 20\% at 80~GeV and significantly overestimated by {\sffamily QGSP\_BERT} for both pions and protons (up to 30\%). The deposition far from the shower axis is underestimated for pions by both physics lists. The underestimation of the halo deposition by the {\sffamily QGSP\_BERT} physics list is larger and amounts up to 10\% for both pions and protons

%%%%%%%%%%%%%%%%%%%%%%%%%%%%%%%%%%%%%%%%%%%%%%%%%%%%%%%%%%%%%%
\section{Parametrisation and fit of hadron shower profiles}
\label{sec:para}

Parametrisation is an instrument for quantitative comparison of the observed shower development with predictions of Monte Carlo models. The available granularity provides a detailed picture of the longitudinal profile with a step size of 0.14$\lambda_{\mathrm{I}}$ up to a depth of $\sim$4.5$\lambda_{\mathrm{I}}$ and of the radial profile a step size of 10~mm up to a width of 340~mm. 

It should be noted that to achieve stable fit results and reliable error estimates, no parameter limits  are applied during the minimisation. Instead, a random variation of initial values for the minimisation procedure has been used. From a sample of 100 attempts, the results with unphysical values have been rejected, and the best fit has been chosen. The obtained $\frac{\chi^2}{NDF}$ is better than 1.5 for the overwhelming majority and does not exceed 2.8 in the worst case. 

%===============================================
\subsection{Fit to longitudinal profiles}
\label{sec:paraLng}

The parametrisation of the longitudinal development of hadronic showers with a sum of two gamma distributions was proposed in ref.~\cite{Bock:1981} as a natural extension of the parametrisation of electromagnetic shower profiles. In previous studies, the application of such a parametrisation to real data required the convolution of this function with the exponential distribution of the shower start, as its position inside the calorimeter was unknown~\cite{ATLASprof:1999}. The fine granularity of the CALICE Fe-AHCAL gives us the opportunity to measure shower profiles from the shower start and parametrise them in the following way as a sum of ``short'' and ``long'' components

\begin{equation}   
\Delta E(z) = A \cdot 
\left\{ \frac{f}{\Gamma(\alpha_{\mathrm{short}})} \cdot \left(\frac{z}{\beta_{\mathrm{short}}}\right)^{\alpha_{\mathrm{short}}-1}
\cdot \frac{e^{-z/\beta_{\mathrm{short}}}}{\beta_{\mathrm{short}}}
+ \frac{1 - f}{\Gamma(\alpha_{\mathrm{long}})} \cdot \left(\frac{z}{\beta_{\mathrm{long}}}\right)^{\alpha_{\mathrm{long}}-1}
\cdot \frac{e^{-z/\beta_{\mathrm{long}}}}{\beta_{\mathrm{long}}}\right\},
\label{eq:lngProf}
\end{equation}

\noindent where $A$ is a scaling factor, $f$ is the fractional contribution of the ``short'' component with the shape parameter $\alpha_{\mathrm{short}}$ and the slope parameter $\beta_{\mathrm{short}}$, $\alpha_{\mathrm{long}}$ and  $\beta_{\mathrm{long}}$ are the shape and the slope parameters of the ``long'' component. 

\begin{figure}
 \centering
 \includegraphics[width=7.5cm]{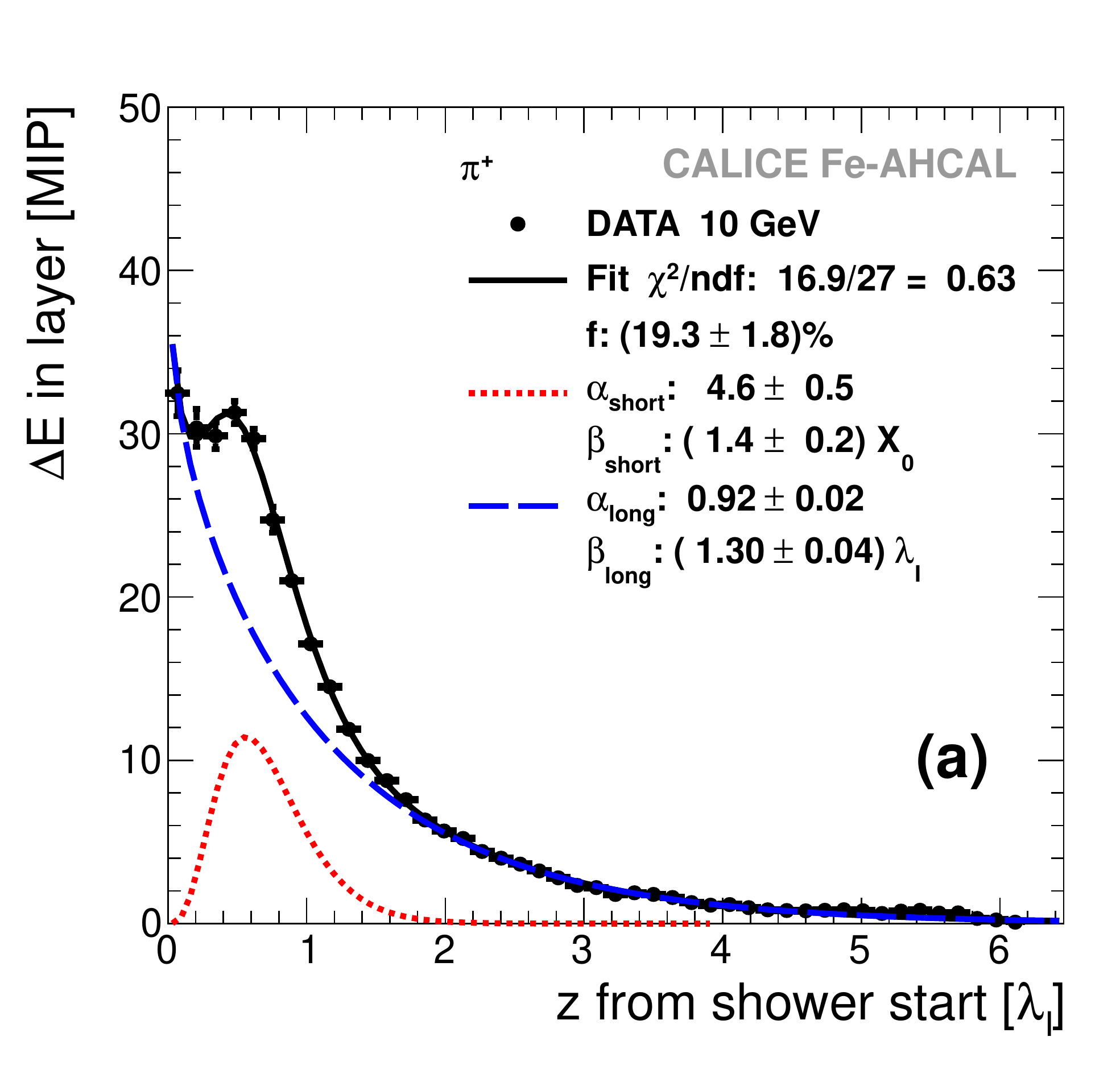}
 \includegraphics[width=7.5cm]{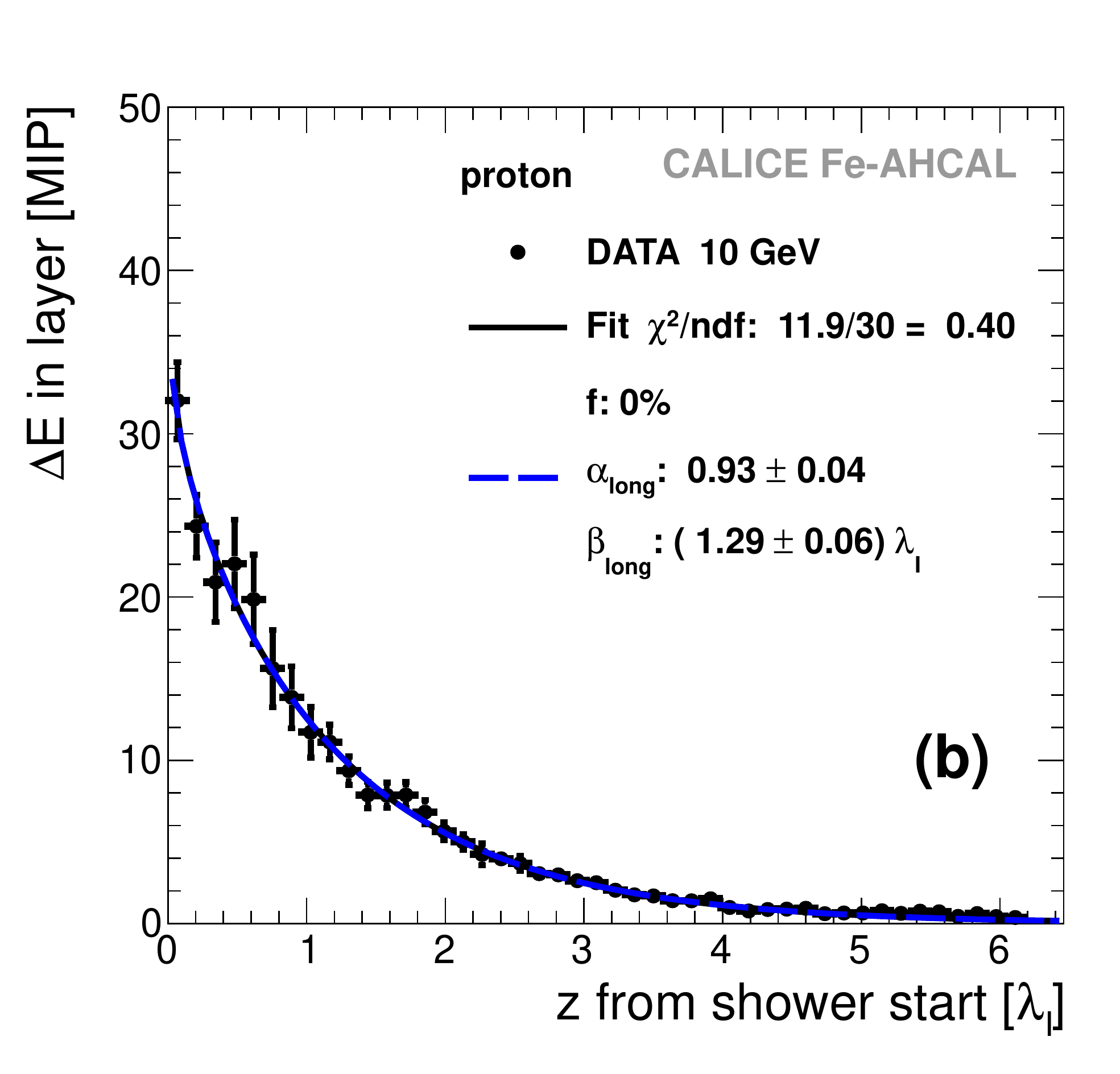}
 \includegraphics[width=7.5cm]{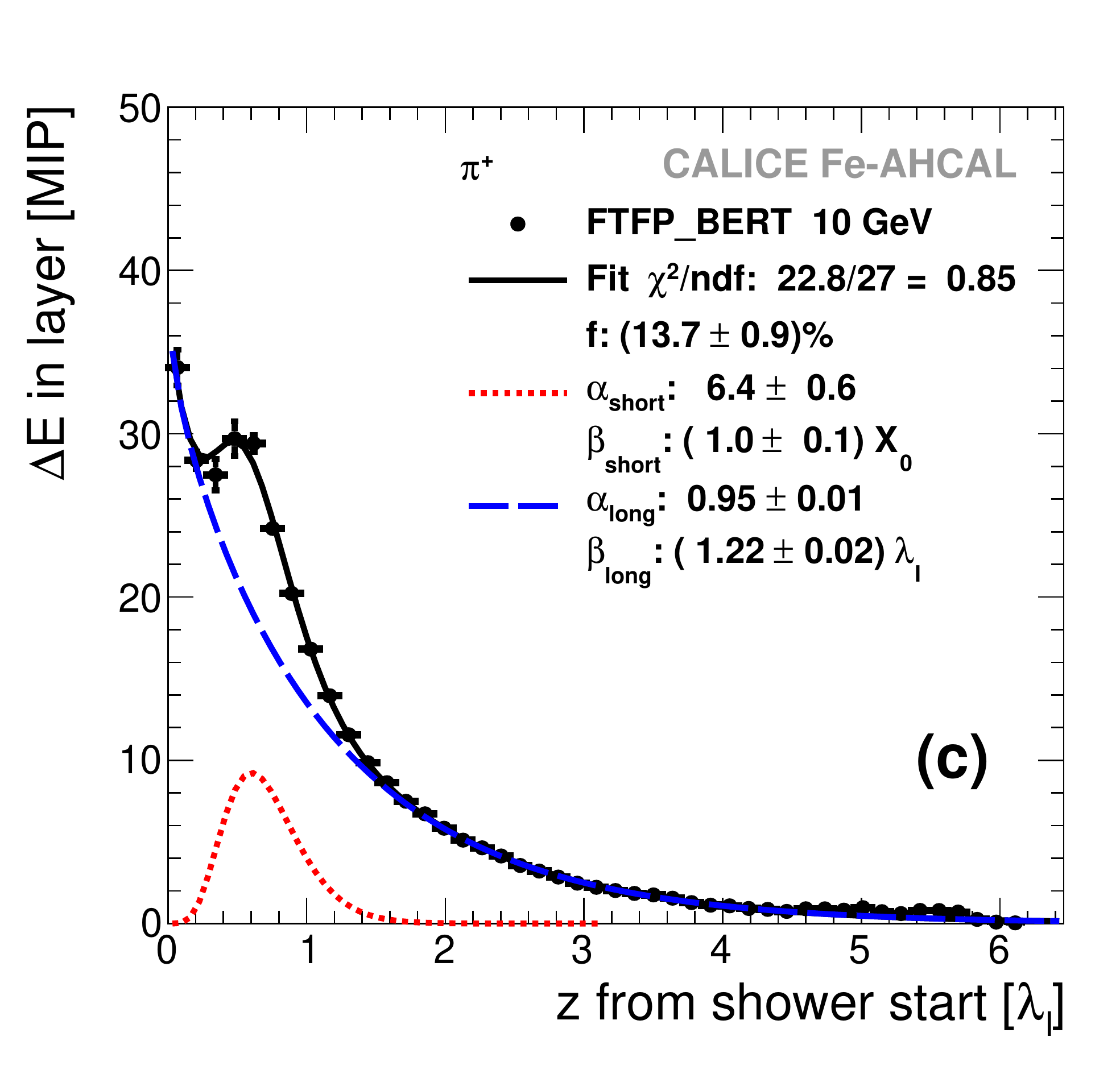}
 \includegraphics[width=7.5cm]{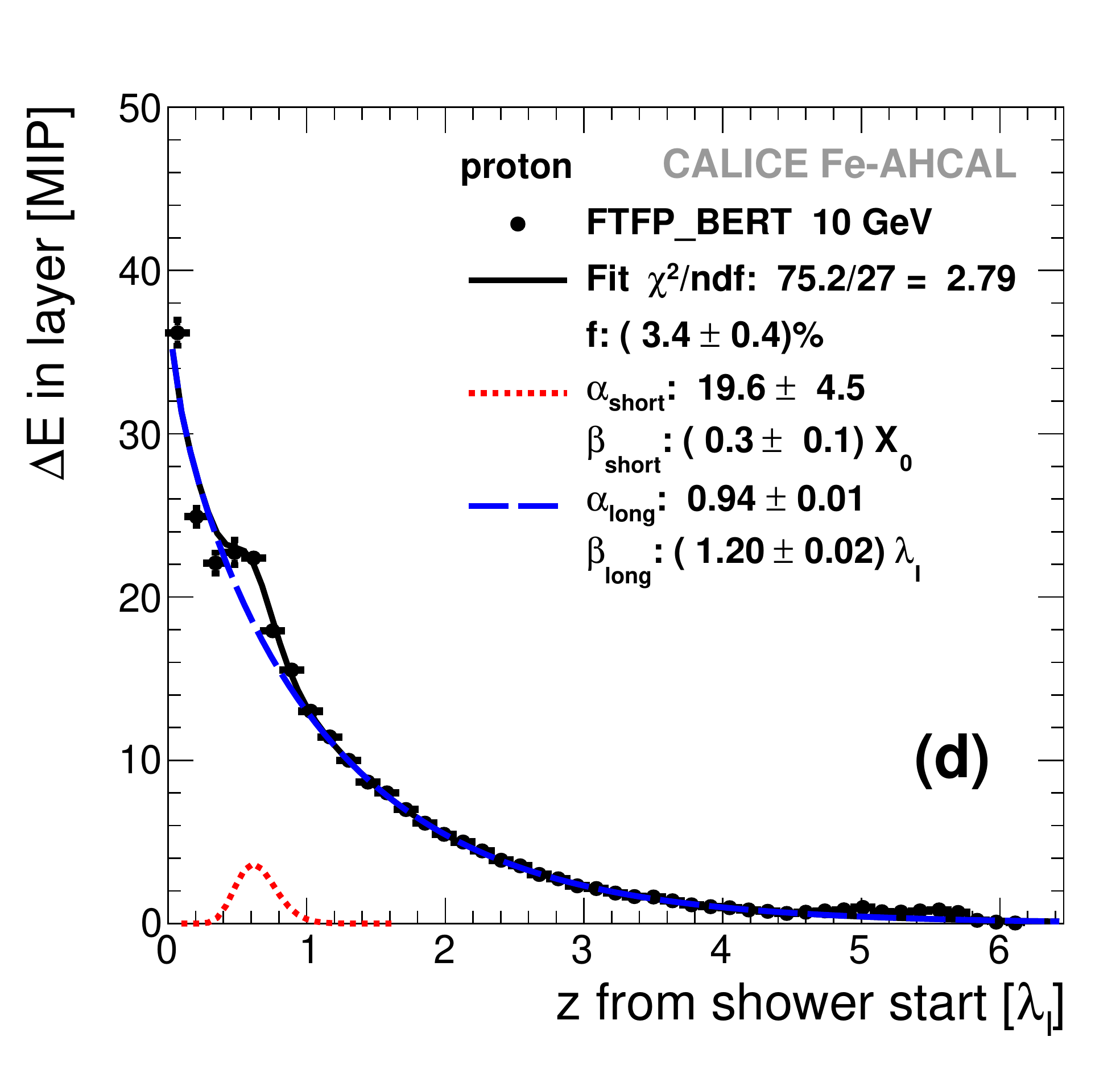}
 \caption{Fit of function (\protect\ref{eq:lngProf}) (black curves) to the longitudinal profiles of showers initiated by (a, c) pions and (b, d) protons with an initial energy of 10~GeV and extracted from (a, b) data and (c, d) simulations with the {\sffamily FTFP\_BERT} physics list. The dotted red and dashed blue curves show the contributions of the ``short'' and ``long'' components, respectively.}
 \label{fig:fitLng10}
\end{figure}

\begin{figure}
 \centering
 \includegraphics[width=7.5cm]{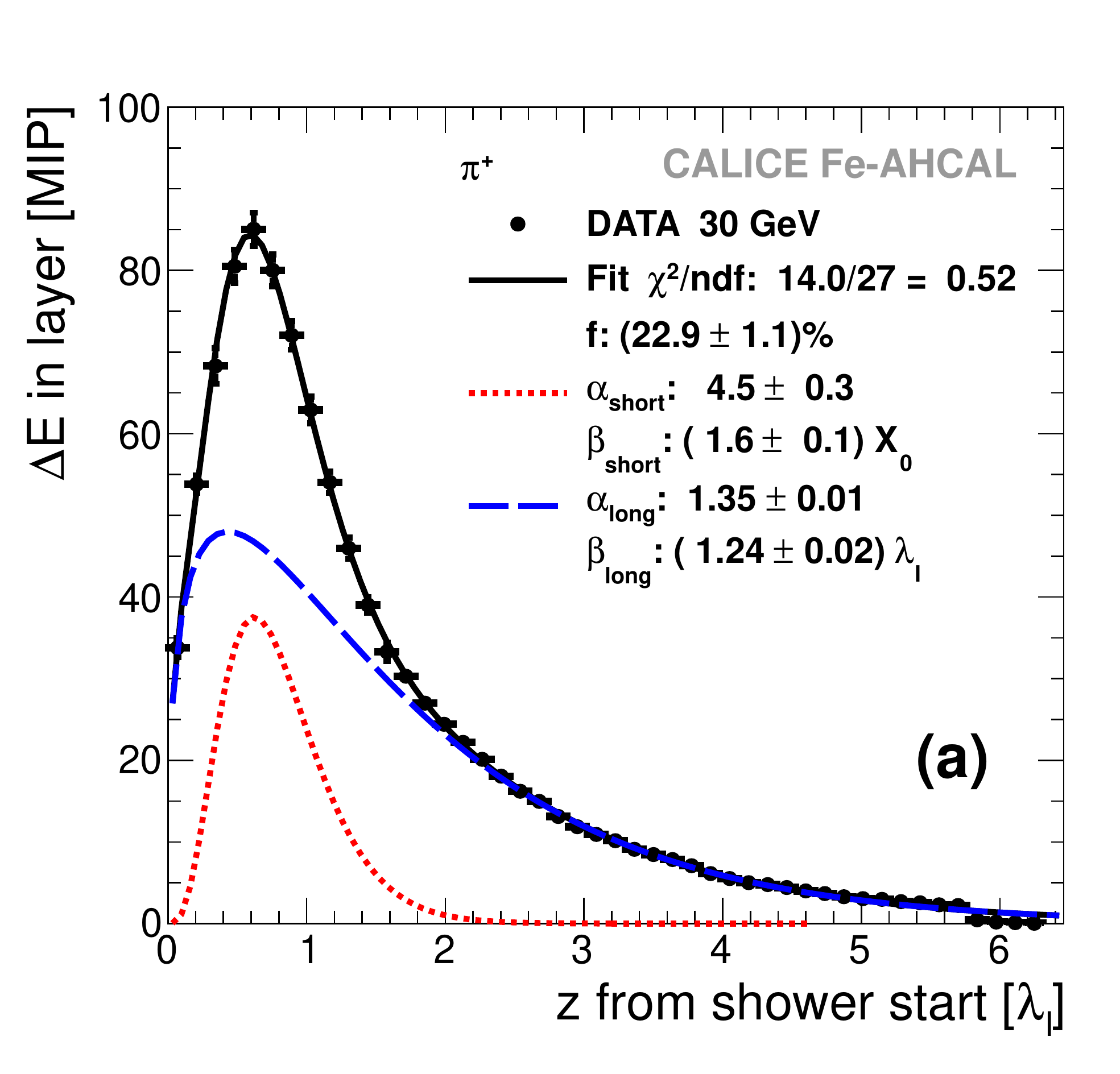}
 \includegraphics[width=7.5cm]{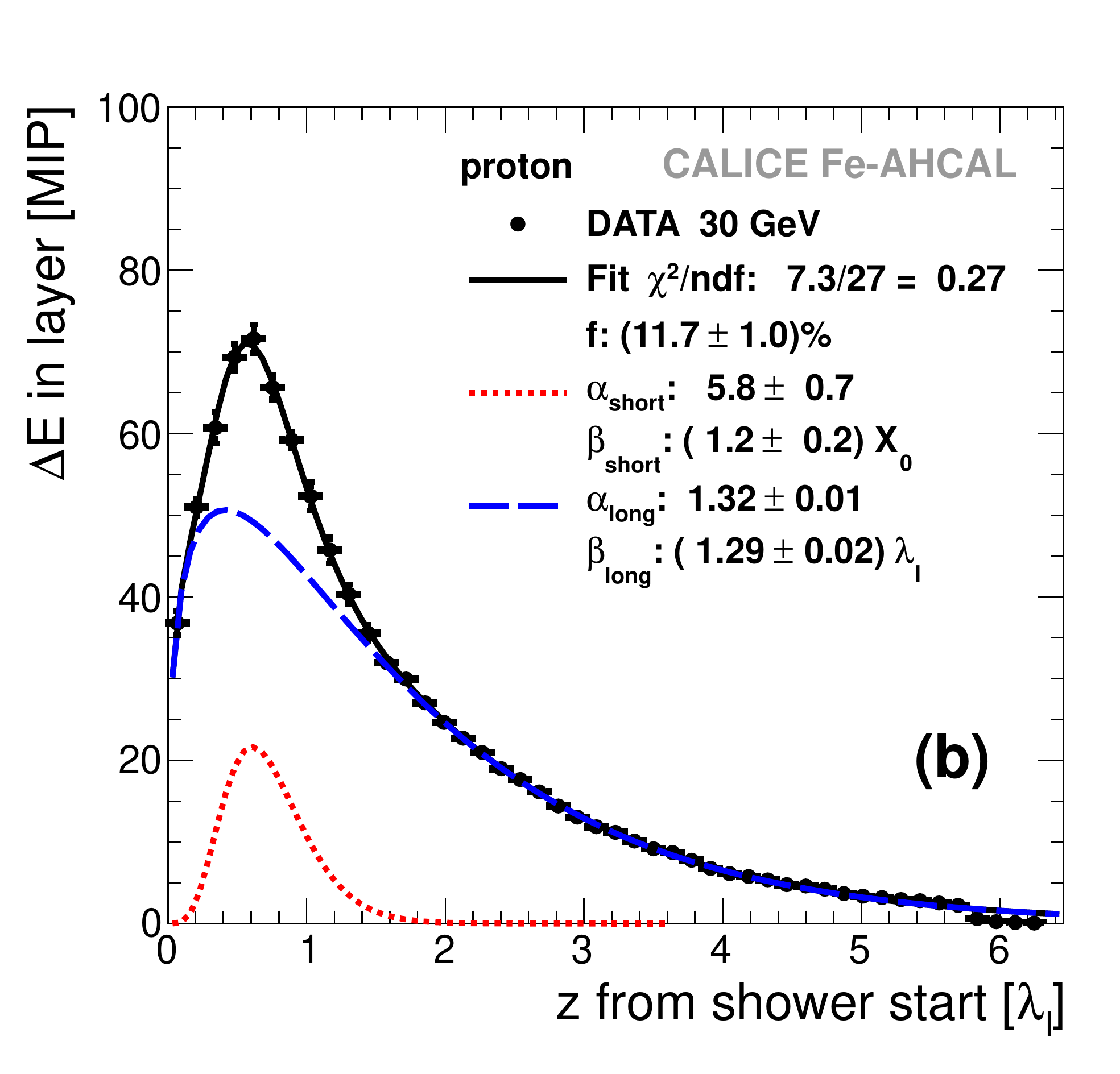}
 \includegraphics[width=7.5cm]{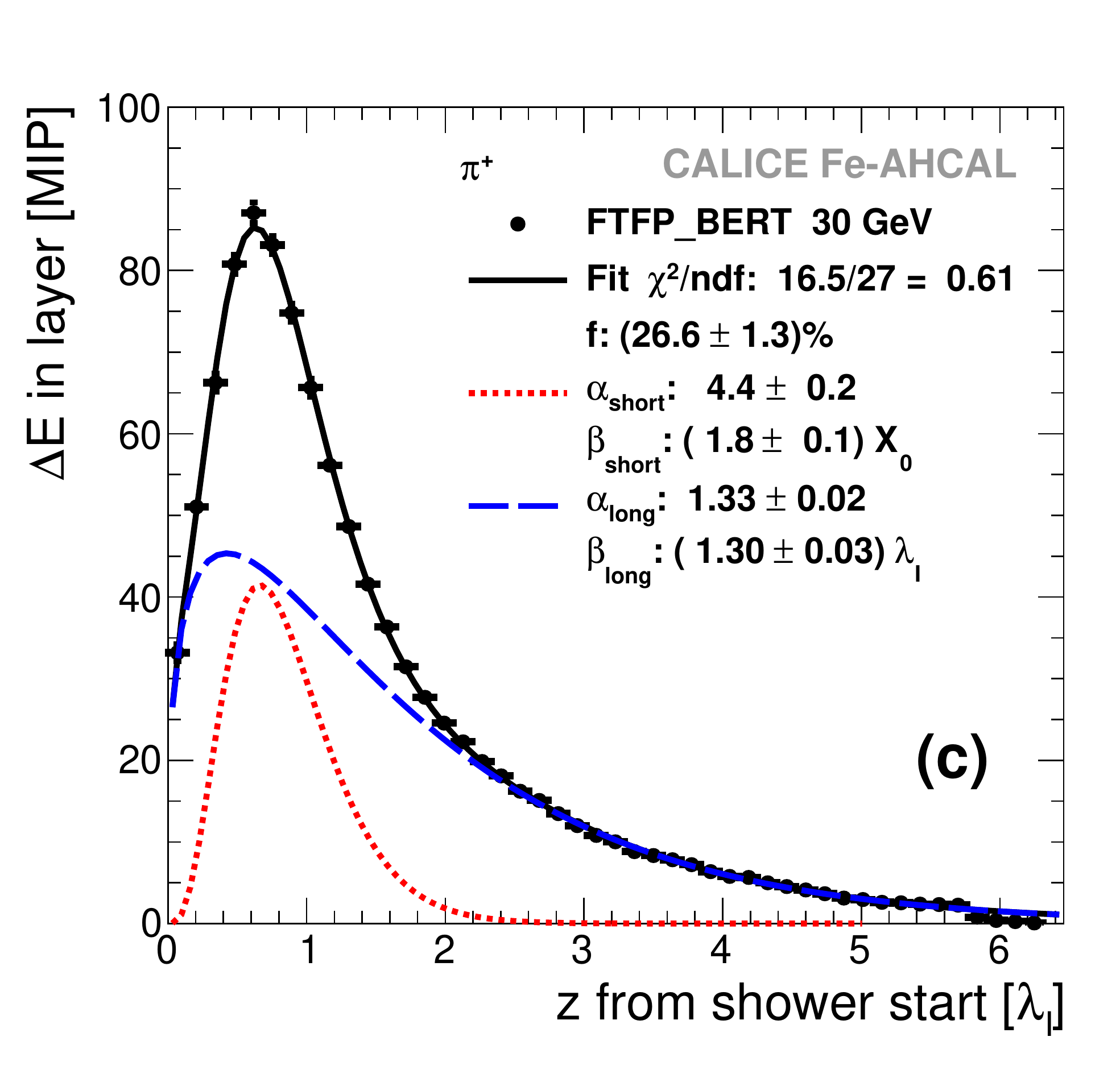}
 \includegraphics[width=7.5cm]{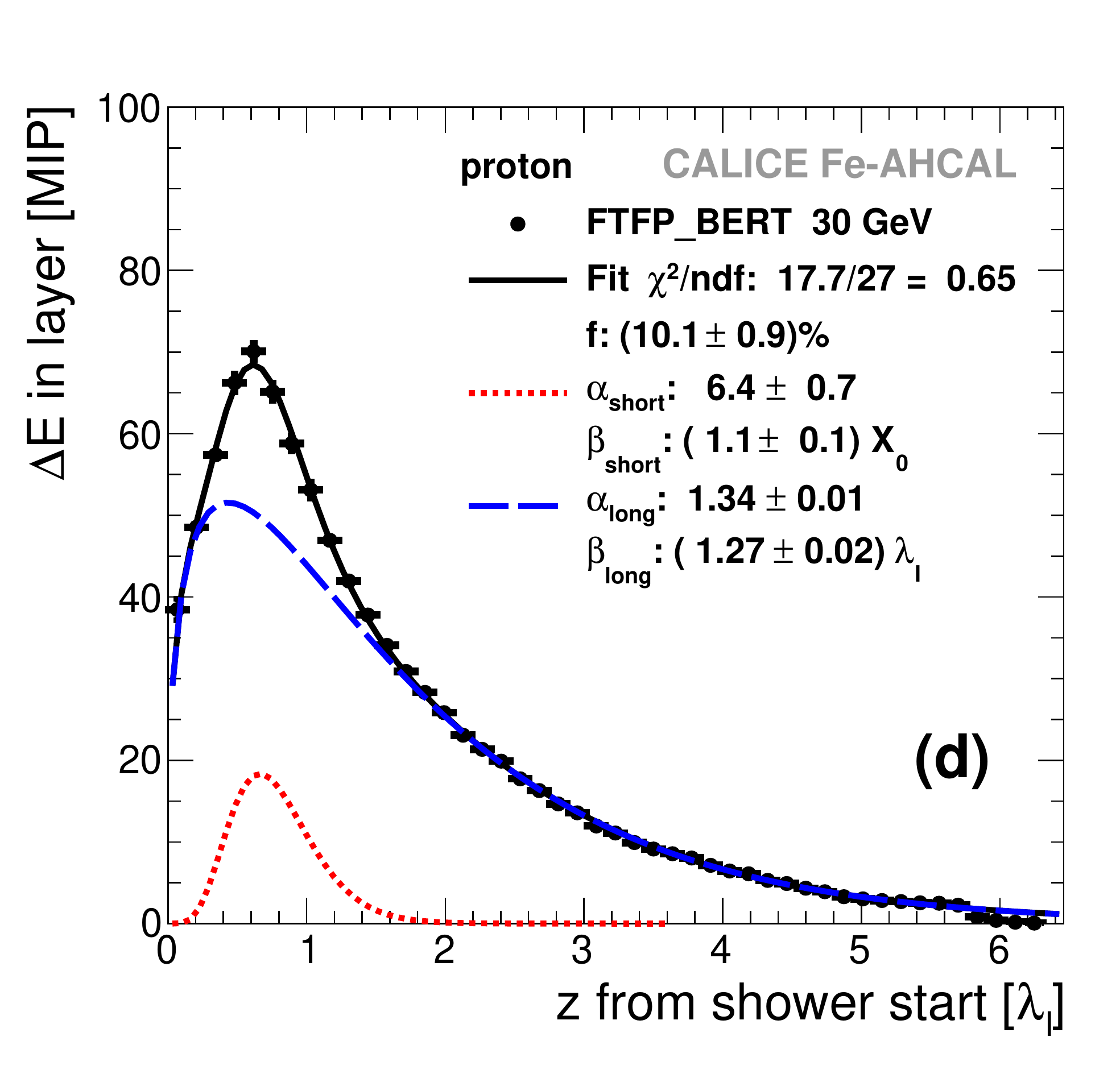}
 \caption{Fit of function (\protect\ref{eq:lngProf}) (black curves) to the longitudinal profiles of showers initiated by (a, c) pions and (b, d) protons with an initial energy of 30~GeV and extracted from (a, b) data and (c, d) simulations with the {\sffamily FTFP\_BERT} physics list. The dotted red and dashed blue curves show the contributions of the "short" and "long" components, respectively.}
 \label{fig:fitLng30}
\end{figure}

\begin{figure}
 \centering
 \includegraphics[width=7.5cm]{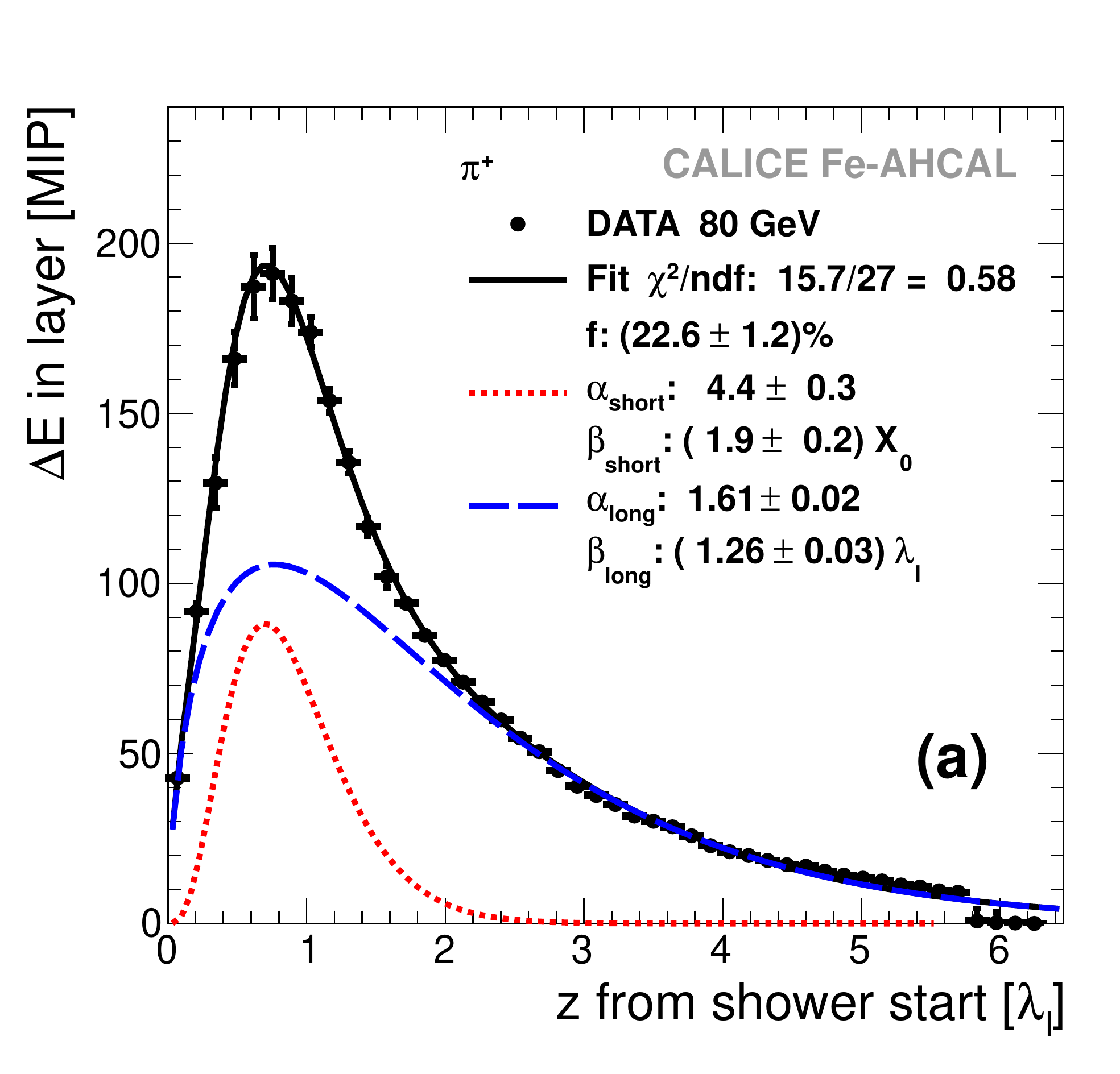}
 \includegraphics[width=7.5cm]{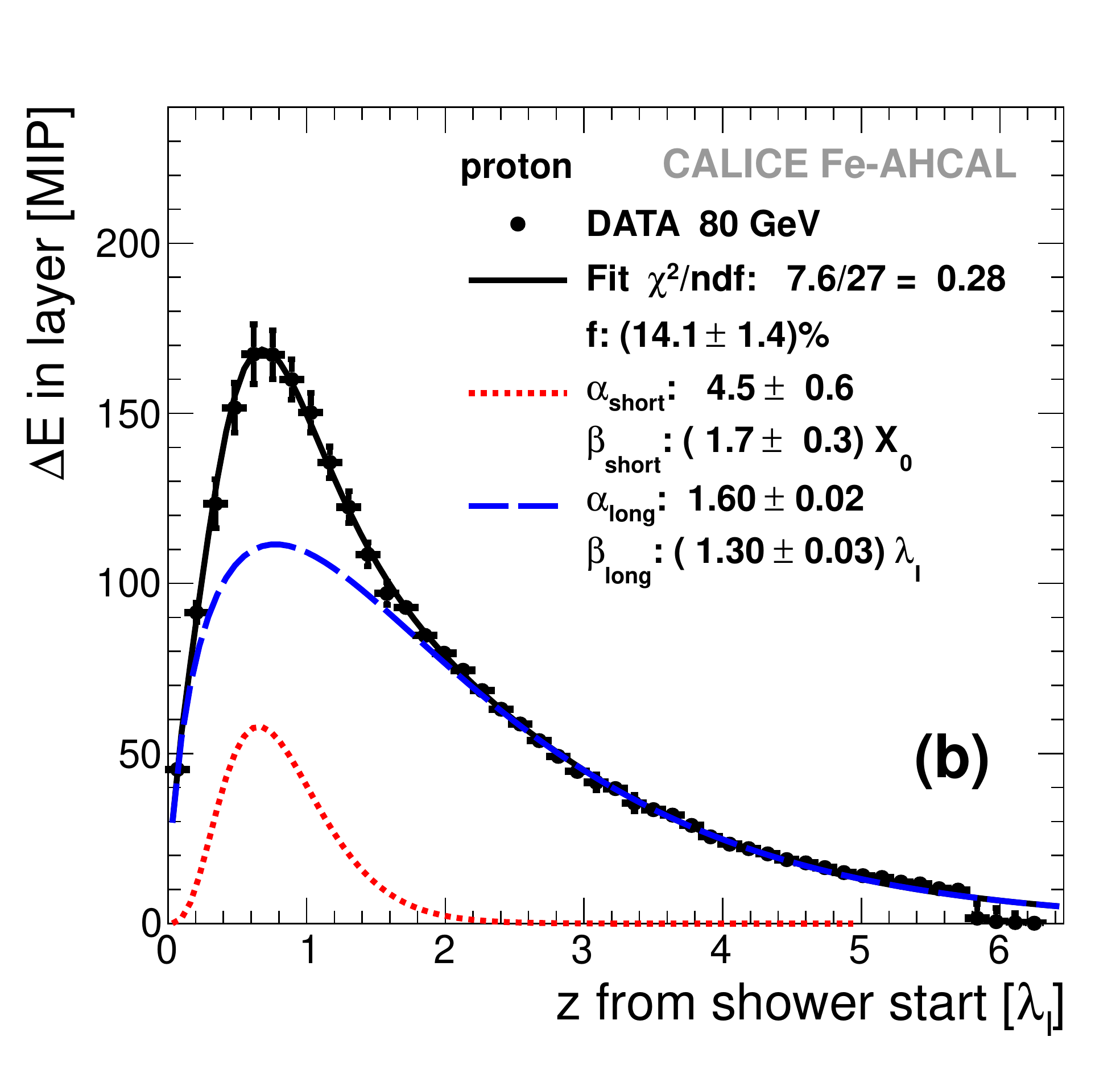}
 \includegraphics[width=7.5cm]{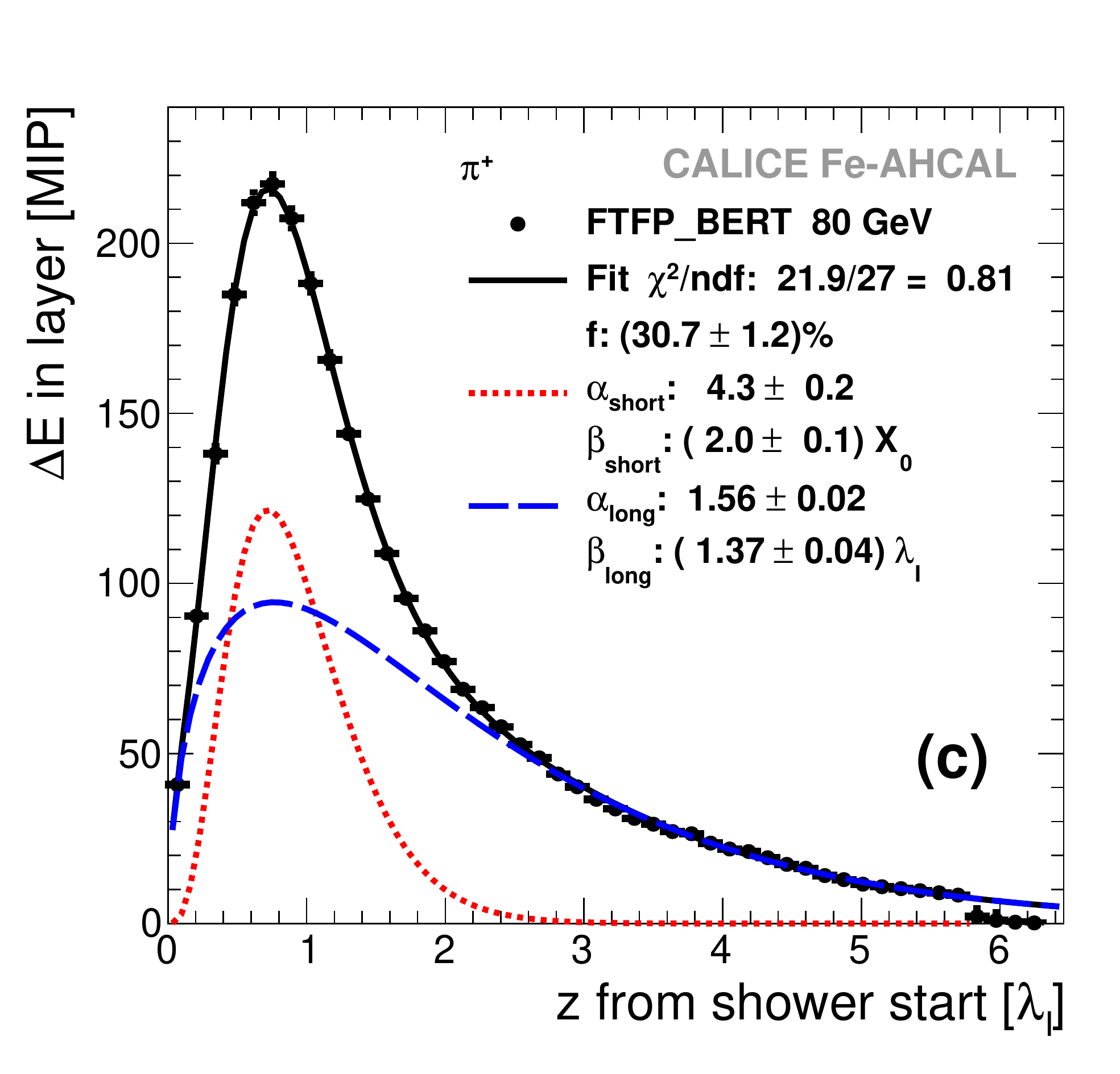}
 \includegraphics[width=7.5cm]{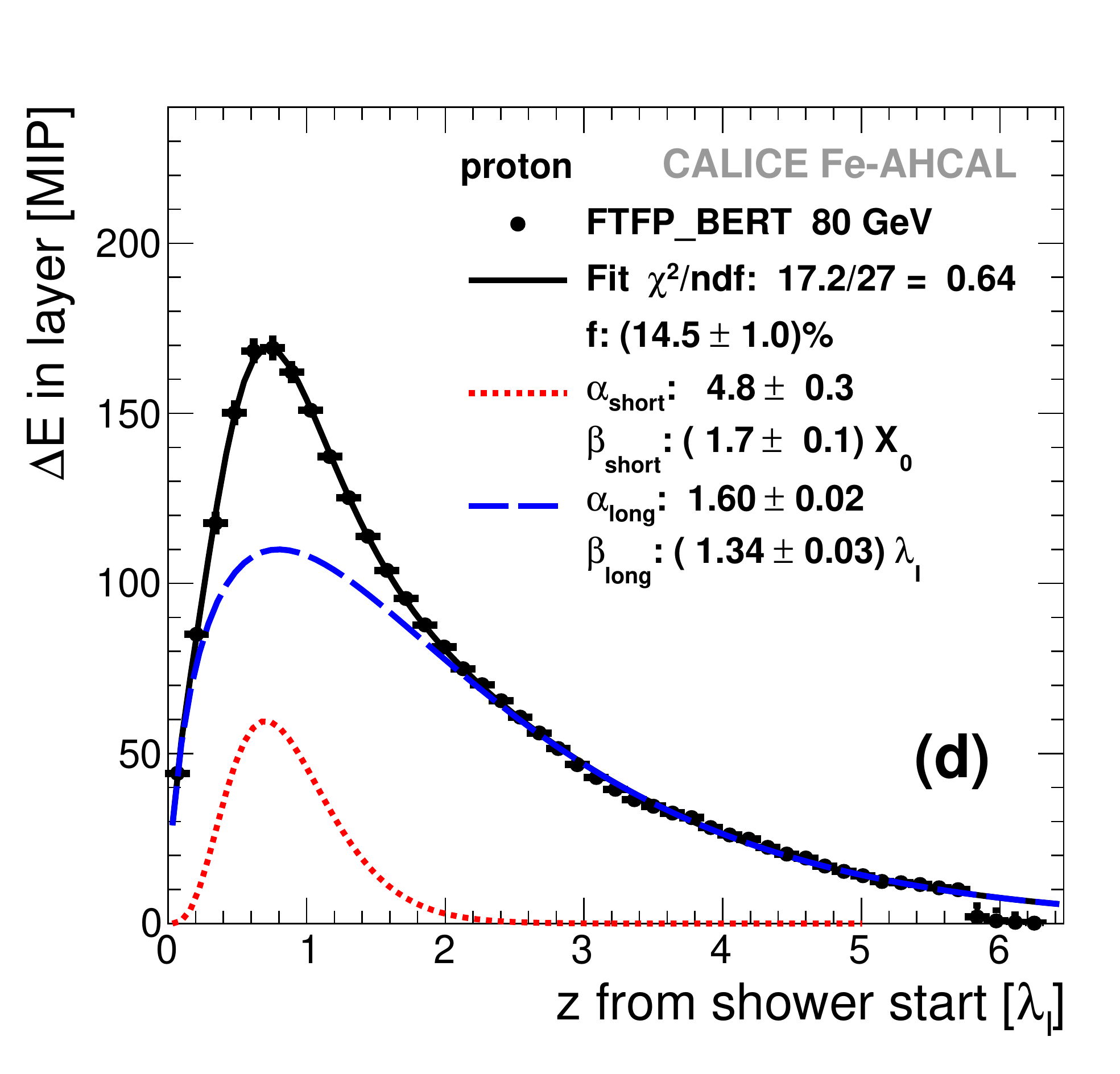}
 \caption{Fit of function (\protect\ref{eq:lngProf}) (black curves) to the longitudinal profiles of showers initiated by (a, c) pions and (b, d) protons with an initial energy of 80~GeV and extracted from (a, b) data and (c, d) simulations with the {\sffamily FTFP\_BERT} physics list. The dotted red and dashed blue curves show the contributions of the "short" and "long" components, respectively.}
 \label{fig:fitLng80}
\end{figure}

The upper limit of the fit range for the longitudinal profile is determined by the chosen range of shower start positions. Since only bins that belong to the Fe-AHCAL are used for the fit, the longitudinal fit range in the current analysis corresponds to a depth of $\sim$4.5$\lambda_{\mathrm{I}}$ from the shower start. The systematic uncertainties are estimated as described  in appendix~\ref{app:sys} and are summed up in quadrature to the statistical uncertainties.
The  slope parameter from the fit with the smaller absolute value is called $\beta_{\mathrm{short}}$ with the corresponding $\alpha_{\mathrm{short}}$ and the fractional contribution $f$.
Examples of fits to the longitudinal profiles are shown in figures~\ref{fig:fitLng10}, \ref{fig:fitLng30} and \ref{fig:fitLng80} for 10, 30 and 80~GeV respectively, for both pions and protons. The  parameters extracted from the fit to longitudinal profiles are listed in tables \ref{tab:parData}, \ref{tab:parFTFP} and \ref{tab:parQGSP} presented in appendix~\ref{app:parametr}. A good fit of function (\ref{eq:lngProf}) to the longitudinal profile of the proton data at 10~GeV can be achieved assuming zero contribution of the ``short'' component due to very large systematic uncertainties. For this reason, the fraction of the ``short'' component as well as the parameters $\alpha_{\mathrm{short}}$ and $\beta_{\mathrm{short}}$ for protons at 10~GeV are not extracted from data and therefore are not compared with simulations.

%===============================================
\subsection{Fit to radial profiles}
\label{sec:paraRad}
     
The transverse distribution of the energy density can be parametrised with the sum of a ``core'' component close to the shower axis and a ``halo'' component distant from the shower axis; 
     
\begin{equation}     
\frac{\Delta E}{\Delta S}(r) = 
A_{\mathrm{core}} \cdot {e^{-r/\beta_{\mathrm{core}}}} + 
A_{\mathrm{halo}} \cdot {e^{-r/\beta_{\mathrm{halo}}}},
\label{eq:radProf}
\end{equation}
    
\noindent where $A_{\mathrm{core}}$ and $A_{\mathrm{halo}}$ are scaling factors, $\beta_{\mathrm{core}}$ and $\beta_{\mathrm{halo}}$ are slope parameters. The slope parameter from the fit with the smaller absolute value is called $\beta_{\mathrm{core}}$. The systematic uncertainties are estimated as described  in appendix~\ref{app:sys} and are added up in quadrature to the statistical uncertainties. The peripheral points corresponding to the 12$\times$12~cm$^2$ cells are excluded from the fit to radial profiles.

\begin{figure}
 \centering
 \includegraphics[width=7.5cm]{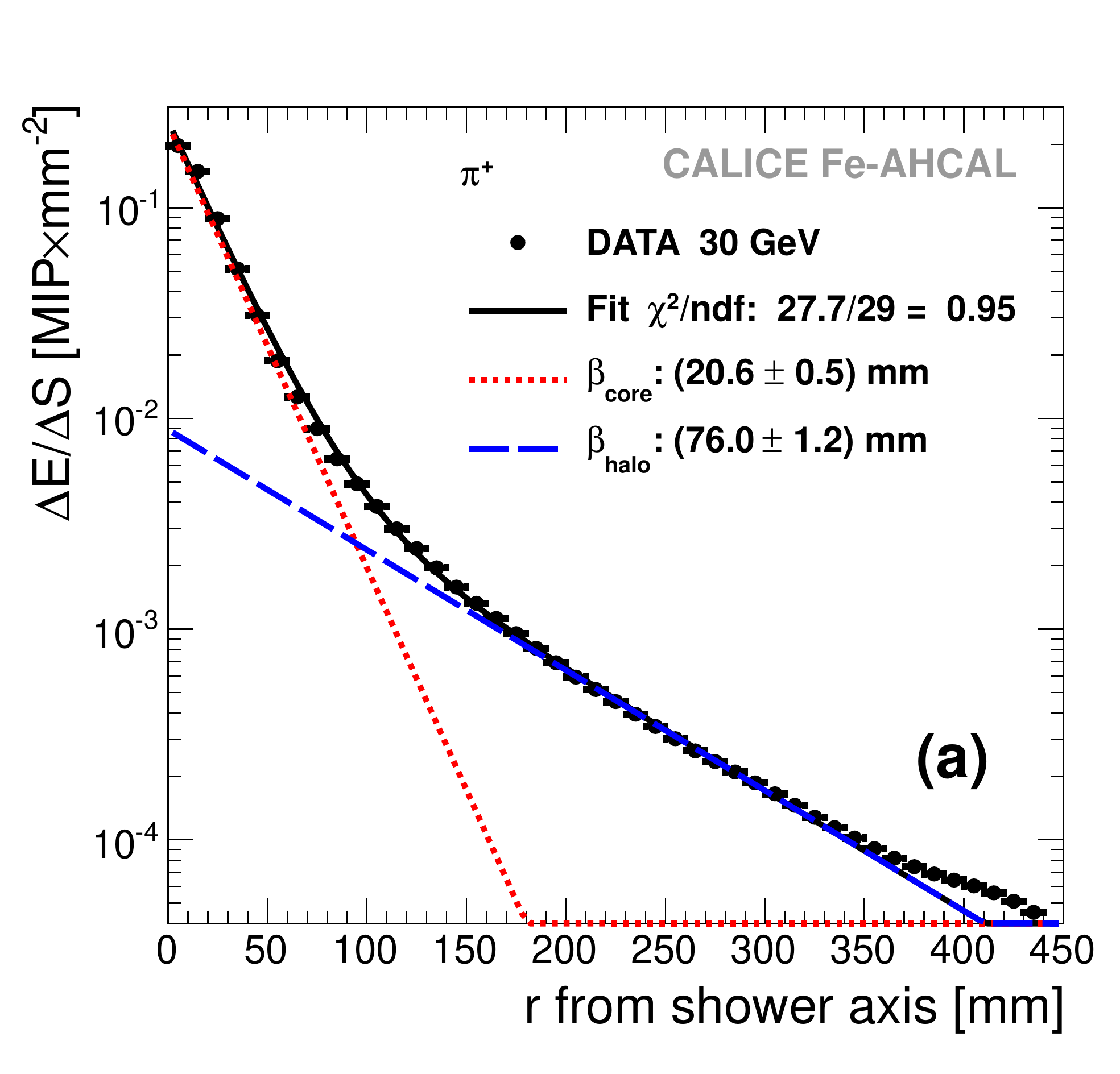}
 \includegraphics[width=7.5cm]{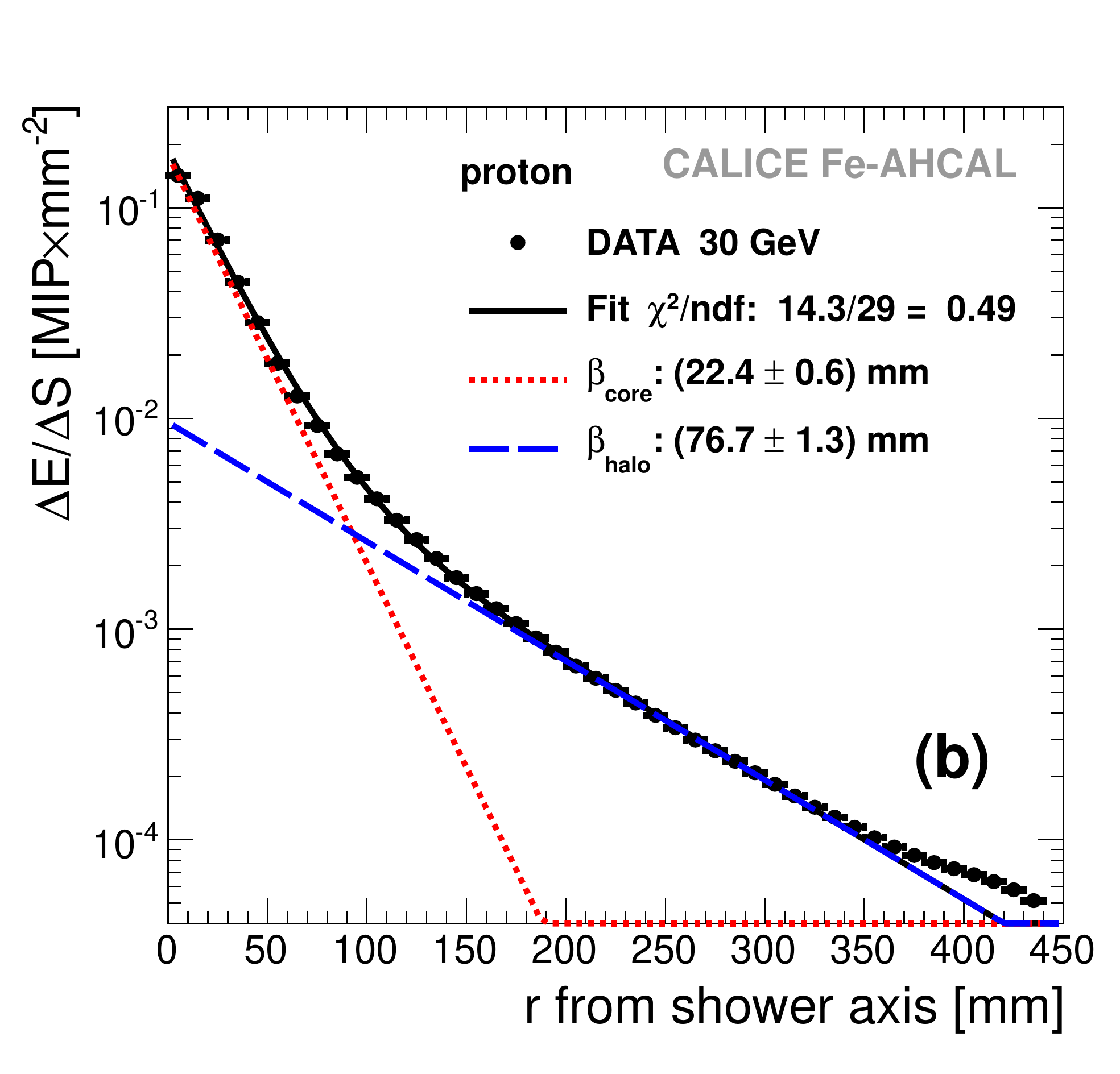}
 \includegraphics[width=7.5cm]{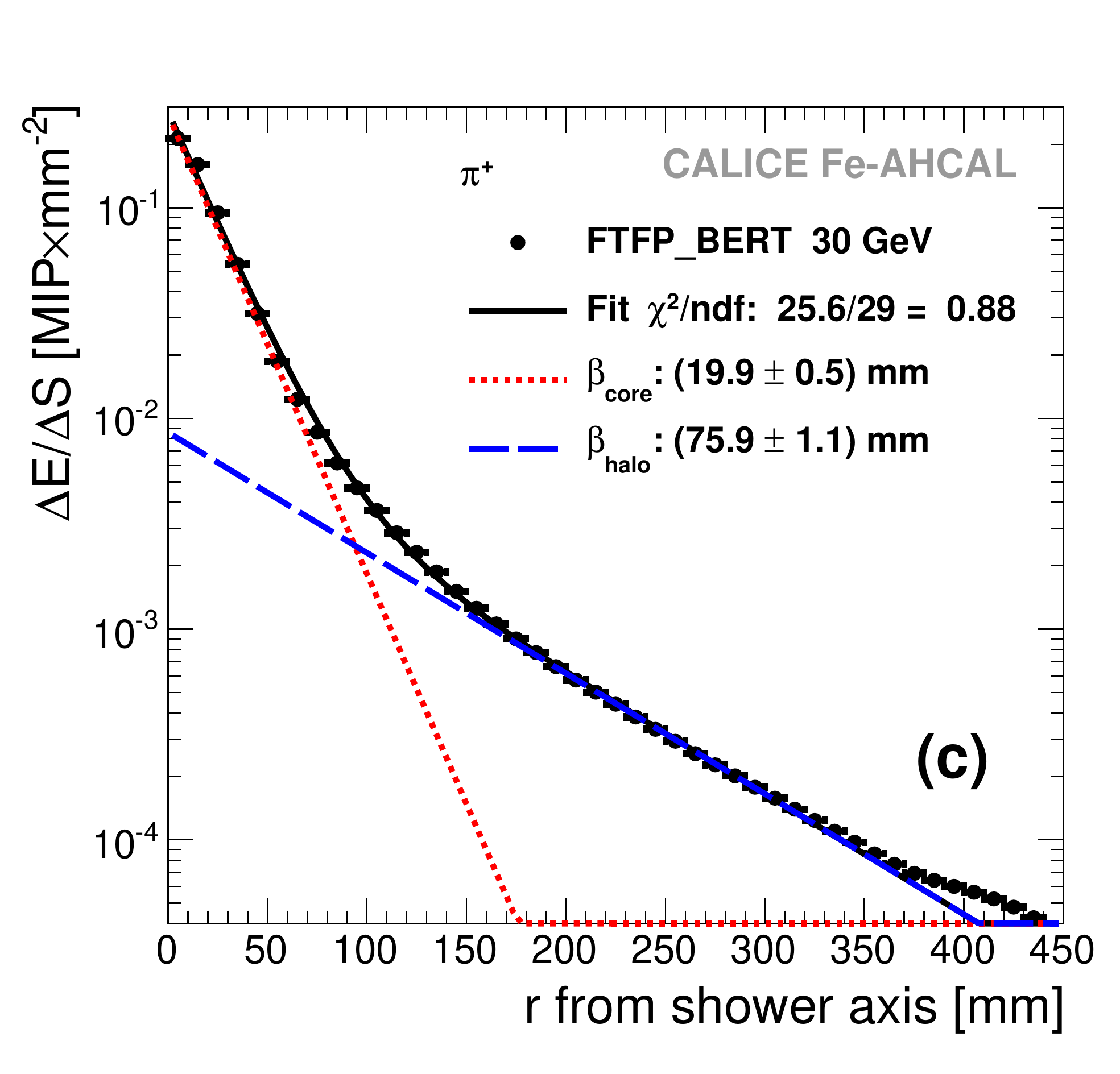}
 \includegraphics[width=7.5cm]{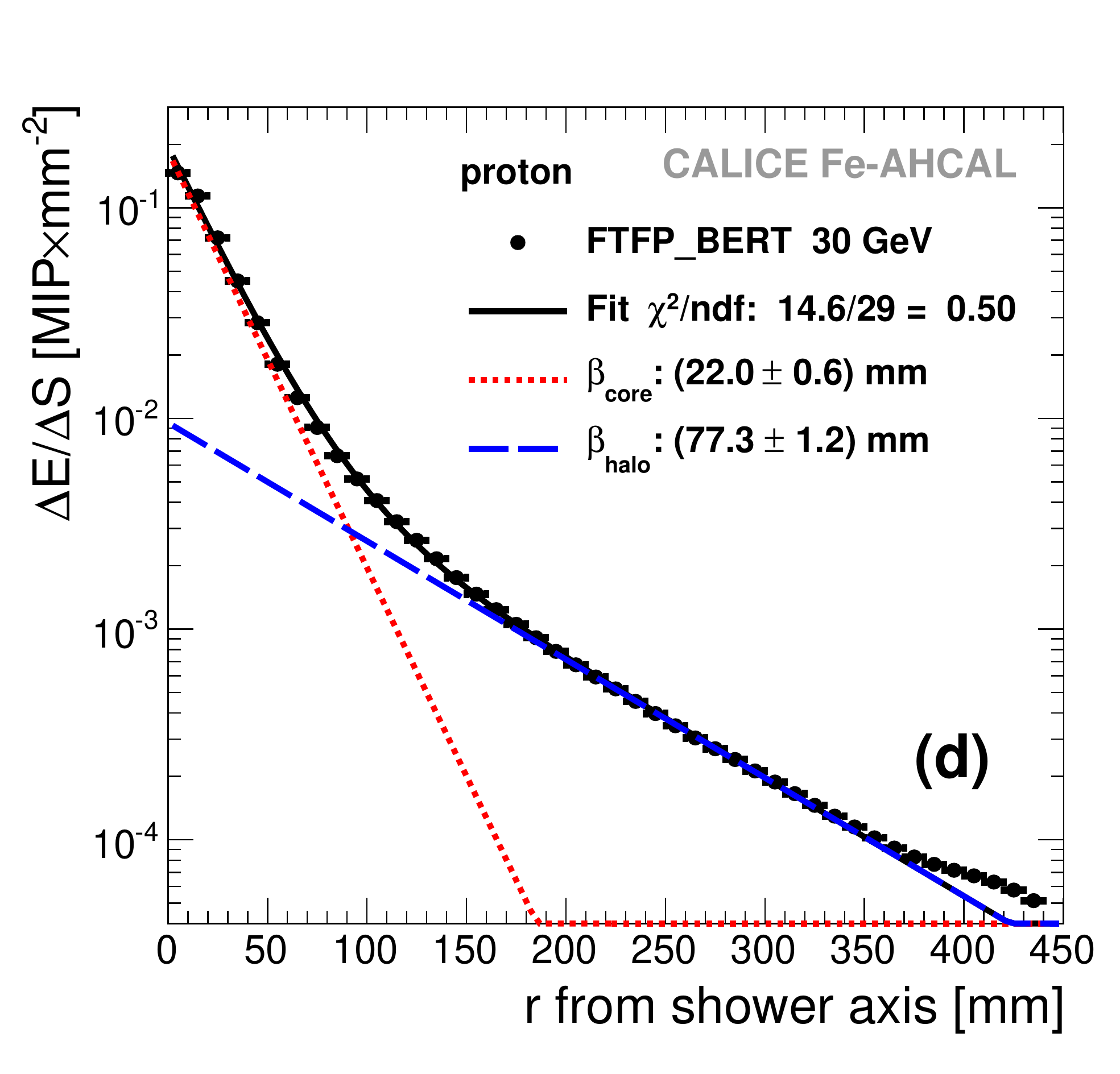}
 \caption{Fit of function (\protect\ref{eq:radProf}) (black curves) to the radial profiles of showers initiated by (a, c) pions and (b, d) protons with an initial energy of 30~GeV and extracted from (a, b) data and (c, d) simulations with the {\sffamily FTFP\_BERT} physics list. The dotted red and dashed blue curves show the contributions of ``core'' and ``halo'' components, respectively.}
 \label{fig:fitRad30}
\end{figure}

Examples of fits to the radial profiles are shown in figure~\ref{fig:fitRad30} for both pions and protons at 30~GeV.
The scale parameters $A_{\mathrm{core}}$ and $A_{\mathrm{halo}}$ indirectly represent the energy scale. 
The values of the slope parameters $\beta_{\mathrm{core}}$ and $\beta_{\mathrm{halo}}$ extracted from the fit to the radial profiles are listed in tables \ref{tab:parData}, \ref{tab:parFTFP} and \ref{tab:parQGSP} presented in appendix~\ref{app:parametr}.

%%%%%%%%%%%%%%%%%%%%%%%%%%%%%%%%%%%%%%%%%%%%%%%%%%%%%%%%%%%%%%
\section{Comparison of shower profile parameters}
\label{sec:paraComp}

The parametrisation of shower profiles provides the possibility for quantitative comparisons of parameters which characterise the shower development. The characteristic values of slope parameters for ``short'' and ``core'' components are $\sim$1.5$X_{0}$ and $\sim$1$R_{\mathrm{M}}$ respectively, comparable with the spatial parameters of electromagnetic showers. For the tail or halo region, the slope parameters are 10 and 4 times larger for longitudinal and radial profiles, respectively.

%===============================================
\subsection{``Long'' and ``halo'' parameters}
\label{sec:paraCompTail}

\begin{figure}
 \centering
 \includegraphics[width=7.0cm]{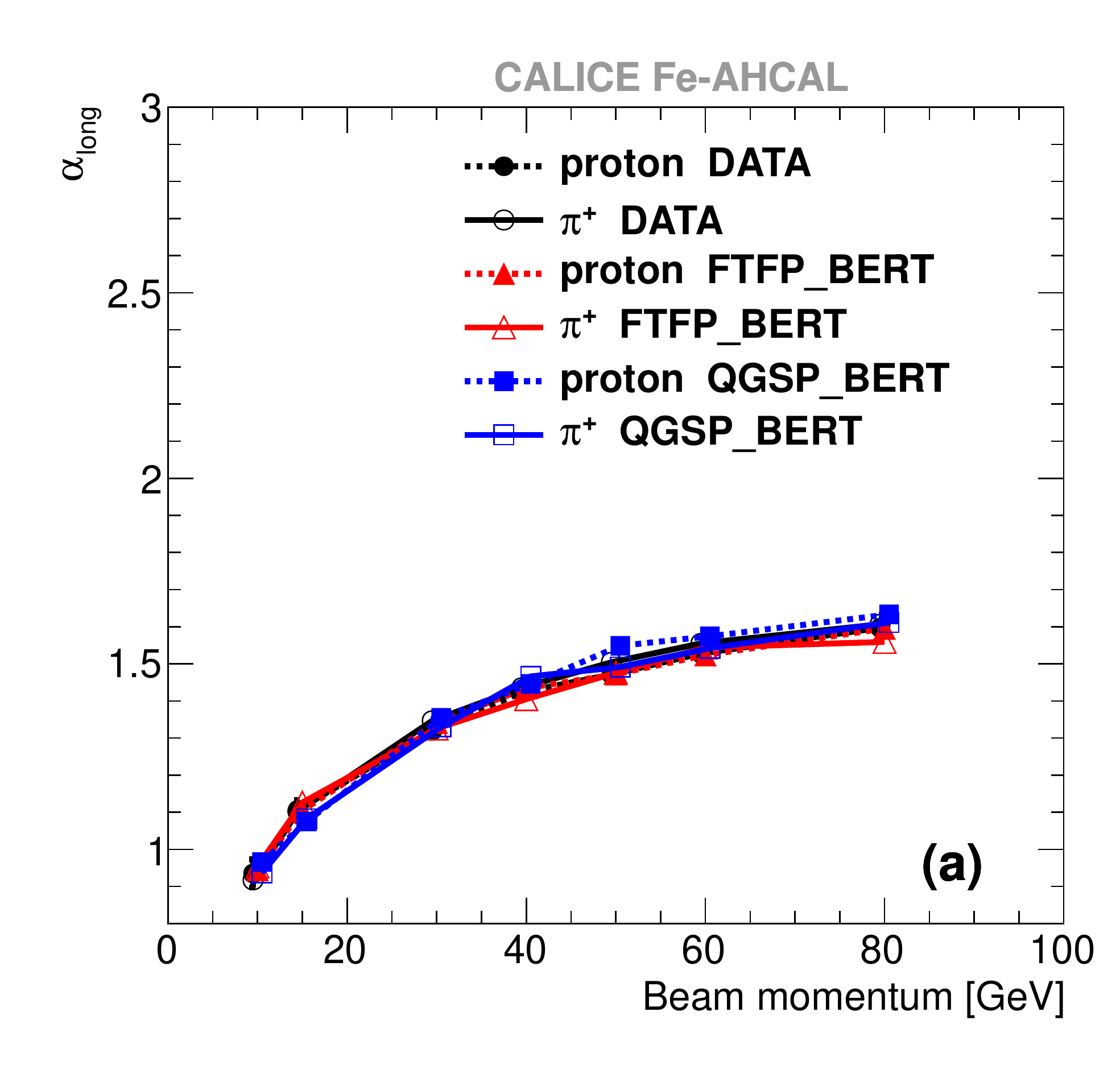} \\
 \includegraphics[width=7.0cm]{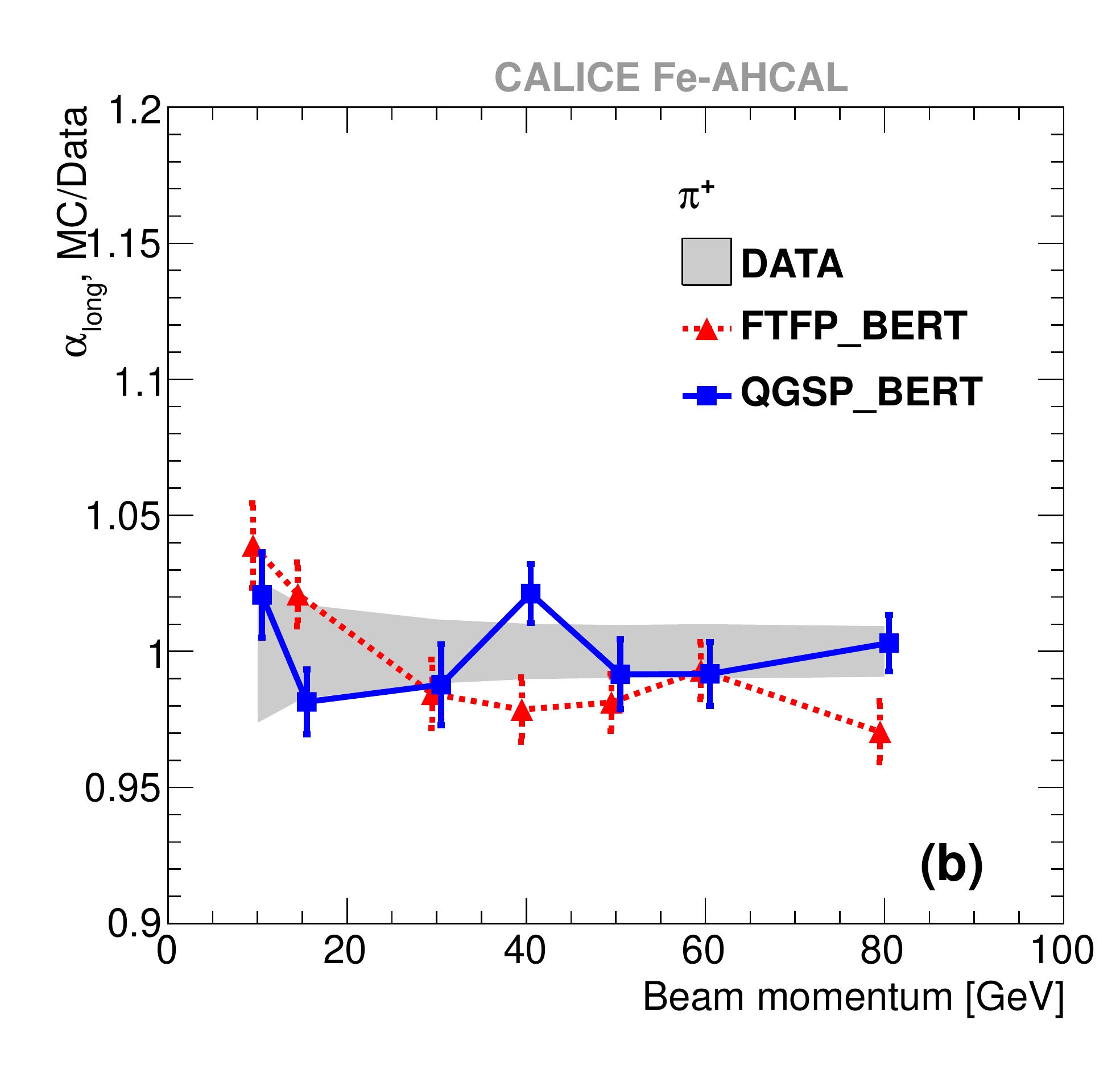}
 \includegraphics[width=7.0cm]{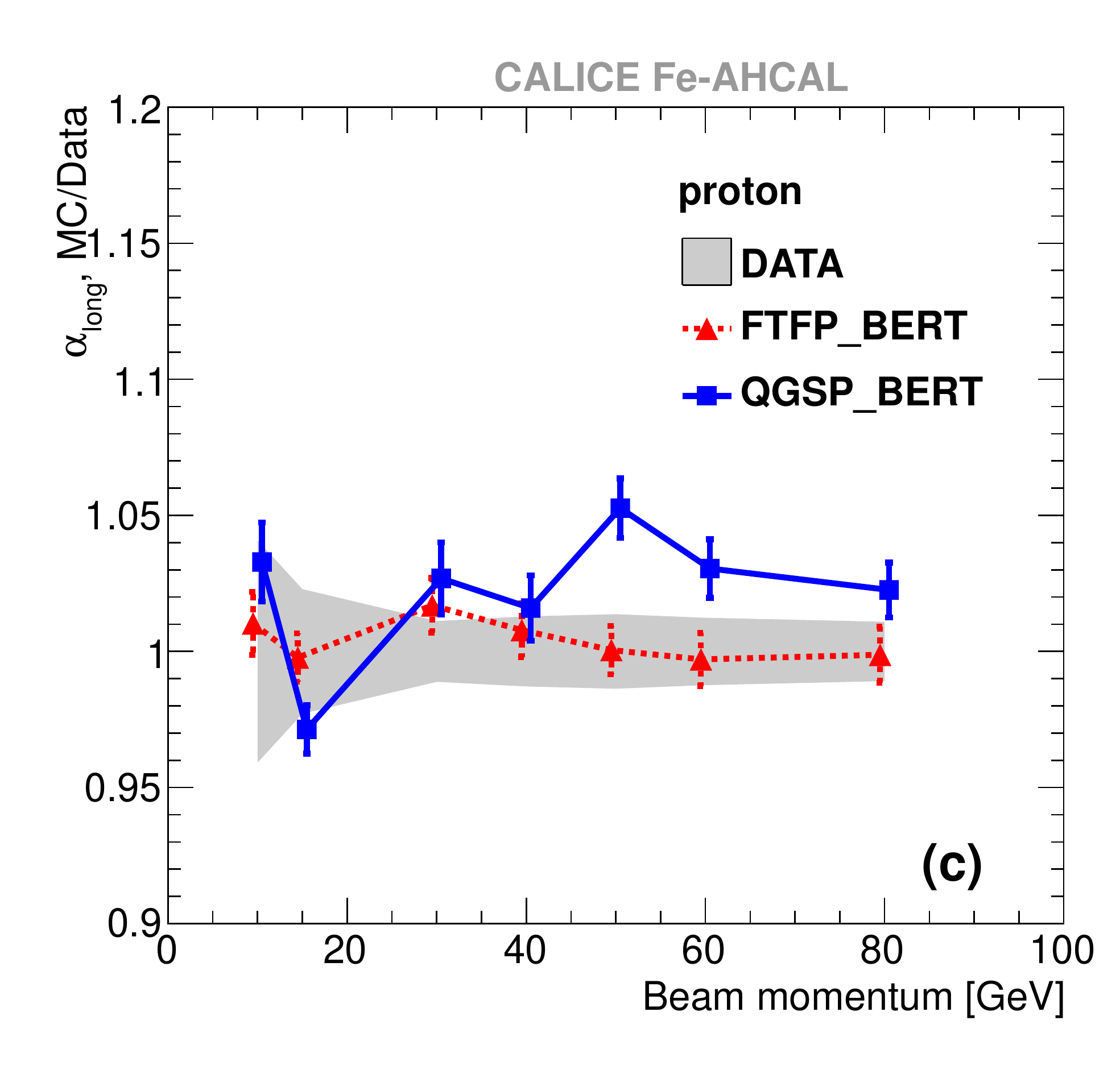}
 \caption{(a) Energy dependence of the shape parameter $\alpha_{\mathrm{long}}$, and the ratio of $\alpha_{\mathrm{long}}$ extracted from simulation to those extracted from data for (b) pions and (c) protons.}
 \label{fig:alphaLong}
\end{figure}

\begin{figure}
 \centering
 \includegraphics[width=7.0cm]{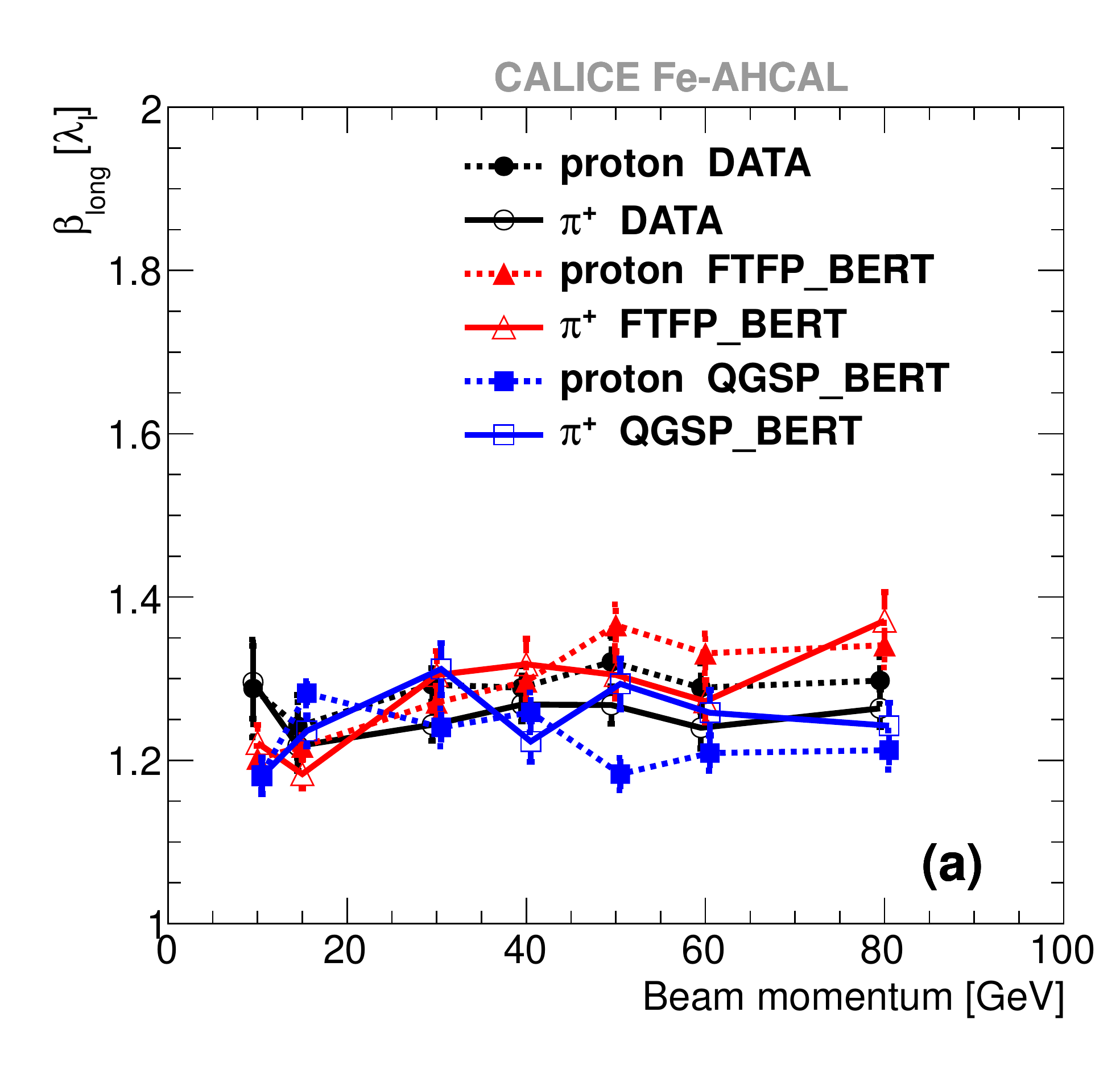}\\
 \includegraphics[width=7.0cm]{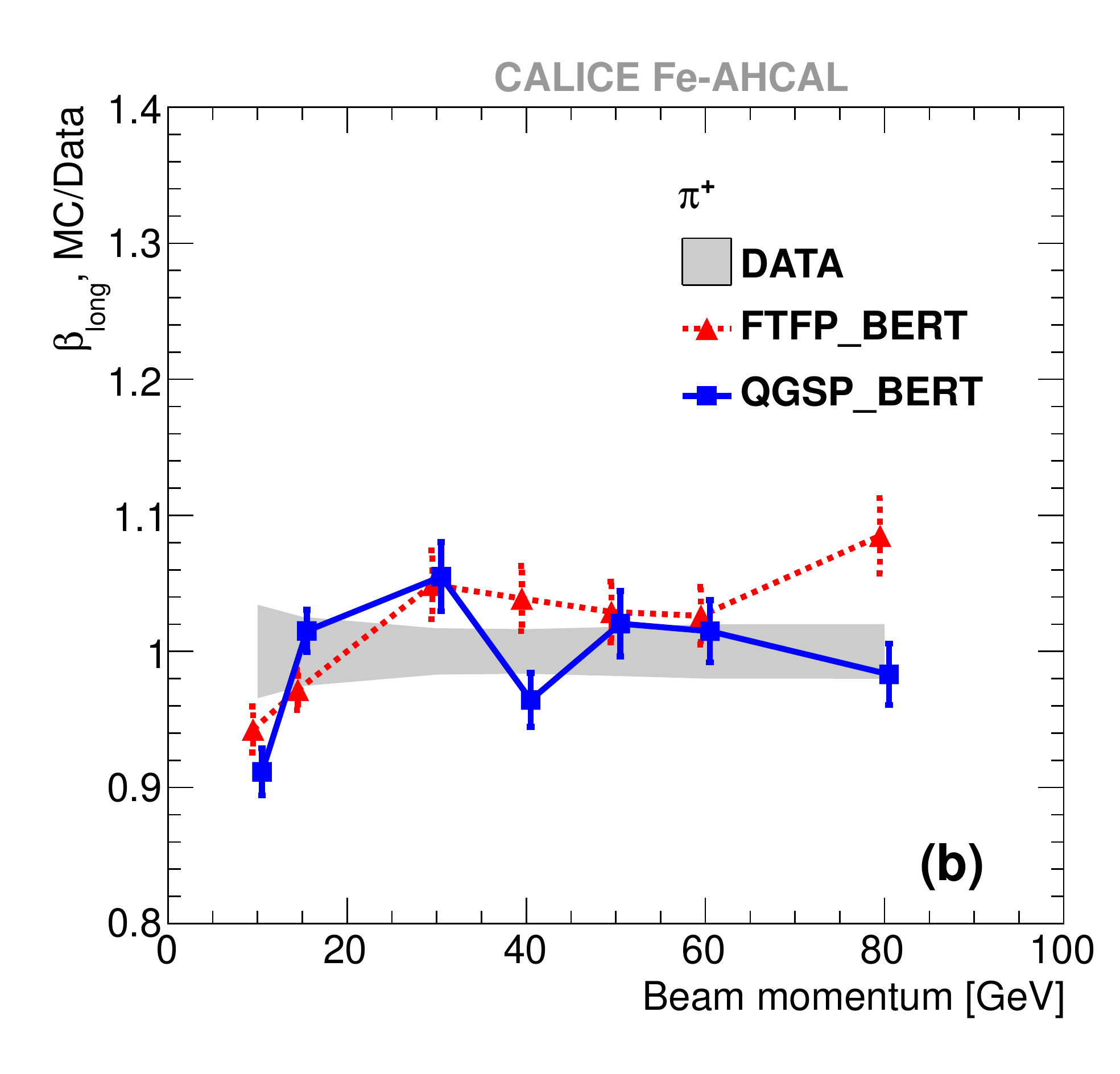}
 \includegraphics[width=7.0cm]{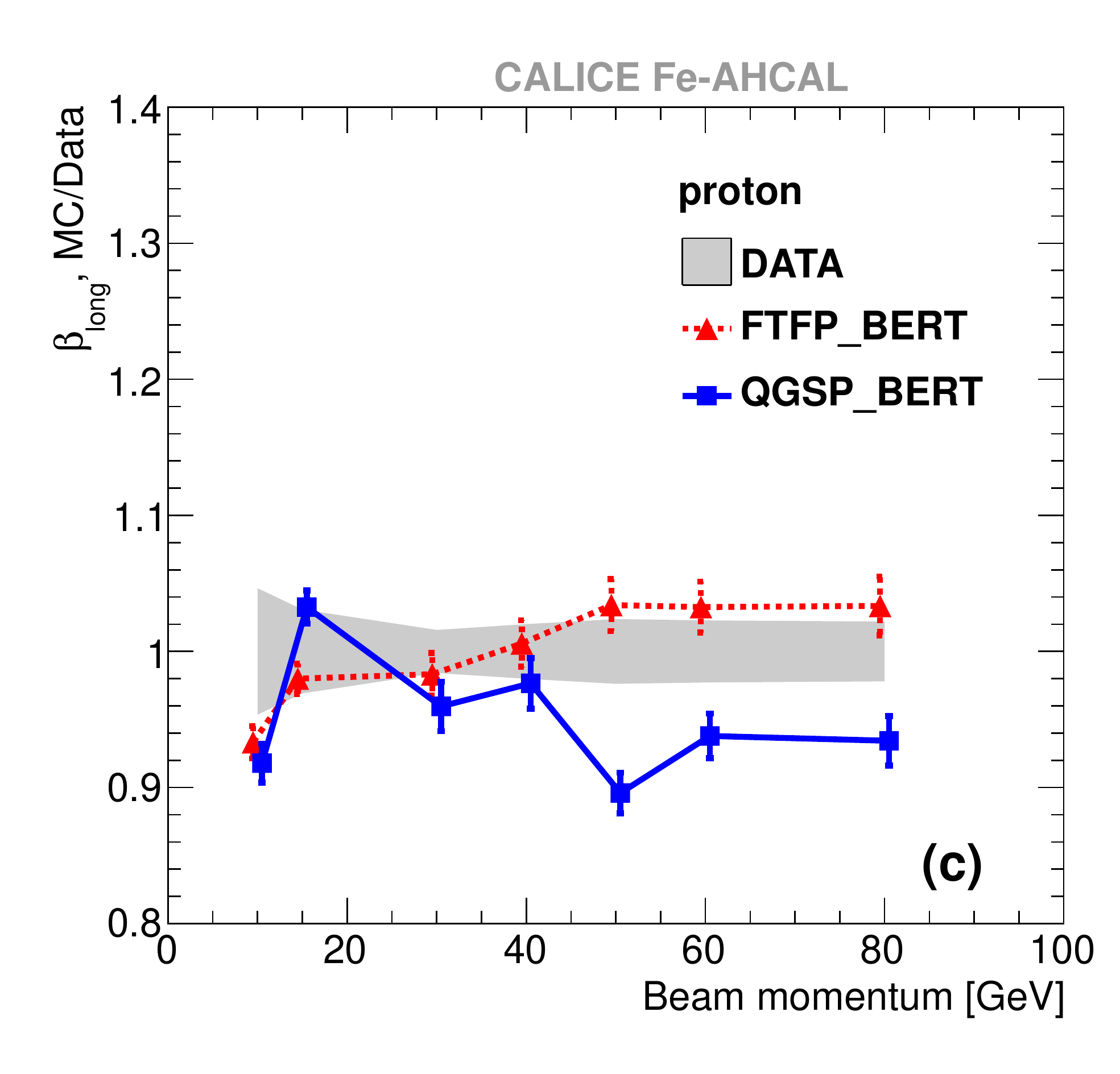}
 \caption{(a) Energy dependence of the tail slope parameter $\beta_{\mathrm{long}}$ and the ratio of $\beta_{\mathrm{long}}$ extracted from simulation to those extracted from data for (b) pions and (c) protons.}
 \label{fig:betaLong}
\end{figure}

\begin{figure}
 \centering
 \includegraphics[width=7.0cm]{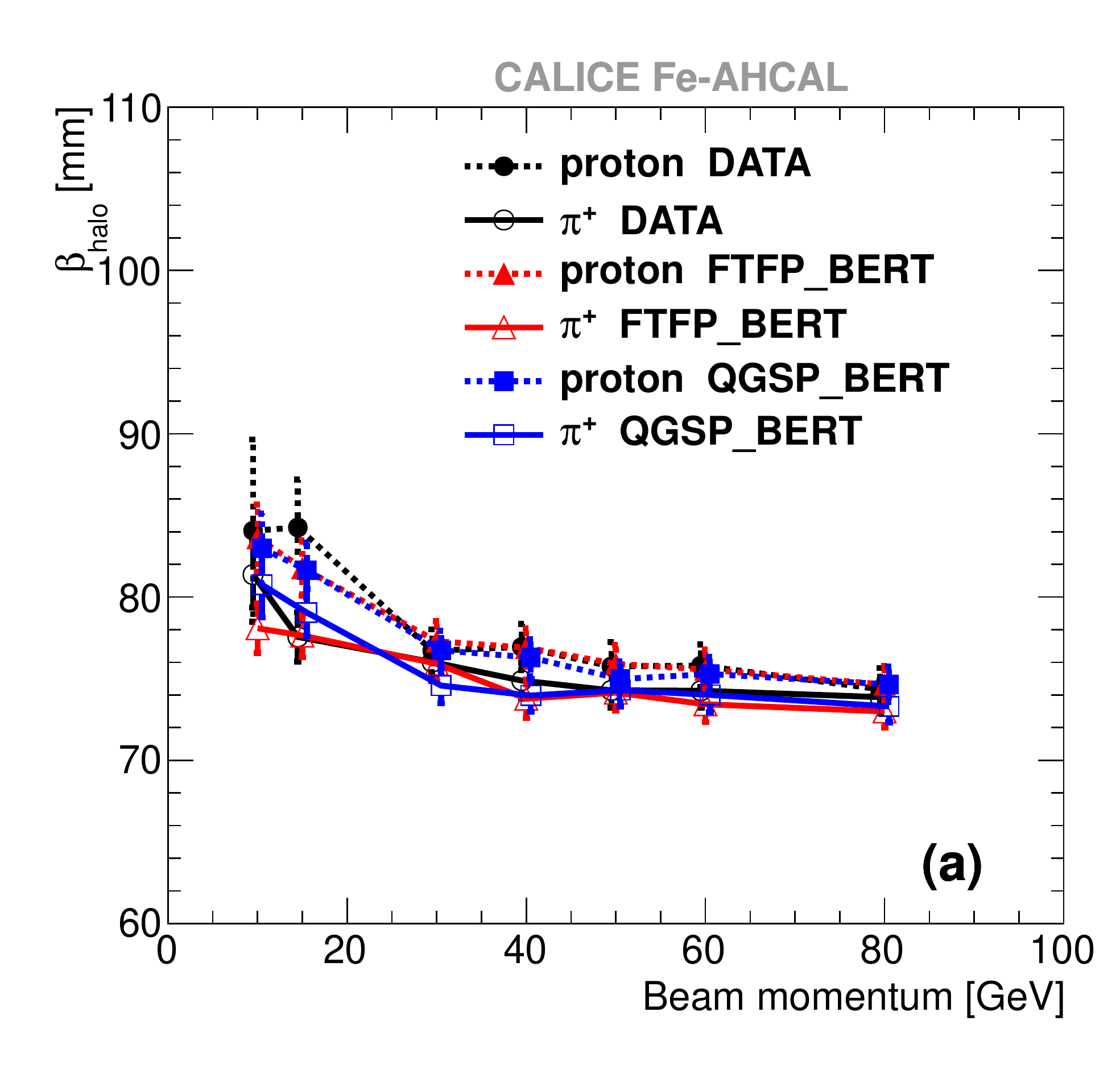}\\
 \includegraphics[width=7.0cm]{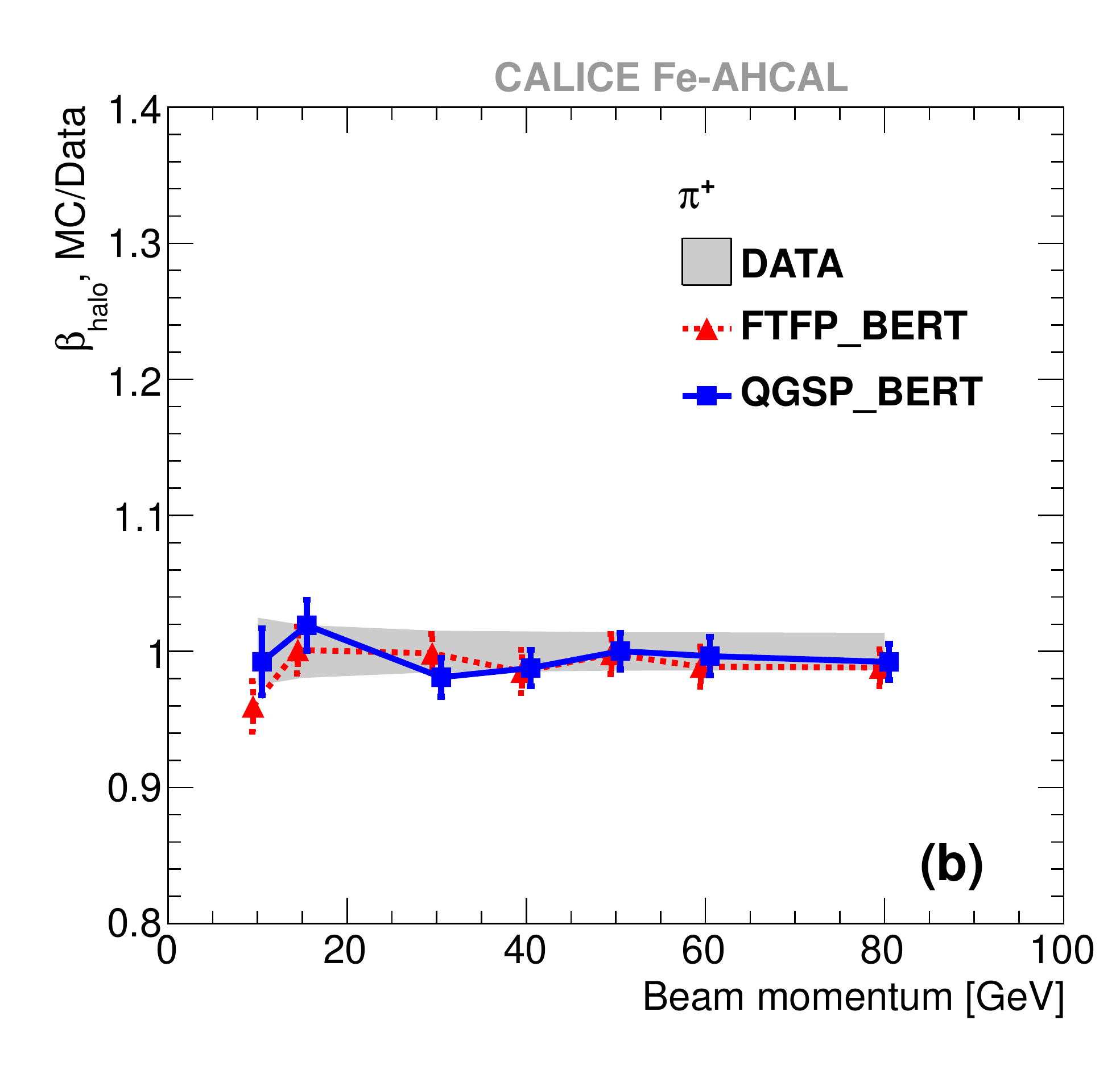}
 \includegraphics[width=7.0cm]{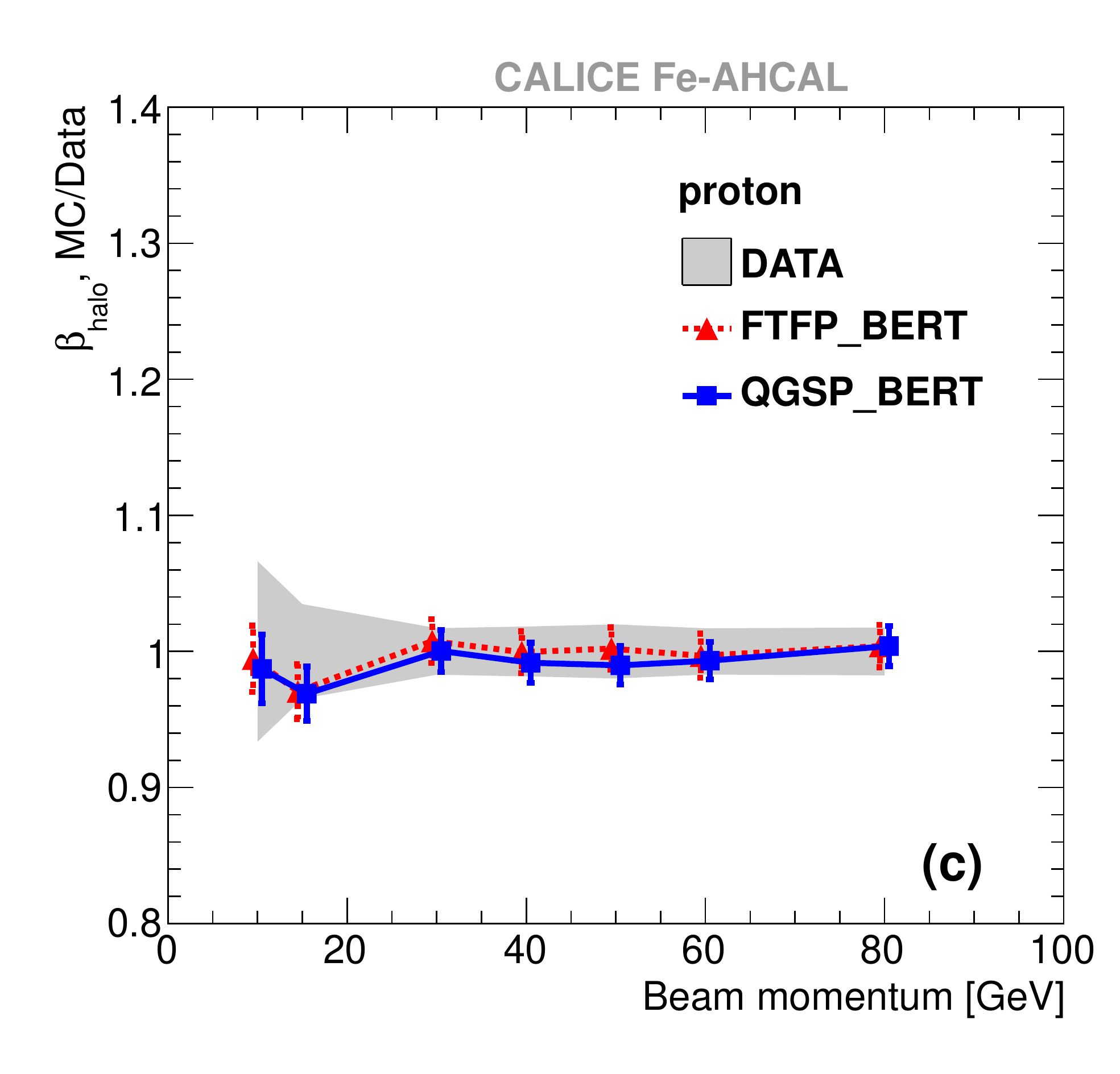}
 \caption{(a) Energy dependence of the halo slope parameter $\beta_{\mathrm{halo}}$ and the ratio of $\beta_{\mathrm{halo}}$ extracted from simulation to those extracted from data for (b) pions and (c) protons.}
 \label{fig:betaHalo}
\end{figure}

The behaviour of the shape parameter $\alpha_{\mathrm{long}}$ shown in figure~\ref{fig:alphaLong} does not depend on the particle type, is well predicted by Monte Carlo and rises logarithmically with energy. The energy dependence  of the ``long'' and ``halo'' slope parameters is shown in figures~\ref{fig:betaLong} and \ref{fig:betaHalo}. These slope parameters are also well predicted by simulations. They demonstrate negligible ($\beta_{\mathrm{long}}$) or weak ($\beta_{\mathrm{halo}}$) energy dependence and are very similar for pions and protons. This observed behaviour supports the general idea that both the shower tail and halo consist of secondary particles which have already forgotten the energy and type of the initial particle. 

%===============================================
\subsection{``Core'' and ``short'' parameters}
\label{sec:paraCompCore}

\begin{figure}
 \centering
 \includegraphics[width=7.0cm]{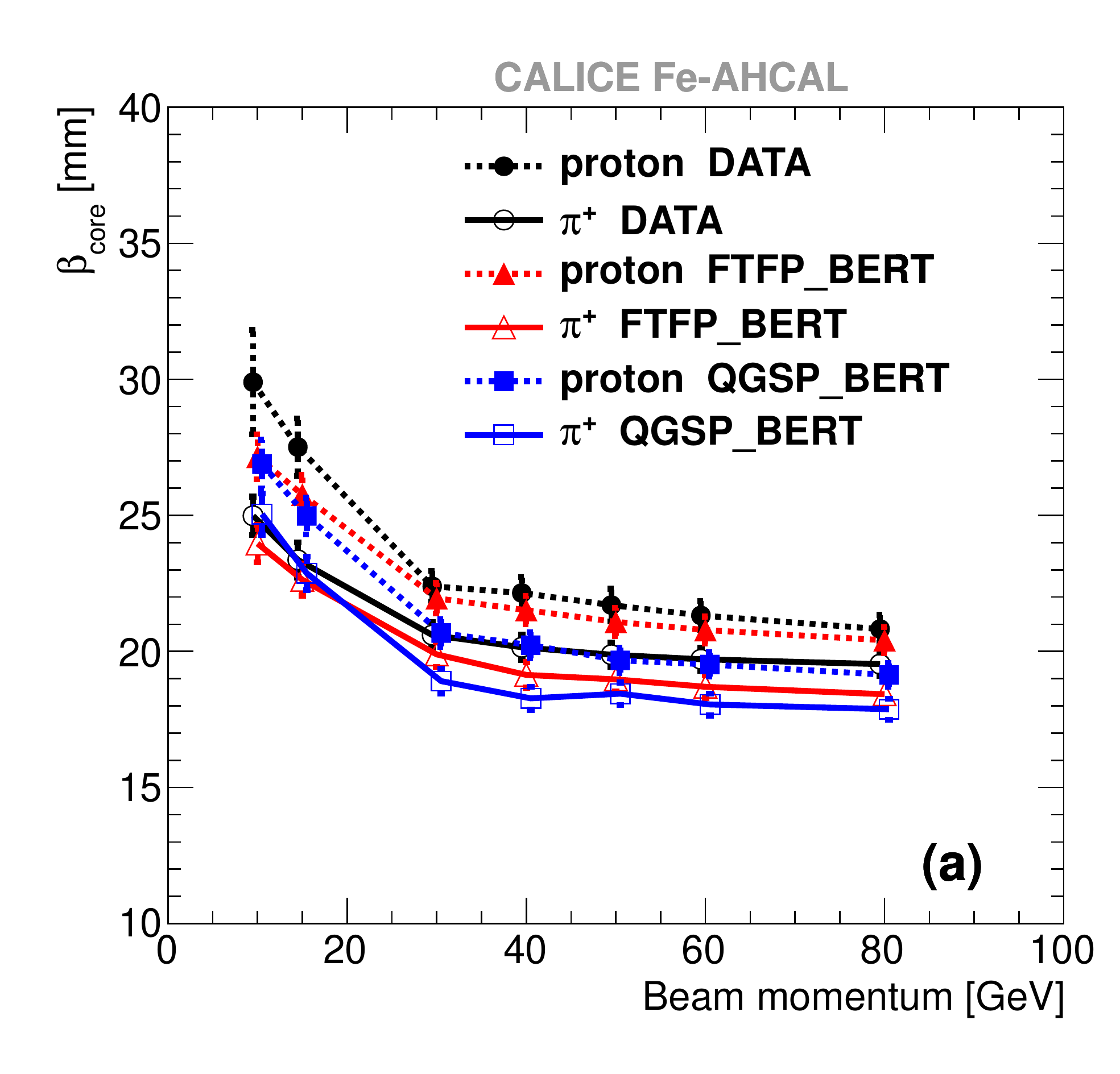}\\
 \includegraphics[width=7.0cm]{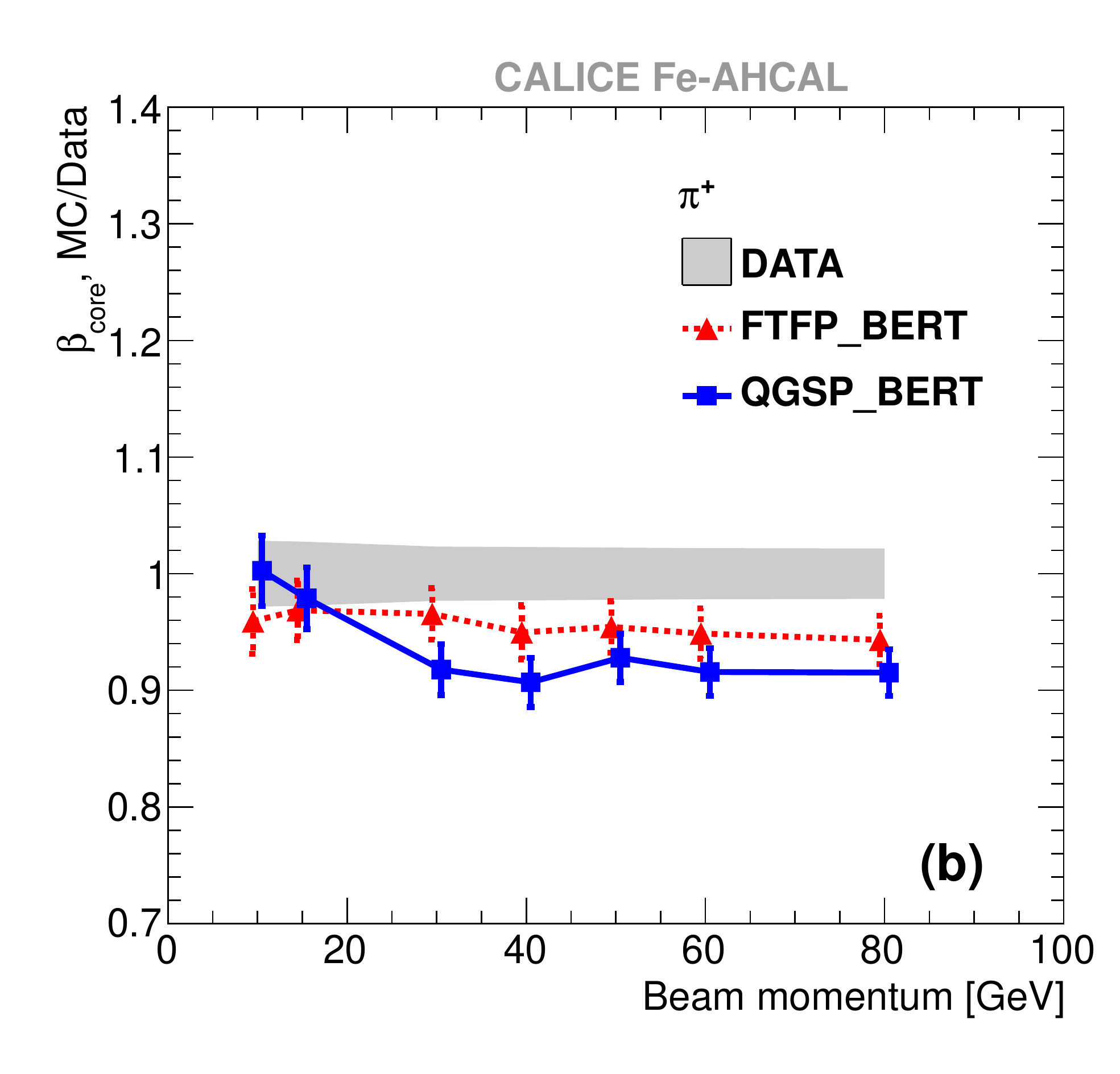}
 \includegraphics[width=7.0cm]{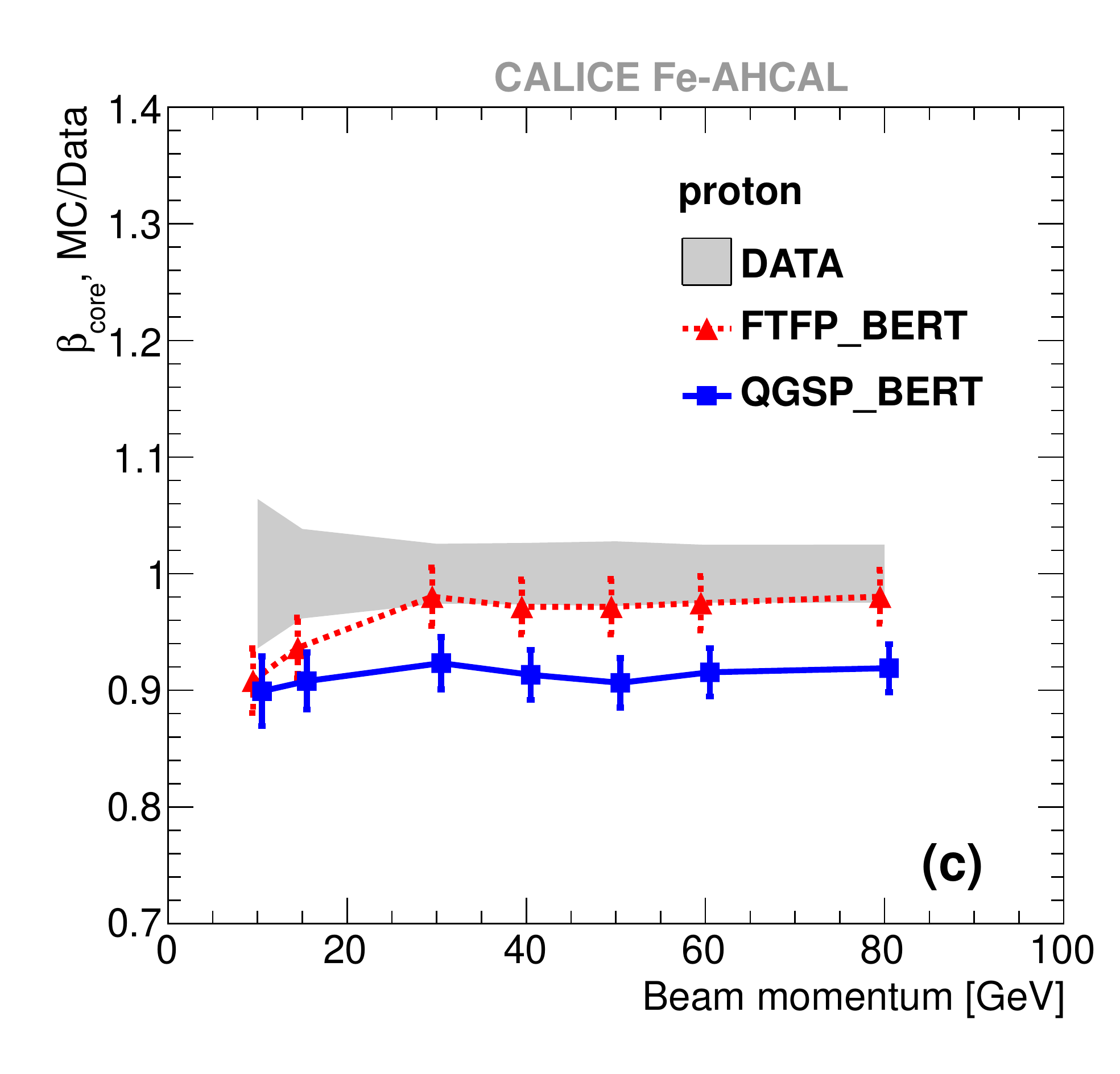}
 \caption{(a) Energy dependence of the slope parameter $\beta_{\mathrm{core}}$ and the ratio of $\beta_{\mathrm{core}}$ extracted from simulation to those extracted from data for (b) pions and (c) protons.}
 \label{fig:betaCore}
\end{figure}

The parameter $\beta_{\mathrm{core}}$ characterises the transverse shower development near the shower axis and is probably related to the angular distribution of secondary $\pi^0$s from the first inelastic interaction. The behaviour of this parameter is shown in figure~\ref{fig:betaCore}. It decreases with energy, the decrease being very slow above 30~GeV. It is well predicted by both physics lists below 30~GeV and for protons by {\sffamily FTFP\_BERT} in the full energy range studied here. The underestimation of the slope in the core region by the {\sffamily FTFP\_BERT} physics list is $\sim$5\% for pions  and  $\sim$10\% by {\sffamily QGSP\_BERT} for both particle types above 30~GeV. 

\begin{figure}
 \centering
 \includegraphics[width=7.0cm]{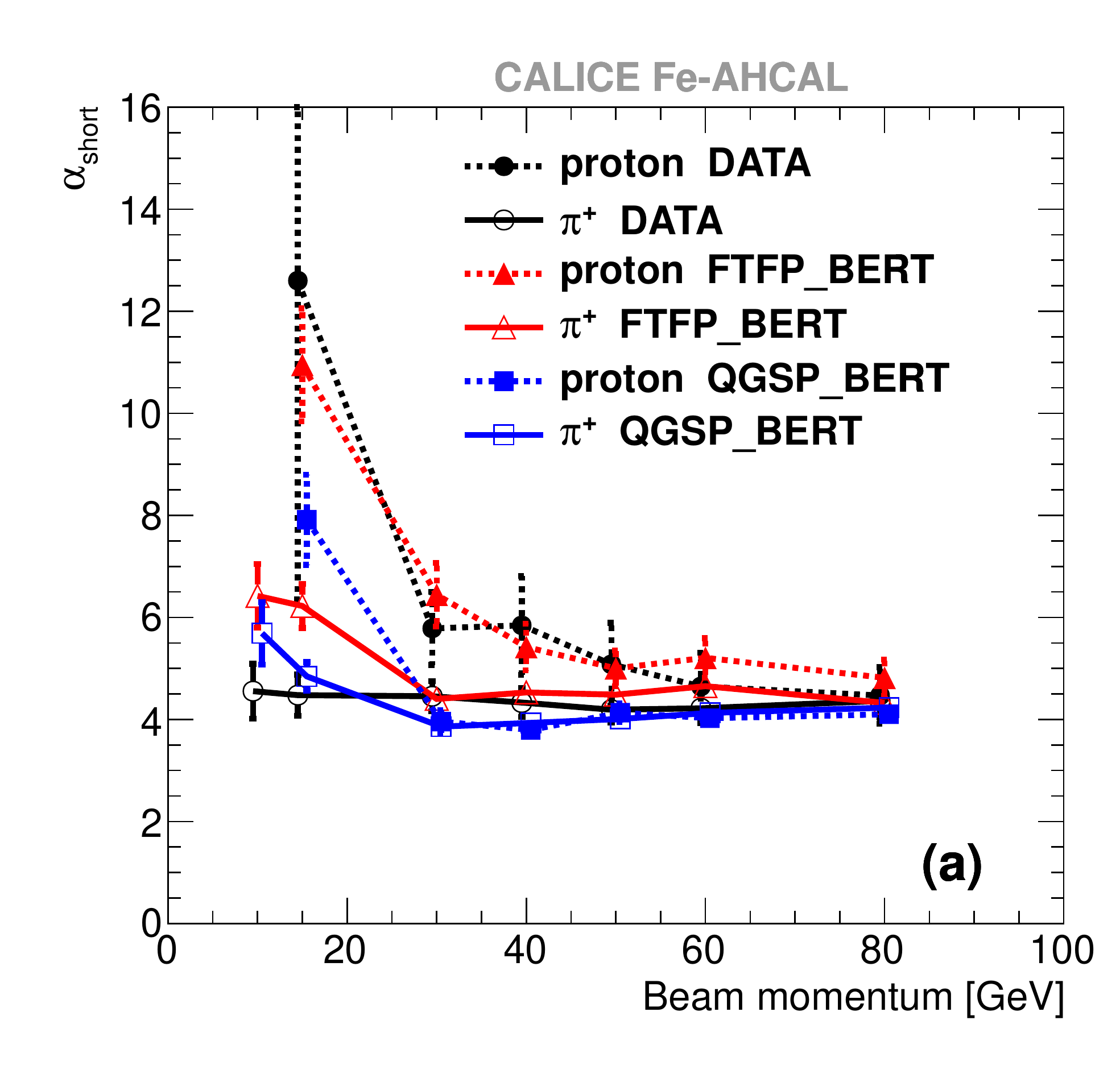}\\
 \includegraphics[width=7.0cm]{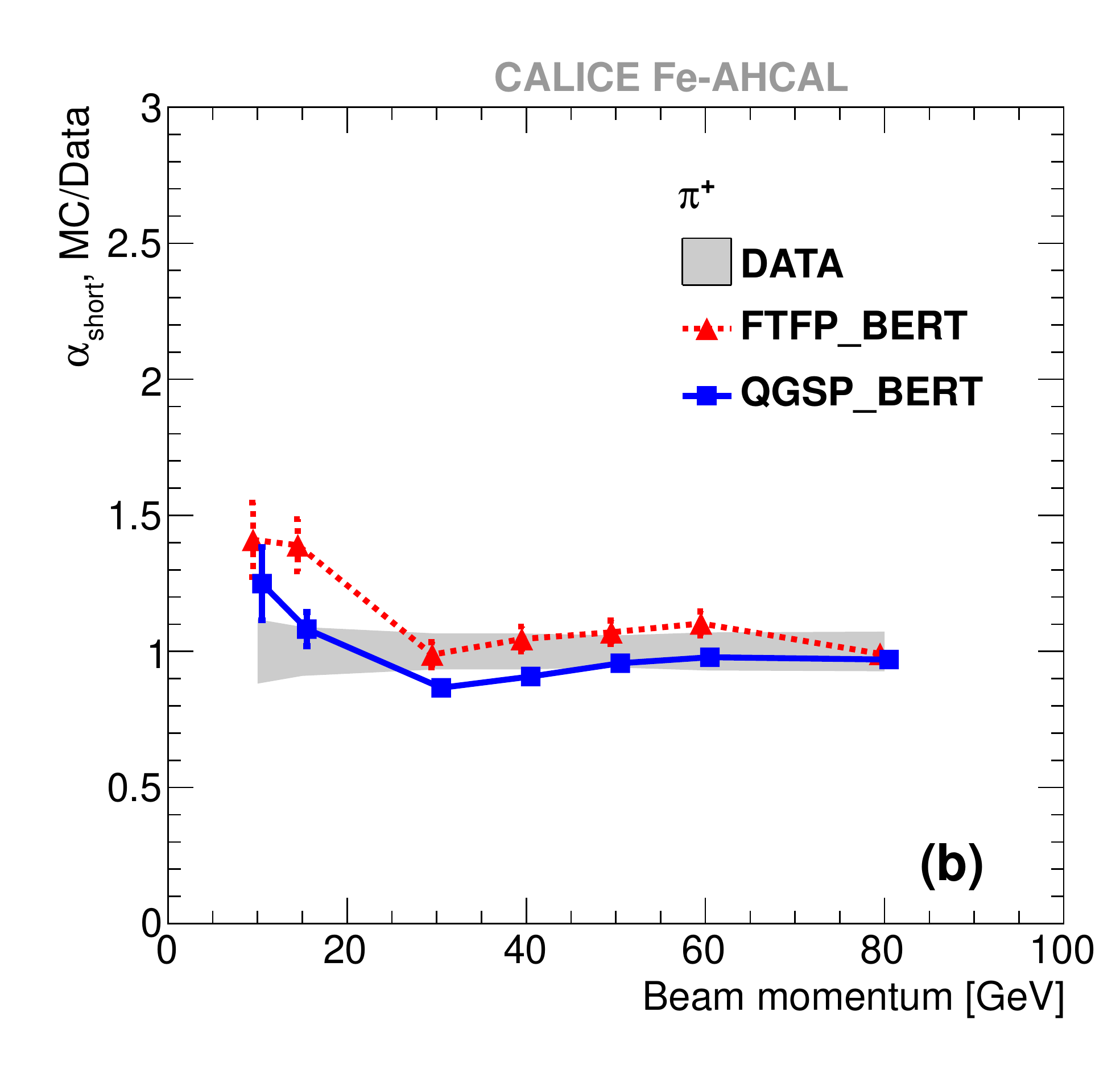}
 \includegraphics[width=7.0cm]{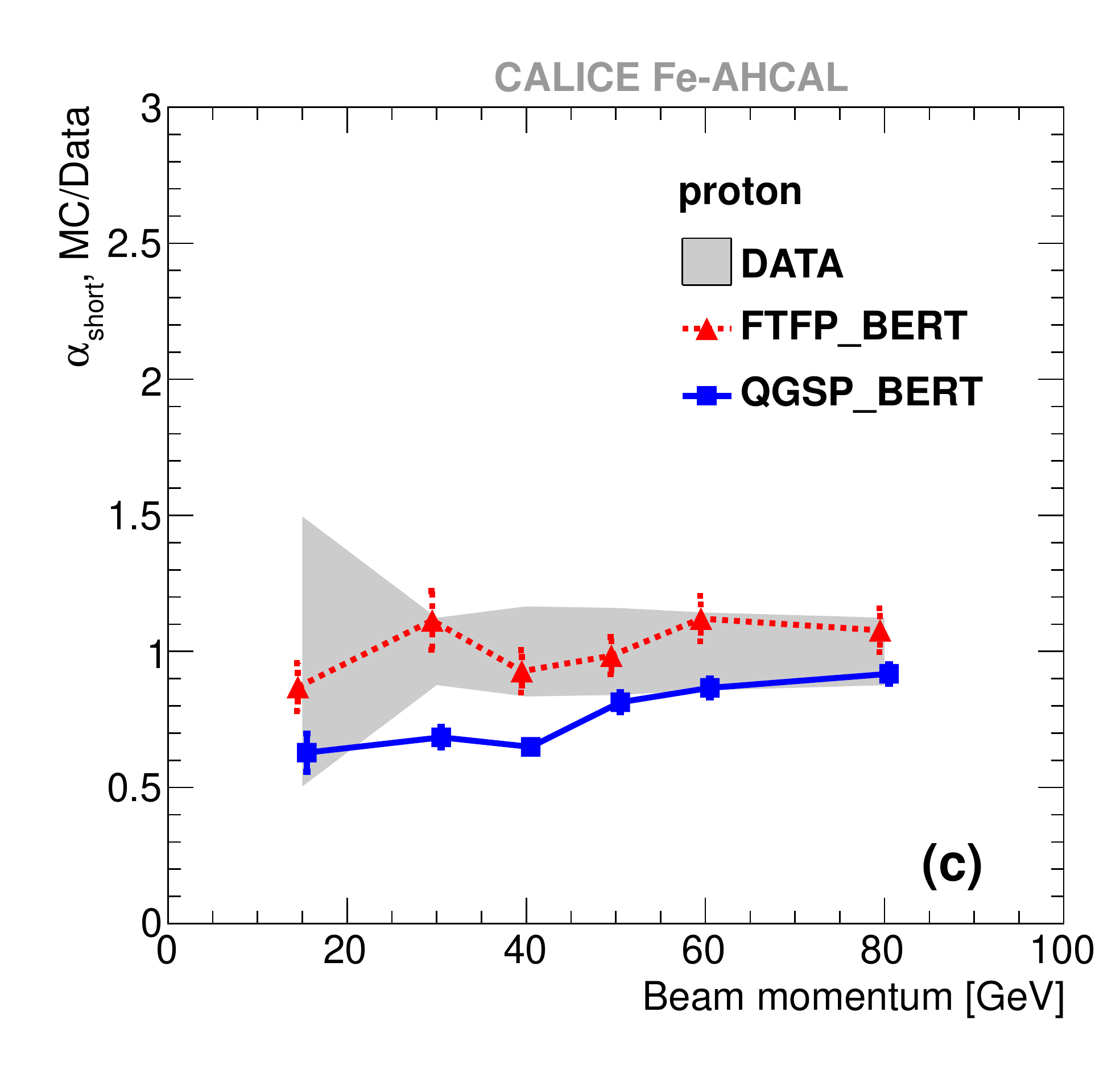}
 \caption{(a) Energy dependence of the shape parameter $\alpha_{\mathrm{short}}$ and the ratio of $\alpha_{\mathrm{short}}$ extracted from simulation to those extracted from data for (b) pions and (c) protons.}
 \label{fig:alphaShort}
\end{figure}

\begin{figure}
 \centering
 \includegraphics[width=7.0cm]{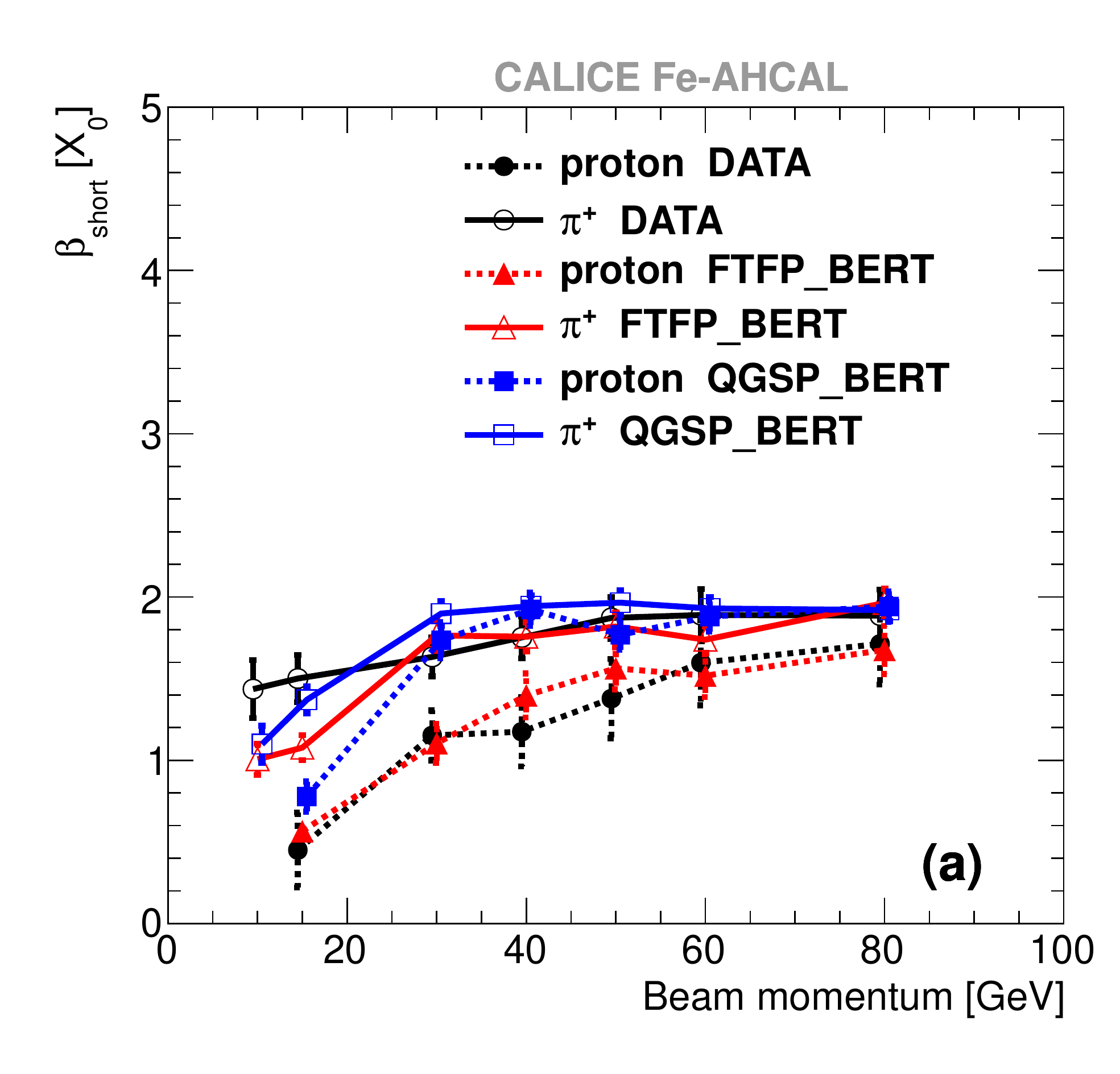}\\
 \includegraphics[width=7.0cm]{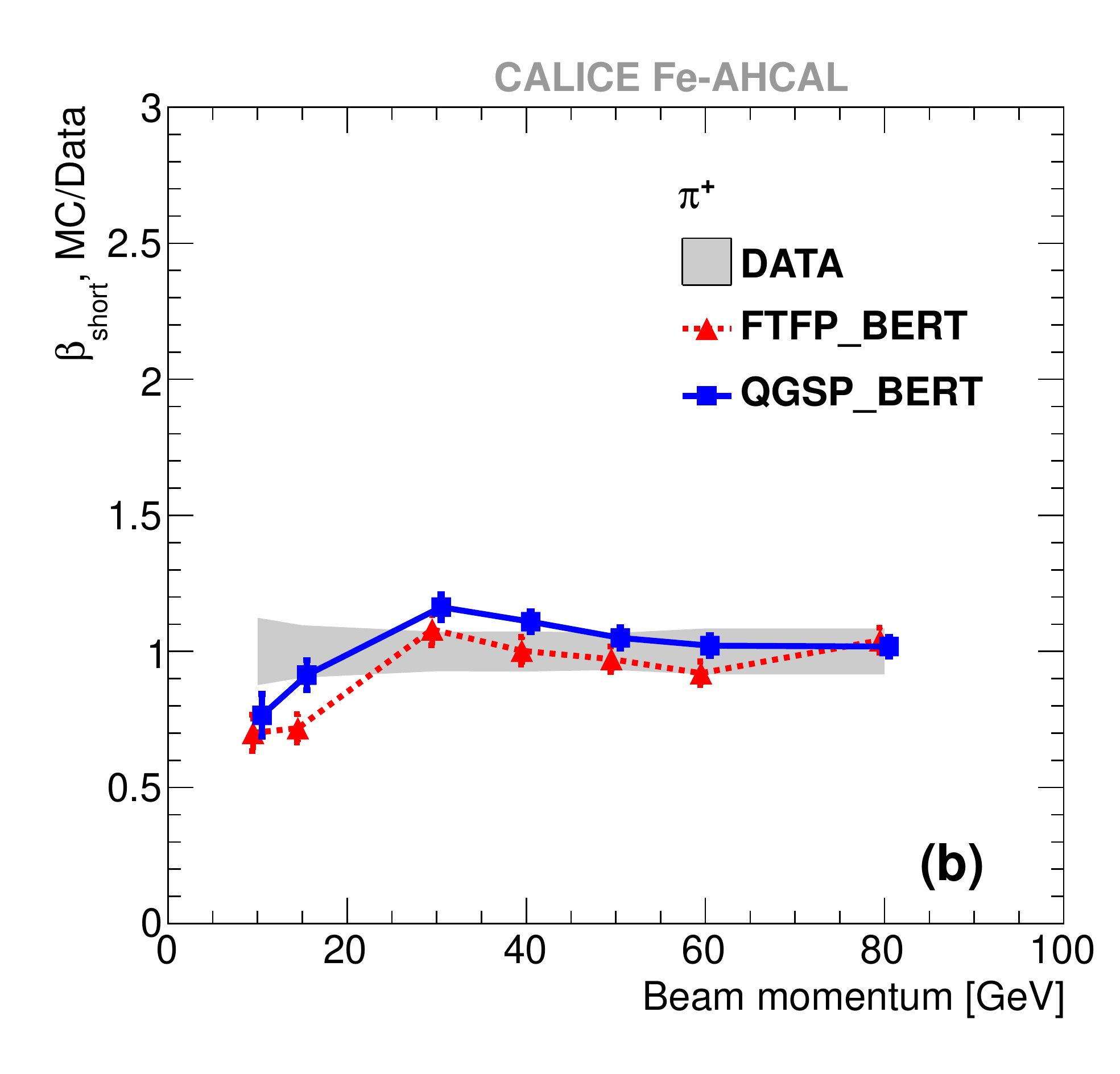}
 \includegraphics[width=7.0cm]{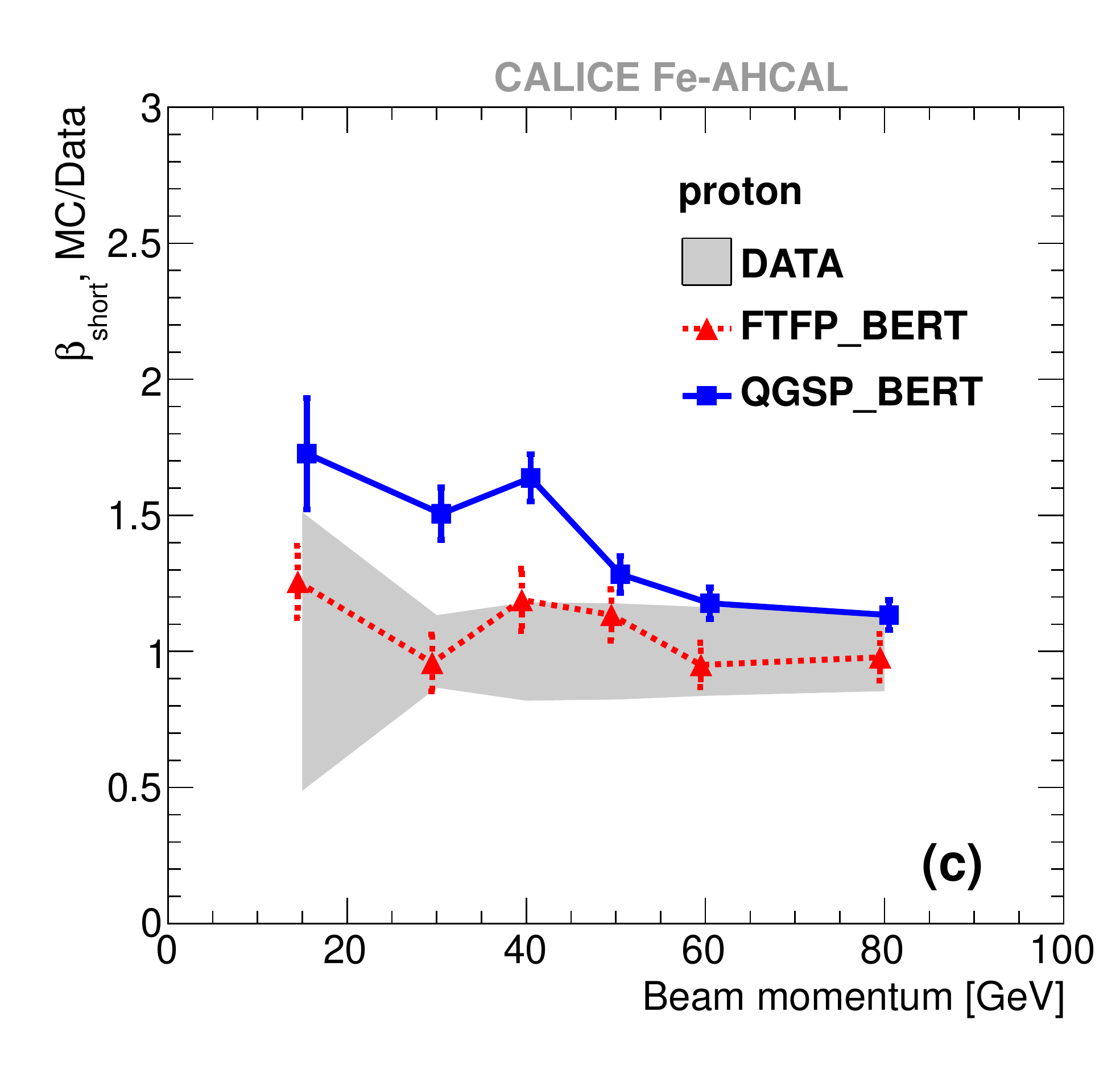}
 \caption{(a) Energy dependence of the slope parameter $\beta_{\mathrm{short}}$ and the ratio of $\beta_{\mathrm{short}}$ extracted from simulation to those extracted from data for (b) pions and (c) protons.}
 \label{fig:betaShort}
\end{figure}

\begin{figure}
 \centering
 \includegraphics[width=14cm]{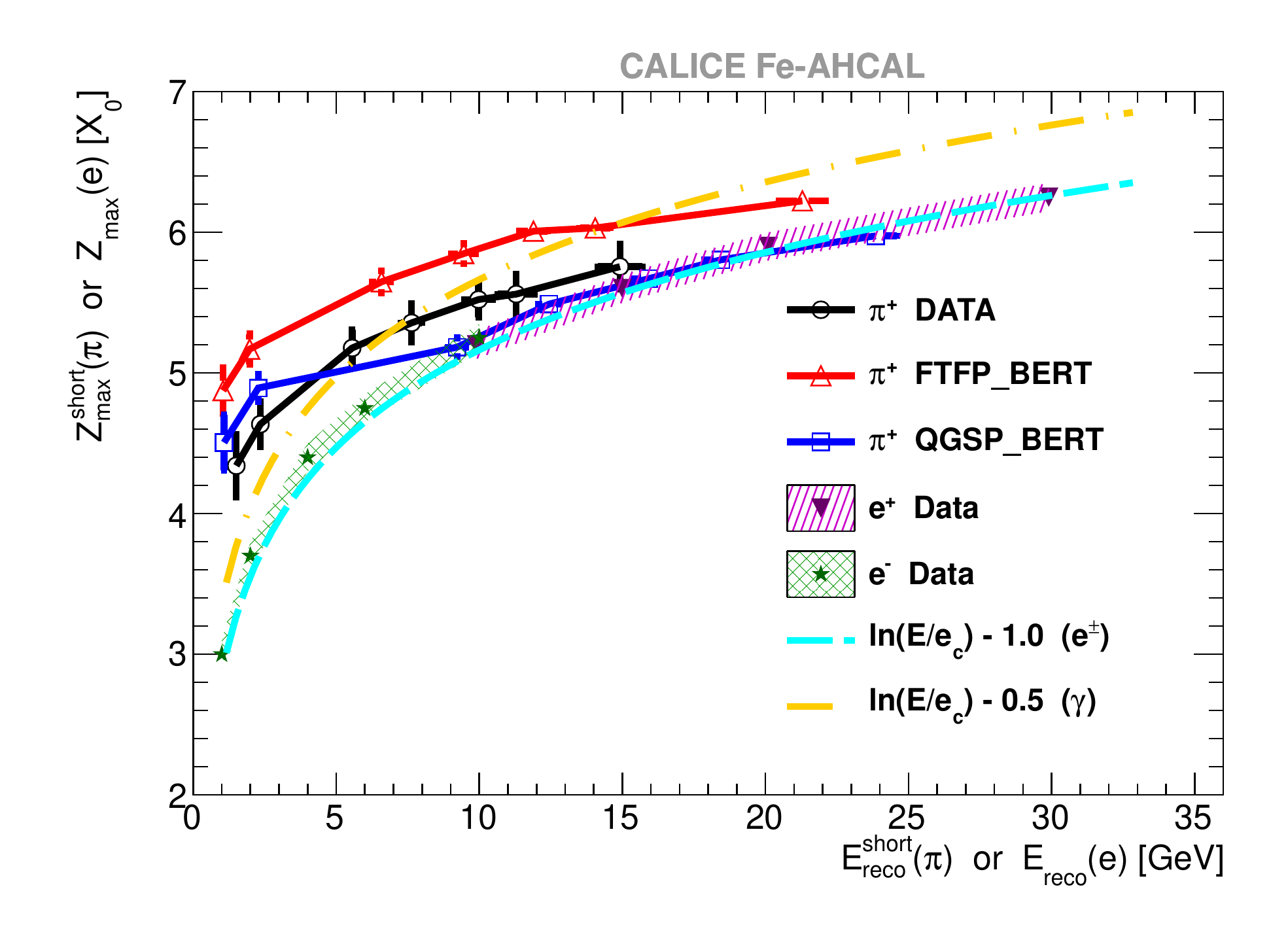}
 \caption{Solid lines: maximum of the ``short'' component versus energy of the ``short'' component estimated from the fit to pion shower profiles from data and simulations with the {\sffamily FTFP\_BERT} and {\sffamily QGSP\_BERT} physics list. Hatched bands: maximum of the longitudinal profile of electromagnetic showers induced by single electrons (stars)~\cite{FeegeDis:2011} or positrons (down triangles)~\cite{AHCAL:2011em} in the Fe-AHCAL versus the mean reconstructed energy. The dashed and dash-dotted curves correspond to the parametrisation with $e_c = 21$~MeV from ref.~\cite{Leroy:2000} for electron-induced and photon-induced showers, respectively. }
 \label{fig:zMax}
\end{figure}

The ``long'' component of the longitudinal profile which dominates in the shower tail, is accompanied  by the ``short'' component around the shower maximum. The energy dependence of the ``short'' parameters $\alpha_{\mathrm{short}}$ and $\beta_{\mathrm{short}}$ is shown in figures~\ref{fig:alphaShort} and \ref{fig:betaShort}. The estimates for protons are rather uncertain due to the small contribution from the ``short'' component and do not allow a comparison of these values at low energies. Both ``short'' parameters for pions are almost energy independent above 20~GeV. They are well predicted by simulations except for the {\sffamily FTFP\_BERT} physics list, which underestimates $\beta_{\mathrm{short}}$ and overestimates $\alpha_{\mathrm{short}}$ below 20~GeV. The position of the maximum of the ``short'' component $Z^{\mathrm{short}}_{\mathrm{max}}$ can be calculated as  

\begin{equation}
Z^{\mathrm{short}}_{\mathrm{max}} = (\alpha_{\mathrm{short}} - 1) \times \beta_{\mathrm{short}} + cov(\alpha_{\mathrm{short}},\beta_{\mathrm{short}}),
\label{eq:zmaxShort}
\end{equation}

\noindent where $cov(\alpha_{\mathrm{short}},\beta_{\mathrm{short}})$ is the covariance between correlated parameters. 
Figure~\ref{fig:zMax} shows the comparison of $Z^{\mathrm{short}}_{\mathrm{max}}$ extracted from the ``short'' component of pion showers with the estimate of the shower maximum position $Z_{\mathrm{max}}$ obtained from the pure electromagnetic showers induced by single electrons or positrons in the Fe-AHCAL~\cite{FeegeDis:2011,AHCAL:2011em}. In the case of pions, the reconstructed energy of the ``short'' component is calculated as the integral under the corresponding ``short'' curve multiplied by the electromagnetic calibration factor for the Fe-AHCAL $C_{\mathrm{em}}=$ 0.02364~GeV/MIP~\cite{AHCAL:2011em}. The maxima of the longitudinal profiles derived for single electrons or positrons are shown versus the mean reconstructed energy which coincides with the beam energy within 1-2\%. The position of the maximum of the ``short'' component for pions is calculated with respect to the shower start that corresponds to the estimates of $Z_{\mathrm{max}}$ from the calorimeter front for single electrons. 

The logarithmic rise of the shower maximum can be well parametrised with a simple logarithmic function, which is different for electron-induced and photon-induced showers~\cite{Leroy:2000}. As follows from figure~\ref{fig:zMax}, the maximum of the ``short'' component, which is more likely produced by photons from $\pi^0$ decay, is closer to that of photon-induced showers, as expected. The difference between $Z^{\mathrm{short}}_{\mathrm{max}}$ and $Z_{\mathrm{max}}$ increases with decreasing energy for both data and simulations. 

\begin{figure}
 \centering
 \includegraphics[width=7.0cm]{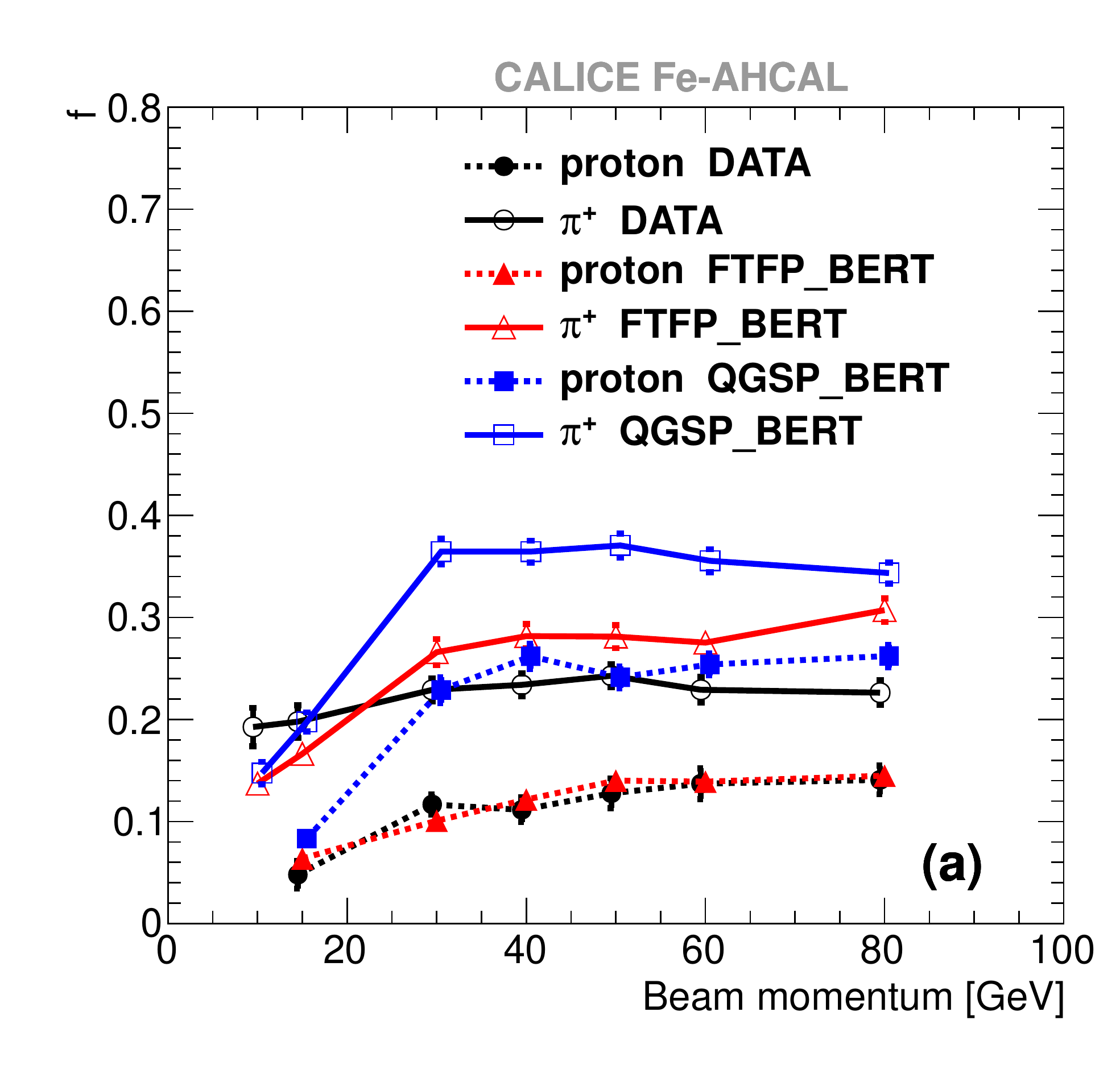}\\
 \includegraphics[width=7.0cm]{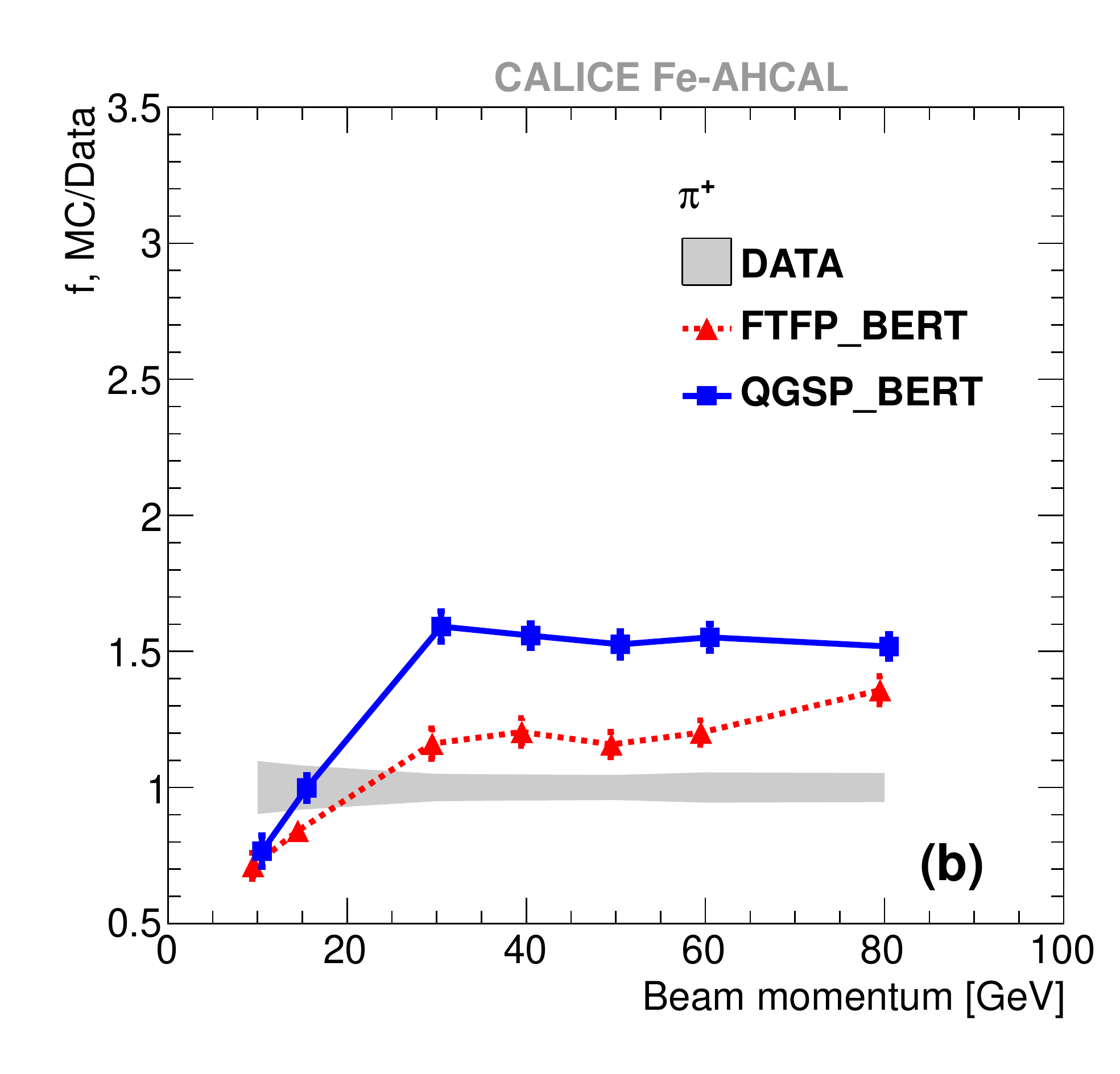}
 \includegraphics[width=7.0cm]{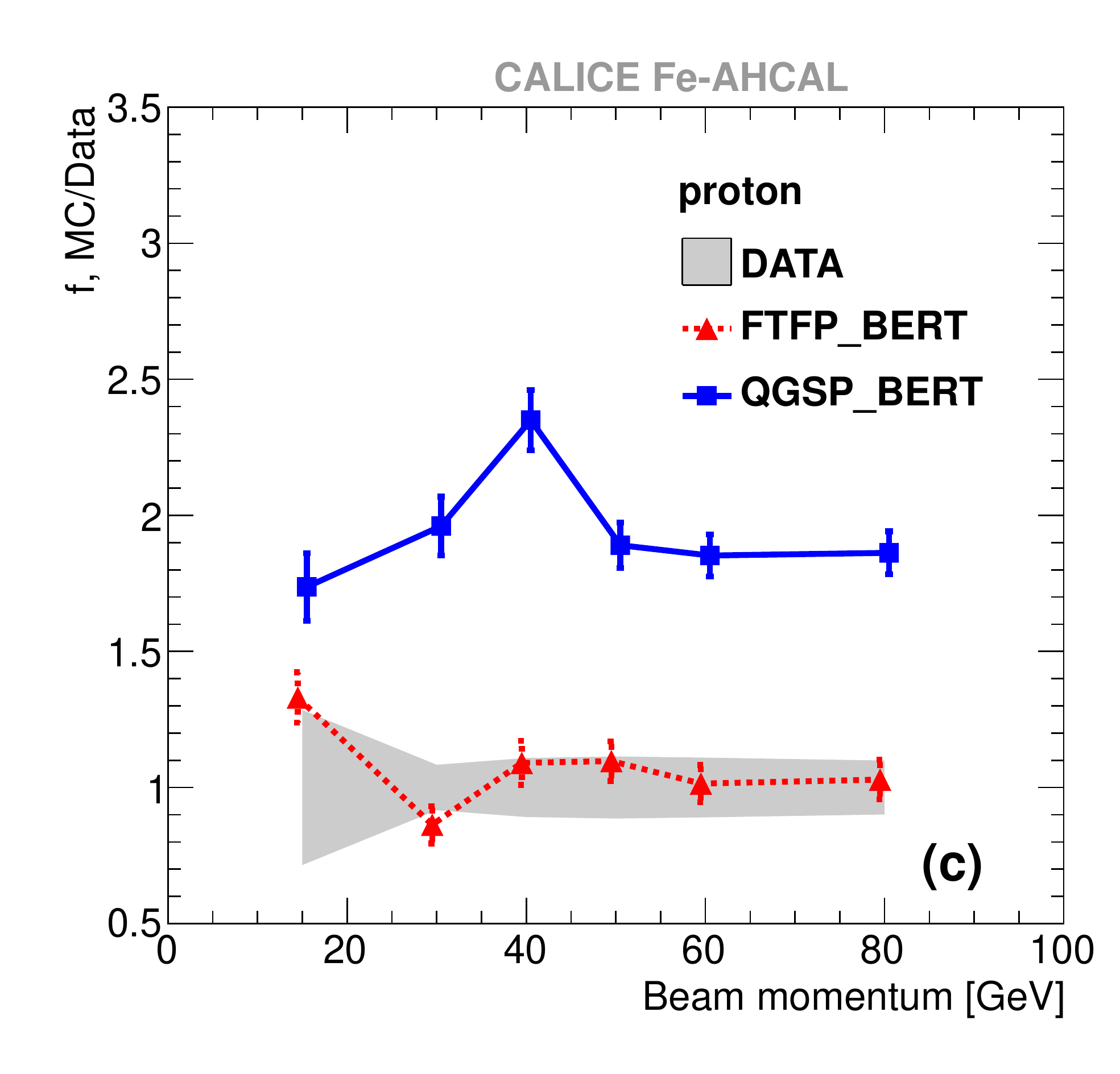}
 \caption{(a) Energy dependence of the ``short'' component fraction $f$ and the ratio of $f$ extracted from simulation to those extracted from data for (b) pions and (c) protons.}
 \label{fig:fraction}
\end{figure}

While the slope and shape parameters extracted from data and simulations coincide within uncertainties, the  fractional contribution $f$ of the ``short'' component is overestimated by both physics lists above 30~GeV for pions and slightly underestimated below 30~GeV. The behaviour of the parameter $f$ is shown in figure~\ref{fig:fraction}. The {\sffamily FTFP\_BERT} physics list gives a good prediction for protons while it underestimates the parameter for pions at 10~GeV and overestimates it at higher energies by 5-25\%. The {\sffamily QGSP\_BERT} physics list significantly overestimates the contribution of the ``short'' component above 20~GeV for both pions and protons, the overestimation exceeding 50\%. The fractional contributions of the core component related to the electromagnetic fraction can also be calculated from the integral under the fit to radial profiles. However, the estimated uncertainties of these integral values are much higher compared to that of the parameter $f$ extracted from the fit to longitudinal profiles. 

The fractional contribution of the ``short'' component in proton showers is approximately half of that in pion showers. This can be explained by a smaller electromagnetic fraction in proton showers due to the baryon number conservation and consequently a smaller quantity of produced $\pi^0$s.

%%%%%%%%%%%%%%%%%%%%%%%%%%%%%%%%%%%%%%%%%%%%%%%%%%%%%%%%%%%%%%
\section{Calorimeter response estimated from the fit to longitudinal profiles}
\label{sec:resp}

The curve obtained from the fit to the longitudinal profile (see section~\ref{sec:paraLng}) can be extrapolated outside the fit range. The value of the scaling parameter $A$ from function (\ref{eq:lngProf}) is equal to the integral under the curve up to infinity and, therefore, corresponds to the mean visible energy, in units of MIP, that would be produced by showers in a calorimeter with infinite depth. To get the deposited energy in units of GeV, $E^{\mathrm{fit}}_{\mathrm{sh}}$, this integral is multiplied by the electromagnetic calibration factor for the Fe-AHCAL, $C_{\mathrm{em}}$. The energy deposited in the incoming track prior to the first inelastic interaction, $E_{\mathrm{track}}$, had to be added to the shower energy to get the total reconstructed energy of a hadron on the electromagnetic scale. The mean reconstructed energy from the fit to the longitudinal profile can be calculated as

\begin{equation}
E^{\mathrm{fit}}_{\mathrm{reco}} = E^{\mathrm{fit}}_{\mathrm{sh}} + \left<E_{\mathrm{track}}\right>,
\label{eq:recoFit}
\end{equation}

\noindent where $\left<E_{\mathrm{track}}\right>$ is the mean energy deposited in the incoming track. It is estimated for two different setup configurations and is found to be 0.40$\pm$0.09~GeV for the setup with the Si-W ECAL and 0.06$\pm$0.02~GeV for the setup without the electromagnetic calorimeter. The quoted uncertainty takes into account the different lengths of incoming tracks in the Fe-AHCAL for different shower start layers.

\begin{figure}
 \centering
 \includegraphics[width=14cm]{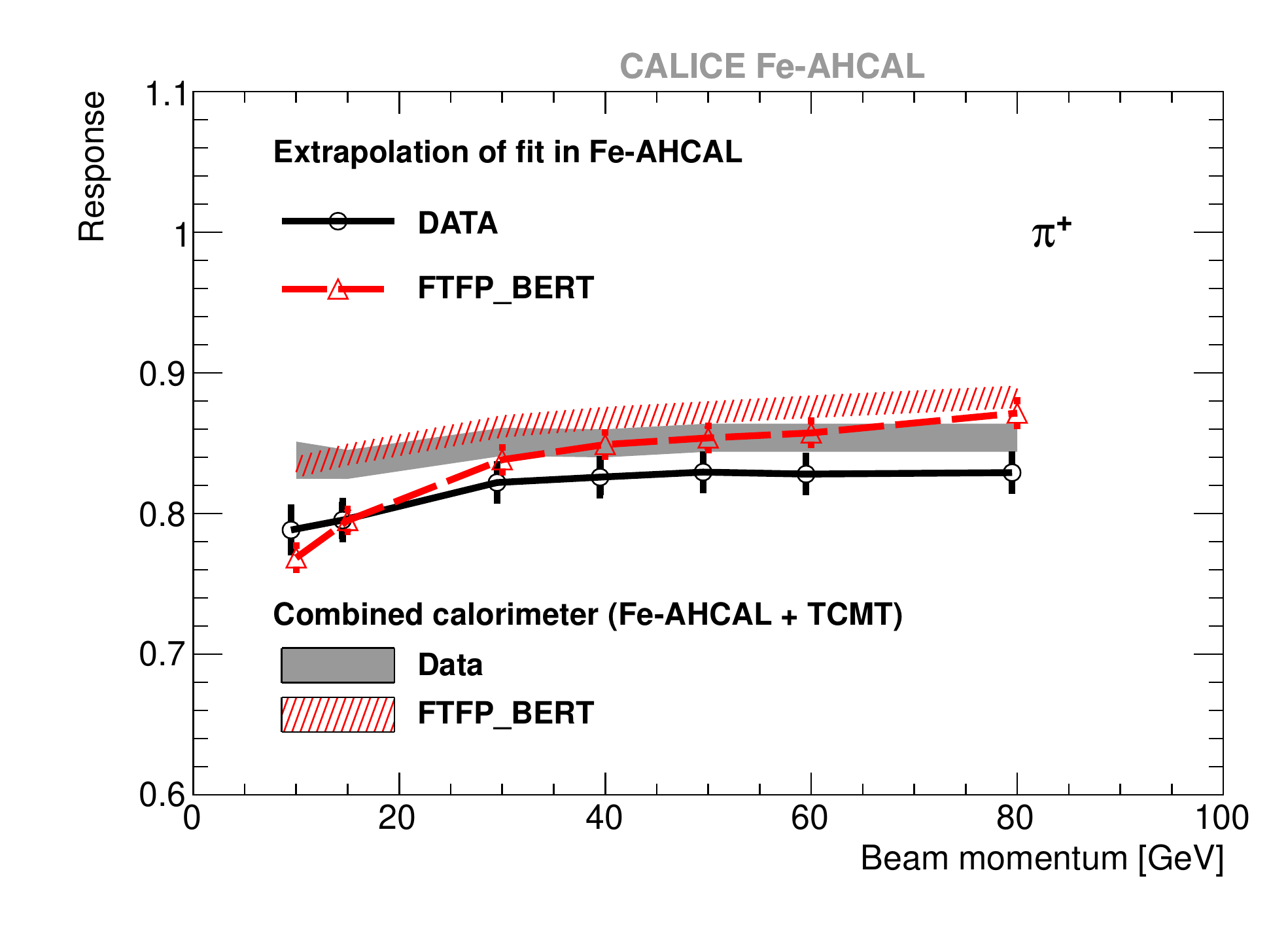}
 \caption{ 
 Response to pions extracted from the fit to longitudinal shower profiles from data (black circles) and simulations with the {\sffamily FTFP\_BERT} physics list; the bands show the response of the combined CALICE calorimeter setup from ref.~\cite{AHCAL:2015pionProton} for data and simulations. See text for details.
 }
 \label{fig:resp}
\end{figure}

Figure~\ref{fig:resp} shows the response $E^{\mathrm{fit}}_{\mathrm{reco}}/E_{\mathrm{beam}}$ of the Fe-AHCAL to pions, estimated from the fit to the longitudinal profiles for both data and simulations with the {\sffamily FTFP\_BERT} physics list. These results can be compared with the value $E_{\mathrm{reco}}/E_{\mathrm{beam}}$ obtained with the combined CALICE calorimeter (Si-W ECAL + Sc-Fe AHCAL + Sc-Fe TCMT)~\cite{AHCAL:2015pionProton}. The selection conditions in both cases are the same including the requirement of a track in the electromagnetic calorimeter placed in front of the Fe-AHCAL. The same electromagnetic calibration factor $C_{\mathrm{em}}$ was applied in both cases. For the combined calorimeter, the reconstructed energy in each event is calculated as the sum of the energies deposited in the Fe-AHCAL and the TCMT plus the energy deposited in the incoming track. The resulting energy distribution is fitted with a Gaussian to obtain the mean reconstructed energy for the given beam energy. The total depth of the Fe-AHCAL and TCMT amounts to $\sim$11$\lambda_{\mathrm{I}}$. The estimates of the response obtained from the combined calorimeter are shown in figure~\ref{fig:resp} by bands, whose widths correspond to the systematic uncertainty. 

The response estimated from the fit to the longitudinal profile tends to be steeper
than the response measured with the combined calorimeter. The observed difference can be largely explained by the constant noise component from the tail catcher (TCMT) at the level of $\sim$0.4--0.6~GeV, which is also added to the simulated samples during the digitisation procedure. The impact of noise decreases with increasing energy, while the impact of leakage is expected to increase. Nevertheless, the results coincide within uncertainties at 80~GeV, though the response from the fit is consistently smaller than that estimated with the combined calorimeter. The estimated uncertainties on the response for data are 2.5\% and 1.5\% at 10 and 80~GeV respectively. 

The difference between the response extracted from the fit and that obtained from the combined calorimeter is $\sim$7\% ($\sim$2.5\%) at 10 (80)~GeV. This is comparable to the aforementioned noise contribution from the TCMT, which amounts to $\sim$5\% and $\sim$0.6\% at 10 and 80~GeV respectively. Therefore, the response extracted from the fit to the longitudinal profiles provides an estimate of the calorimeter response at the percent level. In both cases simulations show a steeper behaviour of the response than observed in the data. 
Given the selection conditions applied in this study, the mean contribution from the TCMT to the overall response is negligible at 10~GeV and amounts to $\sim$5\% at 80~GeV. The majority ($\sim$4\%) of this leakage effect can be taken into account by applying the technique of the mean response estimate based on the extrapolation of the longitudinal profiles in the Fe-AHCAL.

%%%%%%%%%%%%%%%%%%%%%%%%%%%%%%%%%%%%%%%%%%%%%%%%%%%%%%%%%%%%%%
\section{Estimates of $h/e$ from the fit to longitudinal profiles}
\label{sec:pheno}

In the phenomenological approach described in ref.~\cite{Groom:2007}, there are three parameters which define the calorimeter response to hadrons. The first is the mean electromagnetic fraction $f_{\mathrm{em}}$, i.e. the average fraction of the initial hadron energy deposited in the form of the electromagnetic component of a shower. The other two parameters, $e$ and $h$, characterise the response to the electromagnetic and non-electromagnetic (hadronic) components of hadron-induced showers. The response $e$ defines the electromagnetic scale for a given calorimeter and is different from $h$ in non-compensating calorimeters. In terms of the traditional phenomenological approach, the mean reconstructed energy of a hadron-induced shower $E_{\mathrm{sh}}$  can be represented as the sum of an electromagnetic component $E_{\mathrm{em}}$ and a hadronic component $E_{\mathrm{had}}$:

\begin{equation}
E_{\mathrm{sh}} = E_{\mathrm{em}} + E_{\mathrm{had}}.
\label{eq:reco}
\end{equation}

Both components, measured in the electromagnetic scale, can be expressed in terms of the mean electromagnetic fraction $f_{\mathrm{em}}$, the mean hadronic fraction $f_{\mathrm{had}} = 1 - f_{\mathrm{em}}$, and the initial hadron energy $E_{\mathrm{ini}}$ as follows:

\begin{equation}
E_{\mathrm{em}} = f_{\mathrm{em}} \cdot E_{\mathrm{ini}}, \quad  
E_{\mathrm{had}} = \frac{h}{e} \cdot f_{\mathrm{had}} \cdot E_{\mathrm{ini}}.
\label{eq:comp}
\end{equation}

The parametrisation of the longitudinal shower profiles with a two-component function allows us to roughly separate the contributions from the electromagnetic and hadronic components within a shower. To a first approximation, the ``short'' and ``long'' components of the fit can be considered as the electromagnetic and hadronic fractions, respectively. The mean visible energy in each component in units of MIP can be calculated as the integral up to infinity under the corresponding curve, the parameters of which are extracted from the fit to the longitudinal profile. The estimates of the deposited energy in units of GeV for each component, $E_{\mathrm{em}}$ and $E_{\mathrm{had}}$, can be obtained by multiplying each integral by the electromagnetic calibration factor $C_{\mathrm{em}}$. The initial hadron energy is $E_{\mathrm{ini}} = E_{\mathrm{beam}} - E_{\mathrm{track}}$, where $E_{\mathrm{track}}$ is the energy deposited in the incoming track before the identified shower start (see section~\ref{sec:resp}). Then the following expression for $h/e$ can be derived from eq.~(\ref{eq:comp})

\begin{equation}
\frac{h}{e} = \frac{E^{\mathrm{fit}}_{\mathrm{had}}}{E_{\mathrm{beam}} - E_{\mathrm{track}} - E^{\mathrm{fit}}_{\mathrm{em}}}, \quad E^{\mathrm{fit}}_{\mathrm{had}} = E^{\mathrm{long}}_{\mathrm{reco}} \cdot C_{\mathrm{em}}, \quad E^{\mathrm{fit}}_{\mathrm{em}} = E^{\mathrm{short}}_{\mathrm{reco}} \cdot C_{\mathrm{em}},
\label{eq:h2e}
\end{equation}

\noindent where $E^{\mathrm{long}}_{\mathrm{reco}}$ is the integral under the ``long'' component of the longitudinal profile, and $E^{\mathrm{short}}_{\mathrm{reco}}$ is the integral under the ``short'' component of the longitudinal profile. The uncertainties of $h/e$ are calculated using standard error propagation technique taking into account the uncertainties of all variables involved.

\begin{figure}
 \centering
 \includegraphics[width=14cm]{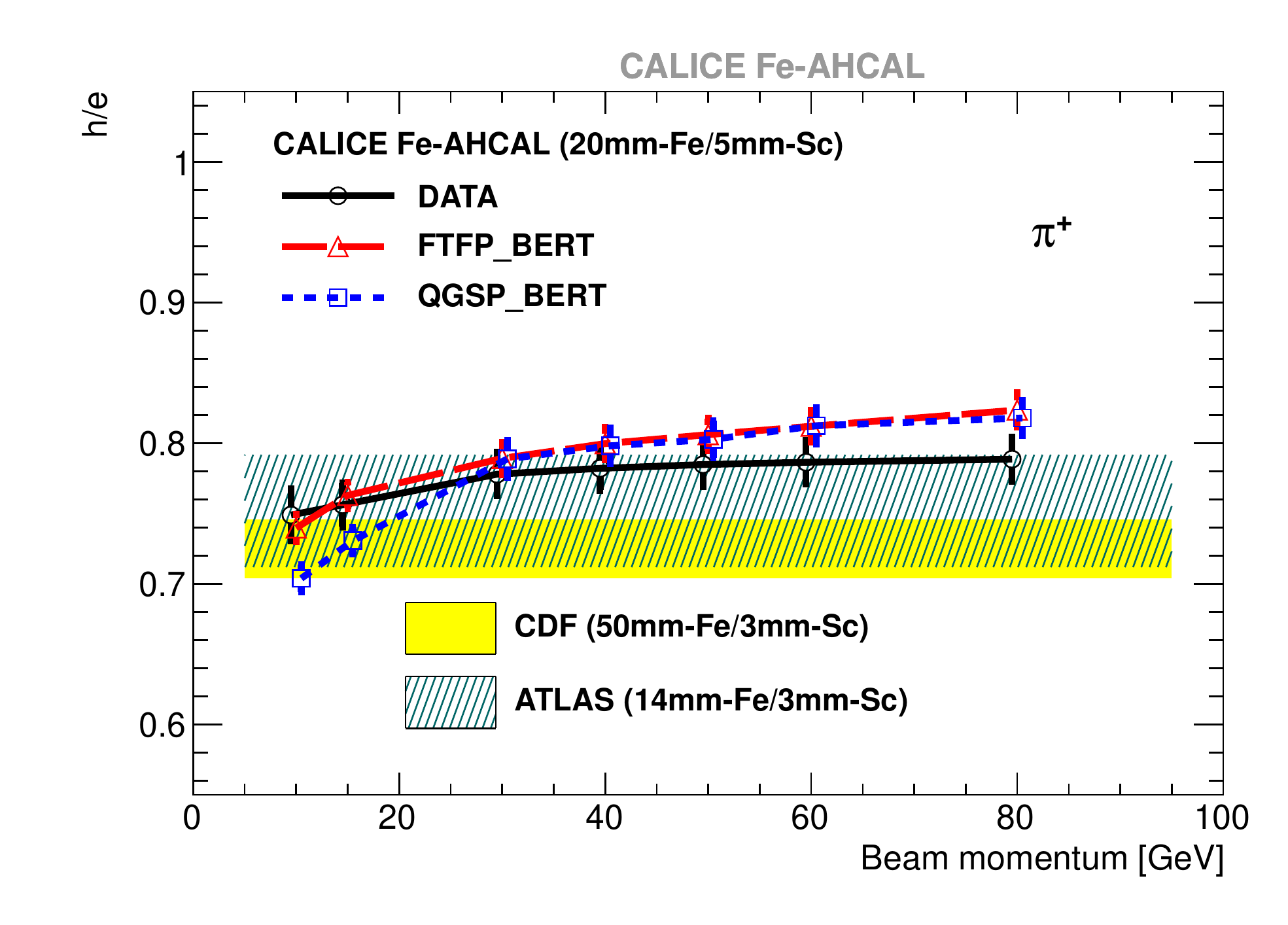}
 \caption{ 
 Energy dependence of the $h/e$ ratio extracted from the fit to longitudinal profiles for data (black circles) and simulations with the {\sffamily FTFP\_BERT} (red) and {\sffamily QGSP\_BERT} (blue) physics lists; the hatched blue and solid yellow bands correspond to the estimates from experimental data of the ATLAS TileCal~\cite{ATLAS:2009} and CDF~\cite{CDF:1997} hadron calorimeters, respectively. 
 }
 \label{fig:h2e}
\end{figure}

Figure~\ref{fig:h2e} shows the values of $h/e$ calculated with eq.~(\ref{eq:h2e}) using the fits to data and simulations. The values of $h/e$ predicted by simulations are in agreement with data within 5\%, though simulations tend to overestimate them with increasing energy. A better agreement with data below 30~GeV is demonstrated by the {\sffamily FTFP\_BERT} physics list. Our results of $h/e$ extracted using the longitudinal profile parametrisation are compared with the estimates obtained using the traditional power-law approximation of the response measured with the ATLAS TileCal~\cite{ATLAS:2009} and CDF~\cite{CDF:1997} hadron calorimeters. The sampling of the CDF calorimeter (50~mm Fe/3~mm Sc) is coarser than that of the CALICE Fe-AHCAL (20~mm Fe/5~mm Sc) and ATLAS (14~mm Fe/3~mm Sc) calorimeters. The estimated value for the CALICE Fe-AHCAL is closer to that of the ATLAS TileCal, as expected due to similar sampling fractions of both calorimeters.

The estimates, based on the power-law approximation of the calorimeter response, rely on the assumption of the energy independence of $h$ and $e$, and therefore of the $h/e$ ratio. The reason for such behaviour is that the energy spectrum of secondaries, which dominate the shower, is almost energy independent. It should be noted that the assumed energy independence of $e$ is supported by the  constant response to electrons observed for the Fe-AHCAL in the energy range studied~\cite{AHCAL:2011em}.

As follows from figure~\ref{fig:h2e}, the $h/e$ ratio, extracted from the fit to longitudinal profiles, exhibits a slow energy dependence. One possible explanation is the simplified representation used in our studies to describe the longitudinal shower development. In reality, the structure of the longitudinal distribution of energy density is more complicated.  With increasing initial hadron energy, the probability of $\pi^{0}$ production in secondary interactions increases. In the given representation, electromagnetic sub-showers which are produced far from the shower starting point contribute more likely to the ``long'' component, so the extracted $h/e$ ratio might be overestimated with increasing energy.

The value of $h/e$ extracted from the fit to longitudinal profiles increases by $\sim$8\% between 10 and 30~GeV and becomes almost energy independent above 30~GeV. It should be noted that hadronic showers become wider with decreasing energy. For instance, the mean radius of pion showers is observed to change from 92~mm at 10~GeV to 76~mm at 30~GeV (by more than 15\%)~\cite{AHCAL:2015pionProton}. Taking into account the energy threshold of 0.5~MIP applied to all calorimeter cells, one can expect a lower efficiency of detecting signals from the soft secondaries with decreasing beam energy.

%%%%%%%%%%%%%%%%%%%%%%%%%%%%%%%%%%%%%%%%%%%%%%%%%%%%%%%%%%%%%%
\section{Conclusion}

We have studied the spatial development of hadronic showers in the CALICE scintillator-steel analogue hadronic calorimeter. The fine longitudinal and radial segmentation of the calorimeter allows a comparison of shower profiles plotted from the shower start position identified on an event-by-event basis. A shower parametrisation is used to perform a detailed comparison with simulated samples as well as to compare the behaviour for different types of hadrons.  We have analysed positive hadron data collected at beam energies from 10 to 80~GeV and  samples simulated using the {\sffamily FTFP\_BERT} and {\sffamily QGSP\_BERT} physics lists from {\scshape Geant4} version 9.6.

The longitudinal profiles have been parametrised with a sum of two contributions (gamma distributions) called ``short'' and ``long''. The parameters of the ``short'' component are comparable with those of electromagnetic showers, therefore this component can be considered to be related to the contribution of electromagnetic showers from $\pi^0$ decays. The spatial parameters of the longitudinal tail are well reproduced by simulations. The behaviour of the tail parameters is very similar for pions and protons and is consistent with the common view that the shower tail is a complex environment of secondaries which have no memory of the primary conditions. Proton profiles are characterised by a smaller fractional contribution $f$ of the so called ``short'' component. The parameter $f$ for pions is overestimated by simulations above 20~GeV and exhibits a steeper rise than observed in data. This leads to a steeper increase with energy of the predicted calorimeter response to pions.  

The radial profiles have been parametrised with the sum of two exponential functions which describe the behaviour near the shower axis (``core'' region) and at the shower  periphery (``halo'' region). While the halo slope parameter is well reproduced by simulations, the core slope parameter is underestimated by $\sim$5\% for pions by {\sffamily FTFP\_BERT} and by $\sim$10\% by {\sffamily QGSP\_BERT} for both types of hadrons resulting in an underestimation of the shower width (shower radius), also observed in previous studies~\cite{Valid:2013}. 

The calorimeter response has been estimated by the integration to infinity of the longitudinal profiles measured up to a depth of $\sim$4.5$\lambda_{\mathrm{I}}$. The comparison of the response extracted from the fit to that obtained with the combined calorimeter (with a depth of $\sim$11$\lambda_{\mathrm{I}}$) shows that the impact of leakage on the shape of the longitudinal profile is relatively small and allows the estimation of the calorimeter response from the fit to longitudinal profiles at the percent level.

The phenomenological calorimeter characteristic $h/e$ (the ratio of the responses to the non-elec\-tro\-mag\-ne\-tic and electromagnetic components of a hadron-induced shower) is estimated for the calorimeter from a fit to the longitudinal profiles using the extracted parameters of the ``short'' and ``long'' components. The {\sffamily FTFP\_BERT} physics list gives better predictions of this value below 30~GeV than the {\sffamily QGSP\_BERT} physics list. Both physics lists tend to overestimate the value of $h/e$ with increasing energy.
The values extracted from the fit to longitudinal profiles are expected to overestimate the $h/e$ ratio with increasing energy due to the simplified representation of the longitudinal shower development. In spite of the slow energy dependence observed in this study, the derived estimates are consistent with the results obtained for the ATLAS and CDF hadron calorimeters using the traditional power-law parametrisation of the calorimeter response. 

Our previous study presented in ref.~\cite{AHCAL:2015pionProton} has shown that the more recent version 9.6 of {\scshape Geant4} better describes the data than the previous version 9.4. At the same time, the behaviour observed in the current analysis for version 9.6 is similar to that shown in ref.~\cite{Valid:2013} for version 9.4: the overestimation of the energy fraction deposited in the shower core and underestimation of the shower radius for pions increase with energy. The {\sffamily FTFP\_BERT} physics list from {\scshape Geant4} version 9.6 predicts the parameters of both longitudinal and radial shower development for protons within uncertainties over the full studied energy range
and demonstrates better agreement with data than the {\sffamily QGSP\_BERT} physics list for both pions and protons.

%%%%%%%%%%%%%%%%%%%%%%%%%%%%%%%%%%%%%%%%%%%%%%%%%%%%%%%%%%%%%%
\acknowledgments

We would like to thank the technicians and the engineers who
contributed to the design and construction of the prototypes.
We also gratefully acknowledge the DESY and CERN managements for their
support and hospitality, and their accelerator staff for the reliable
and efficient beam operation.
The authors would like to thank the RIMST (Zelenograd) group for their
help and sensors manufacturing.
This work was supported by the 
Bundesministerium f\"{u}r Bildung und Forschung (BMBF), Germany; 
by the Deutsche Forschungsgemeinschaft (DFG), Germany; 
by the Helmholtz-Gemeinschaft (HGF), Germany; 
by the Alexander von Humboldt Stiftung (AvH), Germany;
by the Russian Ministry of Education and Science contracts 4465.2014.2 and 14.A12.31.0006 and the Russian
Foundation for Basic Research grant 14-02-00873A;
by MICINN and CPAN, Spain;
by CRI(MST) of MOST/KOSEF in Korea;
by the US Department of Energy and the US National Science
Foundation;
by the Ministry of Education, Youth and Sports of the Czech Republic under the projects AV0~Z3407391, AV0~Z10100502, LG14033 and 7E12050;  
and by the Science and Technology Facilities Council, UK.

%%%%%%%%%%%%%%%%%%%%%%%%%%%%%%%%%%%%%%%%%%%%%%%%%%%%%%%%%%%%%%

%%%%%%%%%%%%%%%%%%%%%%%%%%%%%%%%%%%%%%%%%%%%%%%%%%%%%%%%%%%%%%
\appendix

\section{Systematic uncertainties}
\label{app:sys}

Typical contributions to the systematic uncertainty from different sources are shown in table~\ref{tab:syst} for the test beam data. They represent relative uncertainties at the maximum of the longitudinal profile and in the first bin of the radial profile extracted at different beam energies. The relatively high values observed for the 10~GeV samples for protons are due to low proton statistics in the data samples and relatively low purities of these samples.   

\begin{table}
 \caption{Relative uncertainties from different sources in the maximum of the longitudinal profiles and in the first bin of the radial profiles extracted from the test beam data.}
 \label{tab:syst}
 \begin{center}
  \begin{tabular}{|c|c|c|c|c|}
   \hline
      & \multicolumn{4}{c|}{Relative uncertainty for } \\
      \cline{2-5}
    Beam  & \multicolumn{2}{c|}{maximum of longitudinal profile} & \multicolumn{2}{c|}{1st bin of radial profile} \\
    momentum,  & \multicolumn{2}{c|}{from} & \multicolumn{2}{c|}{from} \\
    \cline{2-5}
    GeV & layer-to-layer  & contami- & identification  & contami-      \\
        & variations  & nation  & of shower axis & nation       \\
    \hline
    \multicolumn{5}{|c|}{$\pi^{+}$}   \\
    \hline
    10 &  4\% & -  &  5\% & - \\
    15 &  1\% & -  &  7\% & - \\
    30 &  2\% & -  &  1\% & - \\
    40 &  3\% & -  &  1\% & - \\
    50 &  4\% & -  &  1\% & - \\
    60 &  4\% & -  &  1\% & - \\
    80 &  4\% & -  &  1\% & - \\
   \hline
    \multicolumn{5}{|c|}{proton}   \\
    \hline
    10 &   4\% &  5\%  &  16\% & 29\% \\
    15 &   3\% &  5\%  &  11\% & 13\% \\
    30 &   2\% &  1\%  &   1\% &  1\% \\
    40 &   2\% &  3\%  &   1\% &  4\% \\
    50 &   3\% &  3\%  &   1\% &  4\% \\
    60 &   4\% &  3\%  &   1\% &  3\% \\
    80 &   4\% &  3\%  &   1\% &  2\% \\
   \hline
  \end{tabular}
 \end{center}
\end{table}

%===============================================
\subsection{Layer-to-layer variations}
\label{app:sysFluct}

\begin{figure}
 \centering
 \includegraphics[width= 7.5cm]{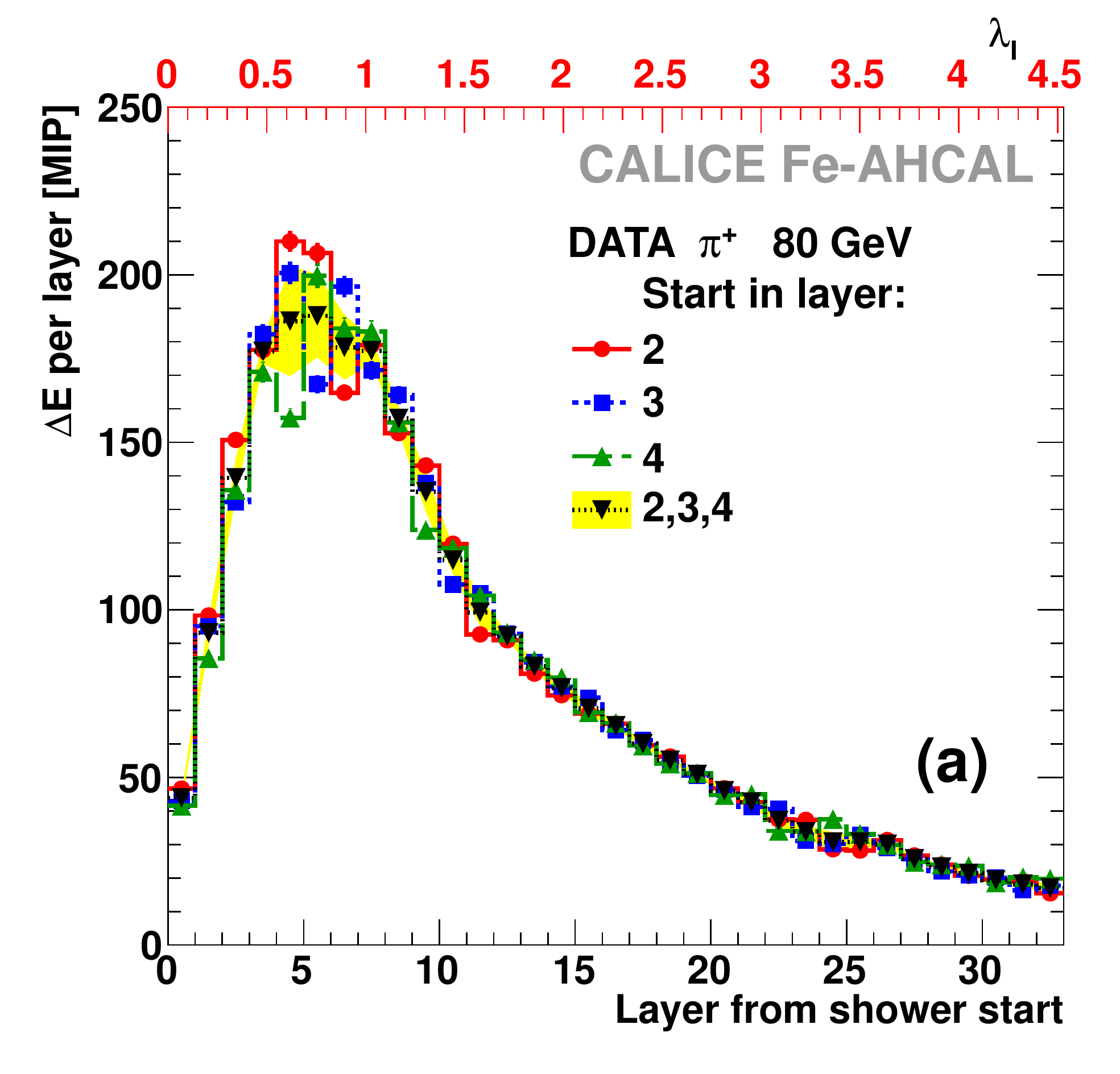}
 \includegraphics[width= 7.5cm]{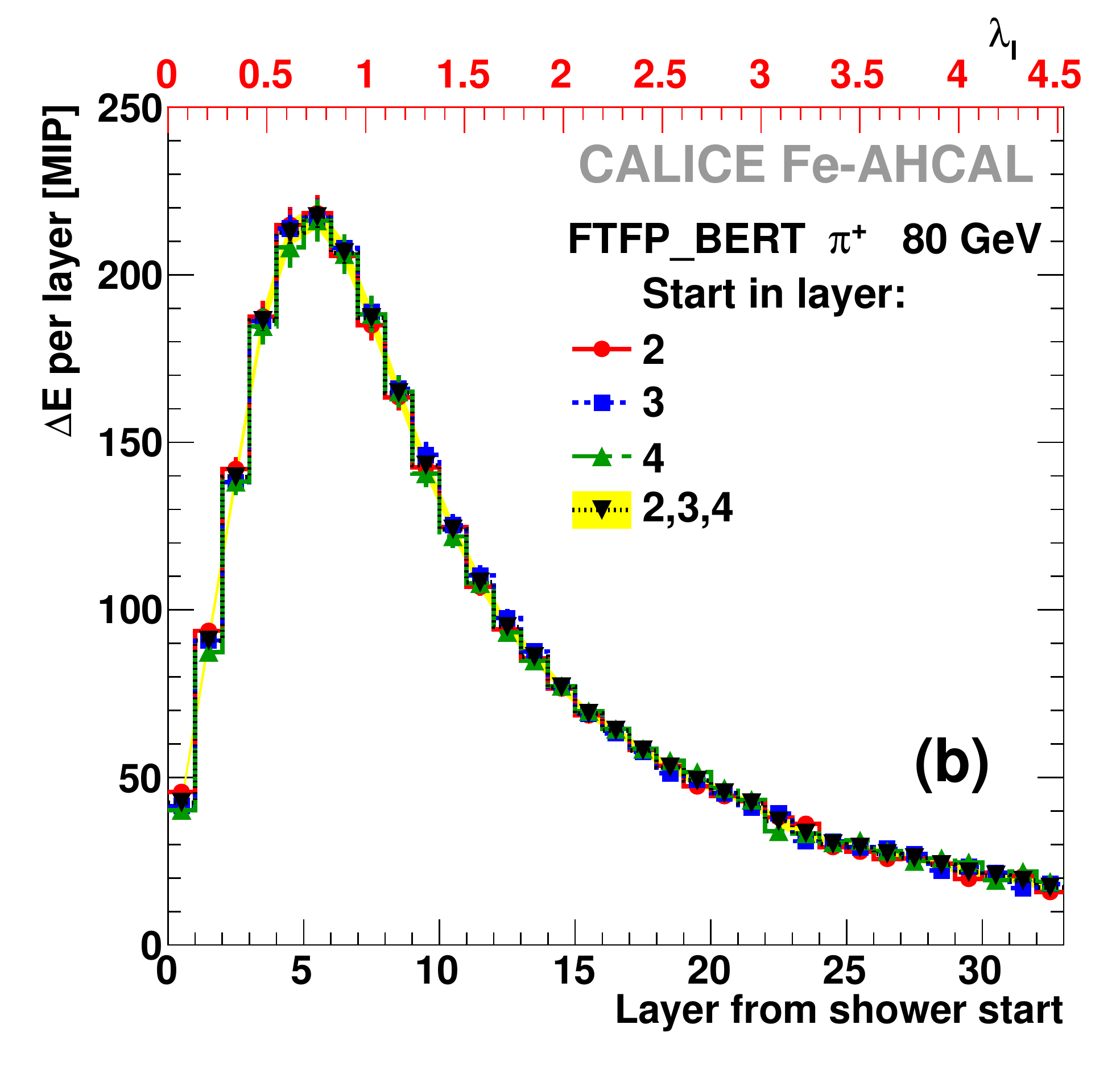}
 \caption{Longitudinal profiles of showers induced by pions with an initial energy of 80~GeV in the AHCAL obtained (a) from data and (b) simulated samples. The profiles for events with identified shower start in physical layer 2 (red circles), physical layer 3 (blue squares) or physical layer 4 (green triangles) are plotted separately. The black histogram shows the mean of the three coloured histograms. The yellow band corresponds to the systematic uncertainty (see appendix~\protect\ref{app:sysFluct} for details).}
 \label{fig:lngSysInter}
\end{figure}

Imperfections in the calibration procedure, temperature correction and saturation estimation as well as the presence of dead and noisy cells lead to variations of the measured response from layer to layer. As a result,  the longitudinal profile plotted from the calorimeter front is not perfectly smooth~\cite{Valid:2013}. The most significant contribution to these variations comes from the saturation correction procedure, which involves uncertainties in the SiPM response function. This statement is supported by the fact that variations increase with beam energy, the largest variations are observed at 80~GeV. The variations also depend on the conditions in a particular run, e.g. the temperature and the beam profile that determines which calorimeter cells are hit. The analysis of energy profiles from shower start helps to minimise layer-to-layer variations due to the averaged contributions from different physical layers. Nevertheless, these profiles still remain non-smooth when a narrow range of starting layers is used in the event selection (e.g. four or five in the current analysis), as shown in figure~\ref{fig:lngSysInter}a.

Figure~\ref{fig:lngSysInter} shows separate longitudinal profiles for events with shower start in a particular physical layer of the AHCAL, as well as the mean of these profiles. All profiles are normalised by the number of events. In case of an ideal calorimeter, these shower profiles should not depend on the shower start position (except for several last layers as the profiles starting later are shorter). Therefore, the difference between profiles with different shower start positions can be interpreted as a systematic uncertainty. To quantify the systematic uncertainty, the following procedure is used. The single profiles for different fixed shower start layers are considered to be a sample of size $n$, where $n$ is the number of separate profiles (e.g. $n =$3 in figure~\ref{fig:lngSysInter}). Each profile has $m$ bins. The content of the i-th bin of the j-th profile is $e_{ij}$, where $1<i<m$ and $1<j<n$. The bin content of the mean profile, averaged over the single profiles, is $E_{i} = \sum_{j=1}^{n}{e_{ij}}/n$. The variance of the bin content for the sample of the profiles is $s_{i} =  \sum_{j=1}^{n}{(e_{ij} - E_{i})^2}/(n-1)$. The mean of profiles is assumed to be an estimate of the true profile and is used for the analysis. The uncertainty in the i-th bin caused by variations between the single profiles is calculated as $\sqrt{s_{i}/n}$ and is shown with the yellow band in figure~\ref{fig:lngSysInter}.     

The same procedure is applied to the simulated samples for which the variations are observed to be smaller than for the data, as shown in figure~\ref{fig:lngSysInter}b. These layer-to-layer variations in the simulated samples appear because of dead cells and cell-wise noise addition in the digitisation procedure. At the same time, the main source of variations observed in the real calorimeter arises from imperfections in the saturation correction, which are not simulated. 

The same approach is applied to estimate the systematic uncertainties for radial profiles. Although the impact of layer-to-layer variations on radial profiles is smaller as they are integrated along the longitudinal coordinate, there is another source of systematic uncertainty, which is related to the determination of the shower axis and is discussed below in appendix~\ref{app:sysAxis}.

%===============================================
\subsection{Identification of the shower start layer}
\label{app:sysStart}

\begin{figure}
 \centering
 \includegraphics[width=7.5cm]{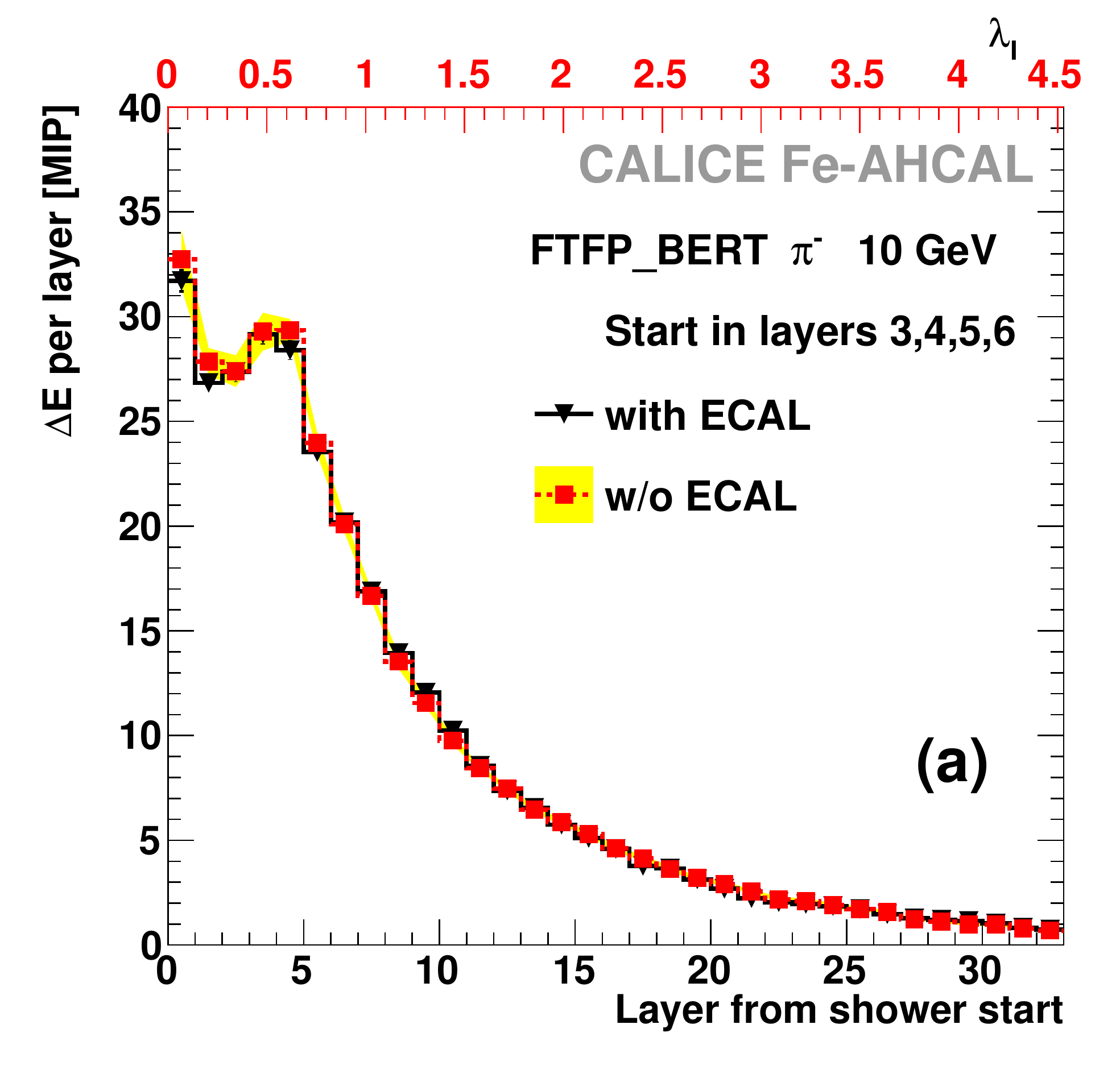}
 \includegraphics[width=7.5cm]{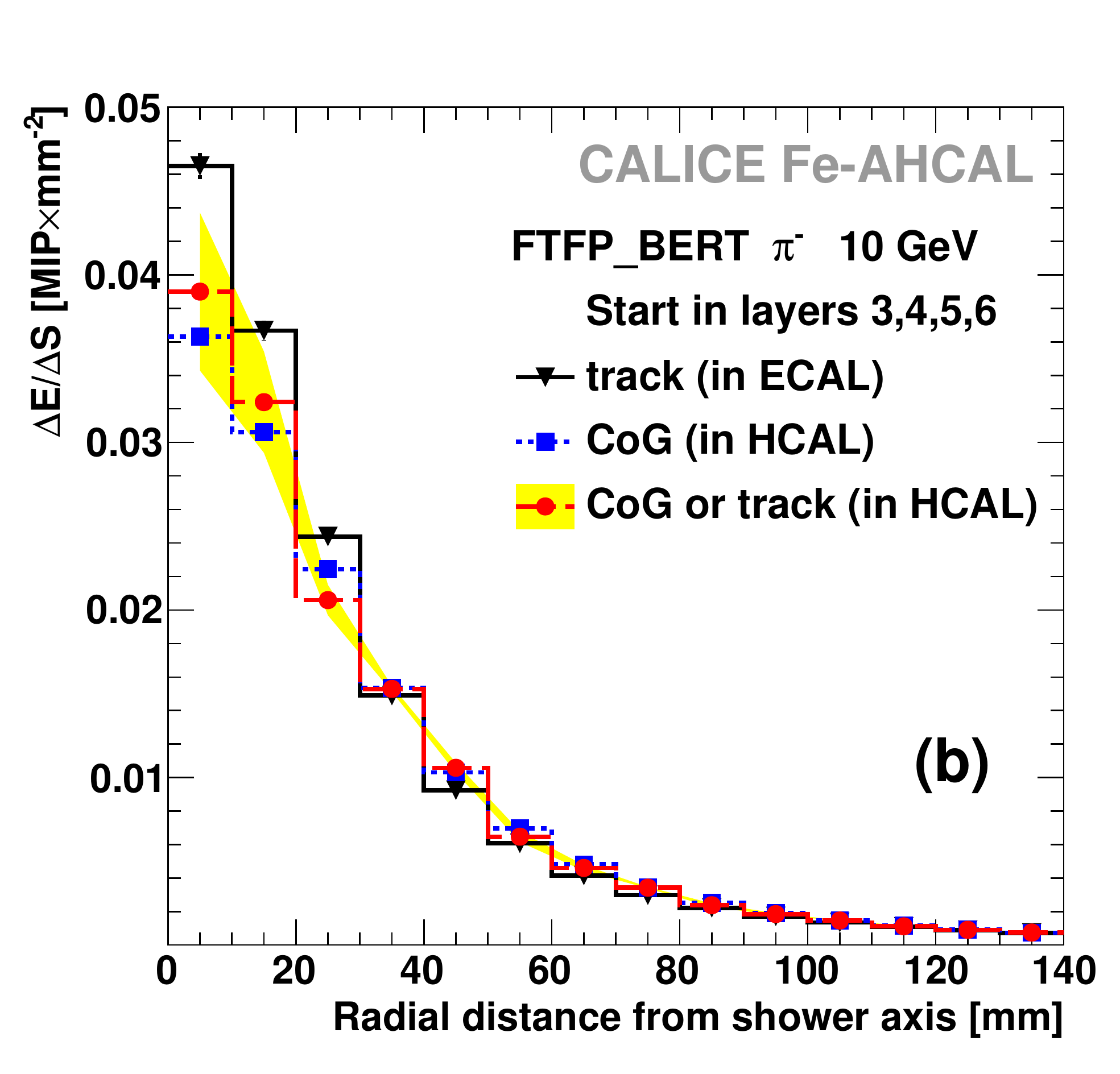}
 \caption{Shower profiles of simulated  10~GeV negative pions which start showering in physical layers 3,4,5 or 6 of the AHCAL: (a) longitudinal profiles obtained with (black triangles) and without (red circles) ECAL, (b) radial profiles with respect to shower axis estimated from track in ECAL (black triangles), shower centre of gravity (blue squares) and combined method (red circles). The yellow band shows the systematic uncertainty, see appendix~\protect\ref{app:sysFluct} and \protect\ref{app:sysAxis} for details.}
 \label{fig:sysStart}
\end{figure}

Uncertainties in the identification of the shower start layer are included in the distributions obtained from both data and simulated showers since the same identification procedure is applied to all samples. The accuracy of the identification algorithm degrades with decreasing energy and  its uncertainty is larger for the very first physical layers of the calorimeter setup.\footnote{ The algorithm includes a calculation of moving average of the energy deposition per layer within a window of 10 layers. For the first nine physical layers, it is assumed that the deposition before the first layer corresponds to the incoming track and equals to 1.3~MIP. The procedure is described in detail in ref.~\cite{AHCAL:2015pionProton}.} Therefore, the uncertainty of the algorithm can distort the energy dependence of shower profile parameters, thereby affecting the comparison of shower profiles at different energies. Thus, one can expect a bigger uncertainty contributed by the shower-start-finder algorithm for samples below 20~GeV taken at FNAL with the system configuration without the ECAL. In order to understand the impact of the algorithm, the simulated samples of negative pions for the two setup configurations have been compared as shown in figure~\ref{fig:sysStart}a, where the longitudinal profiles are plotted for selected events with a found shower start behind the second physical layer of the AHCAL. The differences between the shown profiles are within the uncertainties estimated in appendix~\ref{app:sysFluct} and are not introduced as additional systematic uncertainties in the analysis.  It should be noted that simulated profiles for positive and negative pions also coincide within statistical uncertainties.  

%===============================================
\subsection{Reconstruction of the shower axis}
\label{app:sysAxis}

The reference axis for the calculation of the radial shower profile is defined by the incoming track whose coordinates are reconstructed on an event-by-event basis. The coordinates of the primary track in the setup configuration with the ECAL are calculated from identified track hits with relatively good precision due to the high granularity of the ECAL (1$\times$1~cm$^2$ cells). This is not possible in the configuration without the ECAL, but two alternative approaches are available: (i) a search for the event's centre of gravity and (ii) identification of the incoming track in the AHCAL. A reliable track identification requires at least four points, hence it is not applicable to showers that start before the fifth physical layer of the AHCAL. Moreover, the transverse size of AHCAL central cells  is three times larger than that of the ECAL cell,  which leads to a lower accuracy of the primary-track position (shower axis) extracted from track hits in the AHCAL. The event's centre of gravity is identified with high precision due to the large number of hits (100 and more) but might be shifted with respect to the primary track due to either asymmetry of the shower or instrumental effects, e.g. dead or noisy cells. 

Figure \ref{fig:sysStart}b shows the radial profiles of simulated pion showers obtained using different methods of shower-axis identification. The reconstruction of the shower axis from tracks in the ECAL results in a higher energy density in the core region compared to the application of the centre of gravity, which underestimates the contribution near the shower axis. The red circles correspond to the combined method used in the current analysis where the shower axis is extracted from the centre of gravity for events with the shower start in physical layers 3 and 4 and from the track hits in the AHCAL for events with the start in physical layers 5 and 6. The systematic uncertainty shown with the yellow band in figure~\ref{fig:sysStart}b also includes the uncertainty related to differences between the two methods of shower-axis reconstruction. As follows from figure~\ref{fig:sysStart}b, the most significant contribution of these uncertainties lies in the region near the shower axis where they amount up to 10\%.         

%===============================================
\subsection{Pion contamination of the proton samples}
\label{app:sysProton}

The inefficiency of the \u{C}erenkov counter leads to a contamination with pions of the proton samples in the test beam data. The estimation of the \u{C}erenkov counter efficiency is based on an independent procedure of identifying muons in the same run. The efficiency estimates are then used to calculate the purity of the proton samples $\eta$, as described in ref.~\cite{AHCAL:2015pionProton}. The values of $\eta$ in the current analysis vary from 74\% to 95\%. We assume that the analysed pion samples are not contaminated and correct the proton profiles taking into account the proton sample purity $\eta$ estimated for the given run. The content of each bin of the proton profile is corrected by subtracting the estimated contribution obtained from the pure pion profile in the following way:

\begin{equation}
 \Delta E^{\mathrm{corr}}_i = \Delta E^{\mathrm{mix}}_i \cdot \frac{1}{\eta} - \Delta E^{\pi}_i \cdot \frac{1-\eta}{\eta},
\label{eq:sysProton}
\end{equation}  

\noindent where
 $\Delta E^{\mathrm{corr}}_i$ is the corrected content of the $i$-th bin of the proton profile,
 $\Delta E^{\mathrm{mix}}_i$ is the content of the $i$-th bin in the profile for the mixed sample and
 $\Delta E^{\pi}_i$ is the content of the $i$-th bin in the profile obtained for pions of the same energy. The same correction procedure has been applied to longitudinal and radial profiles of proton-induced showers in test beam data.
The resulting uncertainty of the corrected energy deposition in the particular bin is calculated using standard error propagation techniques taking into account the estimated statistical and systematic uncertainties of all variables involved: $\Delta E^{\mathrm{mix}}_i$, $\Delta E^{\pi}_i$ and $\eta$.  

%===============================================
\subsection{Positron contamination in the samples taken without ECAL} 
\label{app:sysPositron}

\begin{figure}
 \centering
 \includegraphics[width=7.5cm]{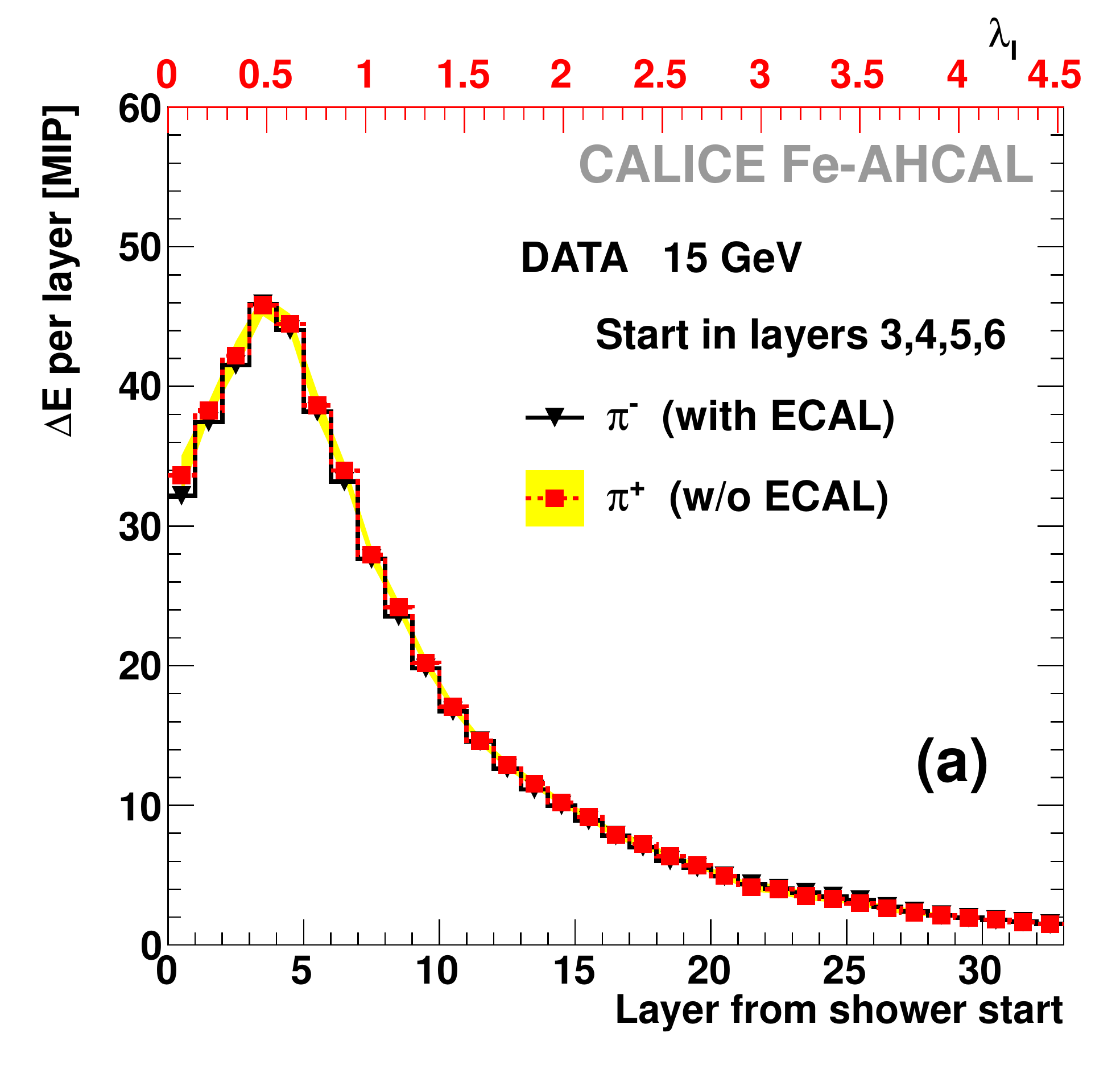}
 \includegraphics[width=7.5cm]{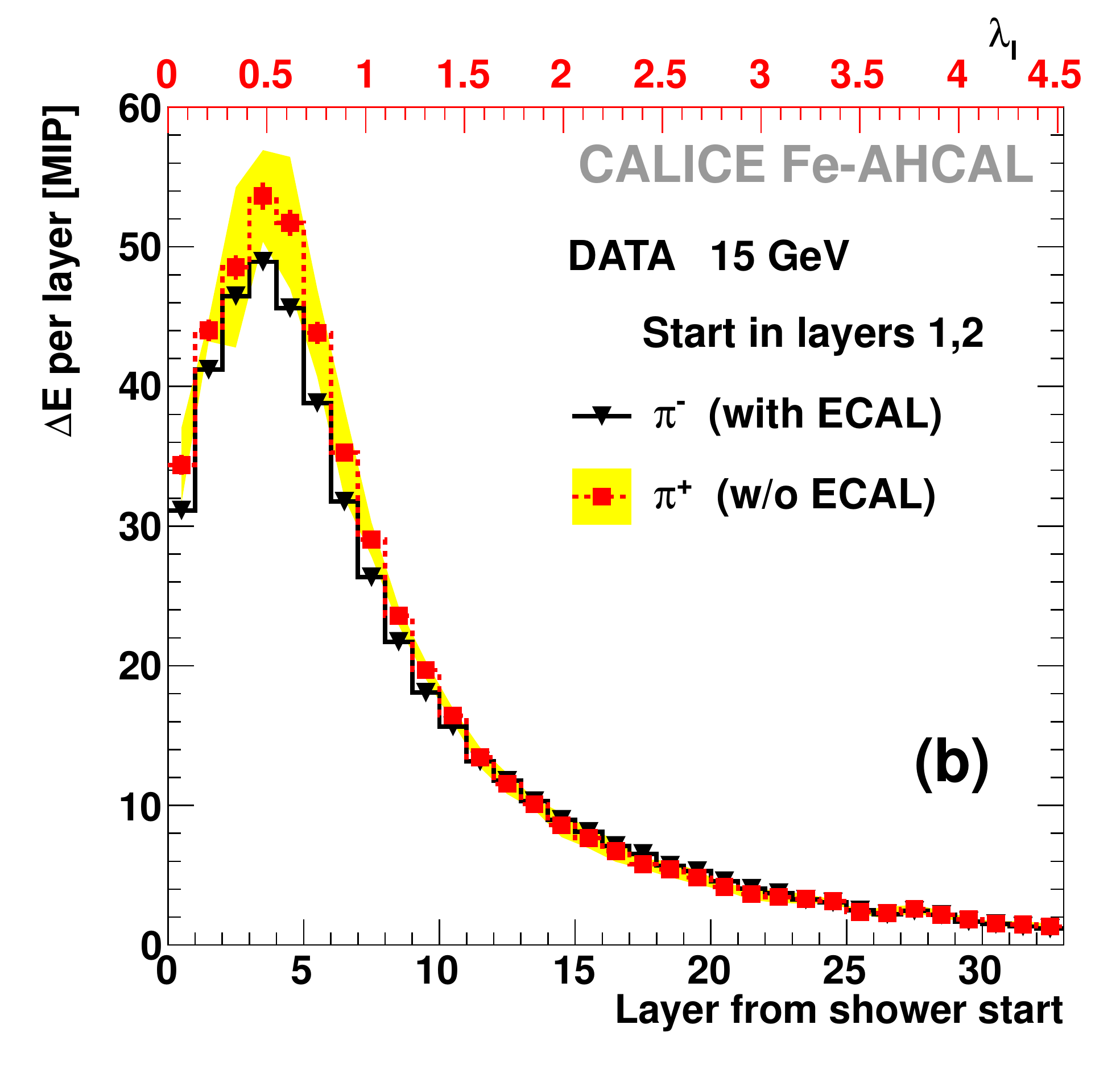}
 \caption{Longitudinal profiles of showers induced by negative pions from test beam data (setup with ECAL, black triangles) and positive pions (setup w/o ECAL, red circles) with initial energy 15~GeV for selected events with the identified shower start in physical layers (a) 3,4,5,6 and (b) 1,2 of the AHCAL. The yellow band corresponds to the systematic uncertainty, see appendix~\protect\ref{app:sysFluct} for details.}
 \label{fig:sysPositron}
\end{figure}

The positron contamination of hadron samples in the runs taken without ECAL can be significantly reduced by the selection procedure described in~\cite{AHCAL:2015pionProton}. The additional selection of events with shower start behind the second AHCAL layer helps to remove remaining positrons.  
Figure~\ref{fig:sysPositron} shows the longitudinal profiles from shower start for pions with an initial energy of 15~GeV for data taken with different setup configurations: with the ECAL in front of the AHCAL for negative pions and without the ECAL for positive pions. Due to the presence of the ECAL, the negative pion samples are assumed to have no electron contamination. 

The difference between profiles shown in figure~\ref{fig:sysPositron}a for selected events with the shower start in physical layers 3, 4, 5, 6 of the Fe-AHCAL lies within the uncertainty estimated in appendix~\ref{app:sysFluct}. For comparison, figure~\ref{fig:sysPositron}b demonstrates the difference between profiles for selected events with the shower start in physical layers 1 and 2. Therefore, one can conclude that the positron contamination does not affect shower profiles for selected events with the shower start positions beyond the second physical layer of the AHCAL and hence will not be introduced as an additional systematic uncertainty.

%===============================================
\subsection{Longitudinal leakage from the AHCAL}
\label{app:leakage}

The selection of the shower start at the beginning of the AHCAL helps to minimise longitudinal leakage. Nevertheless, the AHCAL depth of $\sim$5.3$\lambda_{\mathrm{I}}$ is not enough to contain the entire hadronic shower, even for initial pion energies as low as 30~GeV. The longitudinal fit range in the current analysis corresponds to a depth of $\sim$4.5$\lambda_{\mathrm{I}}$ from the shower start as only bins that belong to the Fe-AHCAL are used.
The restriction of the range used to fit the shower profile can affect the estimation of parameters. 

The uncertainty related to the cut on the fit range of the longitudinal profiles was estimated by the following procedure. First, the histogram was generated from an analytical function used for the profile parametrisation (see section~\ref{sec:paraLng}) with the parameters obtained from the fit to data or simulations. The bin contents were then smeared within real uncertainties, these uncertainties being assigned to bin errors. Then the generated histogram was fit with different upper limits of the fit range (from 4$\lambda_{\mathrm{I}}$ to 11$\lambda_{\mathrm{I}}$) and the fit results were compared. The differences in the obtained parameters were found to be more than 3 times smaller than the uncertainties of the fit.

%%%%%%%%%%%%%%%%%%%%%%%%%%%%%%%%%%%%%%%%%%%%%%%%%%%%%%%%%%%%%%
\section{Values of fit parameters}
\label{app:parametr}

The values of parameters extracted from the fit to the longitudinal and radial profiles of showers induced by pions and protons at different energies in the range from 10 to 80~GeV are listed in the tables below. The fit results for the data samples are shown in table \ref{tab:parData}. Tables \ref{tab:parFTFP} and \ref{tab:parQGSP} represent the fit results for the samples simulated using the {\sffamily FTFP\_BERT} and {\sffamily QGSP\_BERT} physics lists from {\scshape Geant4} version 9.6. 
 
\begin{table}
 \caption{Parameters from the fit to the longitudinal and radial profiles extracted from the data samples.}
 \label{tab:parData}
 \begin{center}
  \begin{tabular}{|c|c|c|c|c|c|c|c|}
   \hline
 $p_{\mathrm{beam}}$, & $\alpha_{\mathrm{short}}$ &  $\beta_{\mathrm{short}}$, & $f$ & $\alpha_{\mathrm{long}}$ & $\beta_{\mathrm{long}}$, & $\beta_{\mathrm{core}}$, & $\beta_{\mathrm{halo}}$, \\
 GeV/$c$ & & $X_{0}$ &  & & $\lambda_{\mathrm{I}}$ & mm & mm \\
   \hline
\multicolumn{8}{|c|}{Proton} \\
   \hline
10 &    -          &    -         &          -     &  0.93$\pm$0.04 &  1.29$\pm$0.06 &  29.9$\pm$1.9 &   84$\pm$6  \\
15 &  12.6$\pm$6.3 &  0.5$\pm$0.2 &  0.05$\pm$0.01 &  1.11$\pm$0.03 &  1.24$\pm$0.04 &  27.5$\pm$1.1 &   84$\pm$3  \\
30 &   5.8$\pm$0.7 &  1.2$\pm$0.2 &  0.12$\pm$0.01 &  1.32$\pm$0.01 &  1.29$\pm$0.02 &  22.4$\pm$0.6 &   77$\pm$1  \\
40 &   5.8$\pm$1.0 &  1.2$\pm$0.2 &  0.11$\pm$0.01 &  1.42$\pm$0.02 &  1.29$\pm$0.03 &  22.1$\pm$0.6 &   77$\pm$1  \\
50 &   5.1$\pm$0.8 &  1.4$\pm$0.2 &  0.13$\pm$0.01 &  1.47$\pm$0.02 &  1.32$\pm$0.03 &  21.7$\pm$0.6 &   76$\pm$2  \\
60 &   4.6$\pm$0.7 &  1.6$\pm$0.3 &  0.14$\pm$0.02 &  1.53$\pm$0.02 &  1.29$\pm$0.03 &  21.3$\pm$0.5 &   76$\pm$1  \\
80 &   4.5$\pm$0.6 &  1.7$\pm$0.2 &  0.14$\pm$0.01 &  1.60$\pm$0.02 &  1.30$\pm$0.03 &  20.8$\pm$0.5 &   74$\pm$1  \\

   \hline
\multicolumn{8}{|c|}{$\pi^{+}$} \\
   \hline
10 &   4.6$\pm$0.5 &  1.4$\pm$0.2 &  0.19$\pm$0.02 &  0.92$\pm$0.02 &  1.30$\pm$0.04 &  25.0$\pm$0.7 &   81$\pm$2  \\
15 &   4.5$\pm$0.4 &  1.5$\pm$0.1 &  0.20$\pm$0.02 &  1.10$\pm$0.02 &  1.22$\pm$0.03 &  23.4$\pm$0.6 &   78$\pm$2  \\
30 &   4.5$\pm$0.3 &  1.6$\pm$0.1 &  0.23$\pm$0.01 &  1.35$\pm$0.02 &  1.24$\pm$0.02 &  20.6$\pm$0.5 &   76$\pm$1  \\
40 &   4.3$\pm$0.3 &  1.8$\pm$0.1 &  0.23$\pm$0.01 &  1.44$\pm$0.01 &  1.27$\pm$0.02 &  20.2$\pm$0.5 &   75$\pm$1  \\
50 &   4.2$\pm$0.2 &  1.9$\pm$0.1 &  0.24$\pm$0.01 &  1.50$\pm$0.01 &  1.27$\pm$0.02 &  19.9$\pm$0.4 &   74$\pm$1  \\
60 &   4.2$\pm$0.3 &  1.9$\pm$0.2 &  0.23$\pm$0.01 &  1.56$\pm$0.02 &  1.24$\pm$0.02 &  19.7$\pm$0.4 &   74$\pm$1  \\
80 &   4.4$\pm$0.3 &  1.9$\pm$0.2 &  0.23$\pm$0.01 &  1.61$\pm$0.01 &  1.26$\pm$0.03 &  19.5$\pm$0.4 &   74$\pm$1  \\

  \hline 
  \end{tabular}
 \end{center}
\end{table}

\begin{table}
 \caption{Parameters from the fit to the longitudinal and radial profiles extracted from the samples simulated using the {\sffamily FTFP\_BERT} physics list.}
 \label{tab:parFTFP}
 \begin{center}
  \begin{tabular}{|c|c|c|c|c|c|c|c|}
   \hline
 $p_{\mathrm{beam}}$, & $\alpha_{\mathrm{short}}$ &  $\beta_{\mathrm{short}}$, & $f$ & $\alpha_{\mathrm{long}}$ & $\beta_{\mathrm{long}}$, & $\beta_{\mathrm{core}}$, & $\beta_{\mathrm{halo}}$, \\
 GeV/$c$ & & $X_{0}$ &  & & $\lambda_{\mathrm{I}}$ & mm & mm \\
   \hline
\multicolumn{8}{|c|}{Proton} \\
   \hline
10 &  19.6$\pm$4.5 &  0.3$\pm$0.1 &  0.034$\pm$0.004 &  0.94$\pm$0.01 &  1.20$\pm$0.02 &  27.2$\pm$0.8 &   84$\pm$2  \\
15 &  10.9$\pm$1.1 &  0.6$\pm$0.1 &  0.064$\pm$0.004 &  1.10$\pm$0.01 &  1.22$\pm$0.01 &  25.8$\pm$0.7 &   82$\pm$2  \\
30 &   6.4$\pm$0.6 &  1.1$\pm$0.1 &  0.10$\pm$0.01 &  1.34$\pm$0.01 &  1.27$\pm$0.02 &  22.0$\pm$0.6 &   77$\pm$1  \\
40 &   5.4$\pm$0.5 &  1.4$\pm$0.1 &  0.12$\pm$0.01 &  1.43$\pm$0.01 &  1.30$\pm$0.02 &  21.5$\pm$0.5 &   77$\pm$1  \\
50 &   5.0$\pm$0.3 &  1.6$\pm$0.1 &  0.14$\pm$0.01 &  1.47$\pm$0.01 &  1.37$\pm$0.03 &  21.1$\pm$0.5 &   76$\pm$1  \\
60 &   5.2$\pm$0.4 &  1.5$\pm$0.1 &  0.14$\pm$0.01 &  1.52$\pm$0.01 &  1.33$\pm$0.02 &  20.8$\pm$0.5 &   76$\pm$1  \\
80 &   4.8$\pm$0.4 &  1.7$\pm$0.1 &  0.14$\pm$0.01 &  1.60$\pm$0.02 &  1.34$\pm$0.03 &  20.4$\pm$0.5 &   75$\pm$1  \\
   \hline
\multicolumn{8}{|c|}{$\pi^{+}$} \\
   \hline
10 &   6.4$\pm$0.6 &  1.0$\pm$0.1 &  0.14$\pm$0.01 &  0.95$\pm$0.01 &  1.22$\pm$0.02 &  24.0$\pm$0.7 &   78$\pm$2  \\
15 &   6.2$\pm$0.4 &  1.1$\pm$0.1 &  0.17$\pm$0.01 &  1.13$\pm$0.01 &  1.18$\pm$0.02 &  22.6$\pm$0.6 &   78$\pm$1  \\
30 &   4.4$\pm$0.2 &  1.8$\pm$0.1 &  0.27$\pm$0.01 &  1.33$\pm$0.02 &  1.30$\pm$0.03 &  19.9$\pm$0.5 &   76$\pm$1  \\
40 &   4.5$\pm$0.2 &  1.8$\pm$0.1 &  0.28$\pm$0.01 &  1.40$\pm$0.02 &  1.32$\pm$0.03 &  19.1$\pm$0.5 &   74$\pm$1  \\
50 &   4.5$\pm$0.2 &  1.8$\pm$0.1 &  0.28$\pm$0.01 &  1.48$\pm$0.02 &  1.30$\pm$0.03 &  19.0$\pm$0.4 &   74$\pm$1  \\
60 &   4.7$\pm$0.2 &  1.7$\pm$0.1 &  0.28$\pm$0.01 &  1.54$\pm$0.02 &  1.27$\pm$0.03 &  18.7$\pm$0.4 &   73$\pm$1  \\
80 &   4.3$\pm$0.2 &  2.0$\pm$0.1 &  0.31$\pm$0.01 &  1.56$\pm$0.02 &  1.37$\pm$0.04 &  18.4$\pm$0.4 &   73$\pm$1  \\
  \hline 
  \end{tabular}
 \end{center}
\end{table}

\begin{table}
 \caption{Parameters from the fit to the longitudinal and radial profiles extracted from the samples simulated using the {\sffamily QGSP\_BERT} physics list.}
 \label{tab:parQGSP}
 \begin{center}
  \begin{tabular}{|c|c|c|c|c|c|c|c|}
   \hline
 $p_{\mathrm{beam}}$, & $\alpha_{\mathrm{short}}$ &  $\beta_{\mathrm{short}}$, & $f$ & $\alpha_{\mathrm{long}}$ & $\beta_{\mathrm{long}}$, & $\beta_{\mathrm{core}}$, & $\beta_{\mathrm{halo}}$, \\
 GeV/$c$ & & $X_{0}$ &  & & $\lambda_{\mathrm{I}}$ & mm & mm \\
   \hline
\multicolumn{8}{|c|}{Proton} \\
   \hline
10 &  10.0$\pm$1.3 &  0.6$\pm$0.1 &  0.07$\pm$0.01 &  0.97$\pm$0.01 &  1.18$\pm$0.02 &  26.9$\pm$0.9 &   83$\pm$2  \\
15 &   7.9$\pm$0.9 &  0.8$\pm$0.1 &  0.08$\pm$0.01 &  1.08$\pm$0.01 &  1.28$\pm$0.02 &  25.0$\pm$0.7 &   82$\pm$2  \\
30 &   4.0$\pm$0.2 &  1.7$\pm$0.1 &  0.23$\pm$0.01 &  1.35$\pm$0.02 &  1.24$\pm$0.02 &  20.7$\pm$0.5 &   77$\pm$1  \\
40 &   3.8$\pm$0.2 &  1.9$\pm$0.1 &  0.26$\pm$0.01 &  1.45$\pm$0.02 &  1.26$\pm$0.02 &  20.2$\pm$0.5 &   76$\pm$1  \\
50 &   4.1$\pm$0.2 &  1.8$\pm$0.1 &  0.24$\pm$0.01 &  1.55$\pm$0.02 &  1.18$\pm$0.02 &  19.7$\pm$0.5 &   75$\pm$1  \\
60 &   4.0$\pm$0.2 &  1.9$\pm$0.1 &  0.25$\pm$0.01 &  1.57$\pm$0.02 &  1.21$\pm$0.02 &  19.5$\pm$0.4 &   75$\pm$1  \\
80 &   4.1$\pm$0.2 &  1.9$\pm$0.1 &  0.26$\pm$0.01 &  1.63$\pm$0.02 &  1.21$\pm$0.02 &  19.1$\pm$0.4 &   75$\pm$1  \\
   \hline
\multicolumn{8}{|c|}{$\pi^{+}$} \\
   \hline
10 &   5.7$\pm$0.6 &  1.1$\pm$0.1 &  0.15$\pm$0.01 &  0.94$\pm$0.01 &  1.18$\pm$0.02 &  25.0$\pm$0.8 &   81$\pm$2  \\
15 &   4.8$\pm$0.3 &  1.4$\pm$0.1 &  0.20$\pm$0.01 &  1.08$\pm$0.01 &  1.24$\pm$0.02 &  22.9$\pm$0.6 &   79$\pm$1  \\
30 &   3.9$\pm$0.1 &  1.9$\pm$0.1 &  0.36$\pm$0.01 &  1.33$\pm$0.02 &  1.31$\pm$0.03 &  18.9$\pm$0.4 &   75$\pm$1  \\
40 &   3.9$\pm$0.1 &  1.9$\pm$0.1 &  0.36$\pm$0.01 &  1.47$\pm$0.02 &  1.22$\pm$0.03 &  18.3$\pm$0.4 &   74$\pm$1  \\
50 &   4.0$\pm$0.1 &  2.0$\pm$0.1 &  0.37$\pm$0.01 &  1.49$\pm$0.02 &  1.29$\pm$0.03 &  18.4$\pm$0.4 &   74$\pm$1  \\
60 &   4.1$\pm$0.1 &  1.9$\pm$0.1 &  0.36$\pm$0.01 &  1.54$\pm$0.02 &  1.26$\pm$0.03 &  18.1$\pm$0.4 &   74$\pm$1  \\
80 &   4.2$\pm$0.1 &  1.9$\pm$0.1 &  0.34$\pm$0.01 &  1.61$\pm$0.02 &  1.24$\pm$0.03 &  17.9$\pm$0.4 &   73$\pm$1  \\
  \hline 
  \end{tabular}
 \end{center}
\end{table}

\end{document}